\documentclass{aimsessay}
\begin{document}
\pagestyle{empty}
\maketitle
\pagenumbering{roman}
\chapter*{\textbf{DECLARATION}}
\addcontentsline{toc}{chapter}{Declaration}

I, the undersigned, hereby declare that the work contained in this M.Sc. dissertation is my original work and that any work done by others or by myself previously has been acknowledged and referenced accordingly. This dissertation is submitted to the University of the Witwatersrand-Johannesburg, in fulfillment of the requirements for the degree of Master of Science. It has not been submitted before for any degree or examination in any other university.

\vspace{1.5cm}
 \includegraphics[height=1.5cm]{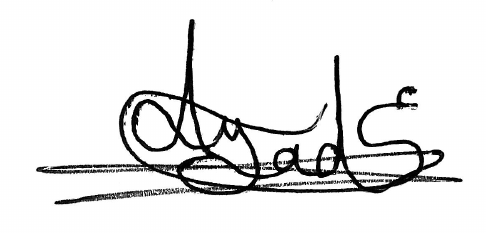}\newline
\hrule
Ahmed Ayad Mohamed Ali, \makebox[2in][r] 27 February 2017
\newpage
\newpage
\chapter*{\textbf{ACKNOWLEDGEMENTS}}
\addcontentsline{toc}{chapter}{Acknowledgements}
\par{This master dissertation has been carried out at the University of the Witwatersrand, since February 2017. After Allah, a number of people deserve thanks for their support and help. It is, therefore, my greatest pleasure to express my gratitude to them all in this acknowledgment.}
\par{First and foremost I would like to thank Allah. In the process of putting this dissertation together, I realized how true this gift of writing is for me. You have given me the power to believe in my passion and pursue my dreams. I could never have done this without the faith I have in you, the Almighty.}
\par{I would like to express my sincere gratitude to my supervisor Prof. Kevin Goldstein, for the patient guidance, encouragement, and advice he has provided throughout my time as his student. I have been extremely lucky to have a supervisor who cared so much about my work, and who responded to my questions and queries so promptly.}  
\par{My special thanks go to Prof. Robert de Mello, Prof. Eslam El-Sheikh, and Dr. Emad Sultan, for their excellent and patient technical assistance, for believing in my potential and for the nice moments we spent together.} 
\par{ Finally, I must express my very profound gratitude to my family, friends, and colleagues, for providing me with unfailing support and continuous encouragement throughout my study duration and through the process of researching and writing this dissertation. This accomplishment would not have been possible without them.}
\par{Thank you all, for always being there for me.}
\newpage
\chapter*{\textbf{DEDICATION}} 
\addcontentsline{toc}{chapter}{Dedication}

\vspace*{\fill}
\begin{center}
\begin{minipage}{.8\textwidth}{\begin{center}
{\large \bf This dissertation is dedicated to my Mother.} \\
{\bf For her endless love, support and encouragement at all times.}\end{center}} 
\end{minipage}
\end{center}
\vfill 

\chapter*{\textbf{Abstract}} 
\addcontentsline{toc}{chapter}{Abstract}
\par{Supersymmetry plays a main role in all current thinking about superstring theory. Indeed, many remarkable properties of string theory have been explained using supersymmetry as a tool. In this dissertation, we review the basic formulation of supersymmetric quantum mechanics starting with introducing the concepts of supercharges and superalgebra. We show that, if there is a supersymmetric state, it is the zero-energy ground state. If such a state exists, the supersymmetry is unbroken otherwise it is broken. So far, there has been no unbroken supersymmetry observed in nature, and if nature is described by supersymmetry, it must be broken. In fact, supersymmetry may be broken spontaneously at any order of perturbation theory, or dynamically due to non-perturbative effects. The goal of this dissertation is to study the methods of supersymmetry breaking. For this purpose, special attention is given to discuss the normalization of the ground state of the supersymmetric harmonic oscillator. Then we explain that perturbation theory gives us incorrect results for both the ground state wave function as well as the energy spectrum and it fails to give an explanation to the supersymmetry breaking. Later in the dissertation, a review of the uses of instantons in quantum mechanics is given. In particular, instantons are used to compute the tunneling effects within the path integral approach to quantum mechanics. As a result, we give evidence that the instantons, which are a non-perturbative effect in quantum mechanics and can not be seen in perturbation theory, leads  to calculate the corrections to the ground state energy and provides a possible explanation for the supersymmetry breaking.} 

\tableofcontents
\listoffigures
\addcontentsline{toc}{chapter}{\listfigurename}
\newpage
\pagenumbering{arabic}
\pagestyle{myheadings}
\chapter{\textbf{Introduction}}

\par{There are two major topics discussed in this dissertation. The first is supersymmetry in quantum mechanics, the second is the path integral realization of instantons in both real-time as well as Euclidean-time.}

\par{Supersymmetry, often abbreviated SUSY, is a quantum mechanical space-time symmetry mapping particles and fields of integer spins, the bosons, between particles and fields of half-integer spins, the fermions, and vice versa \cite{bertolini2011lectures}. In 1971, supersymmetry was first introduced by Gel’fand and Likhtman, Raymond, and Neveu and Schwartz, and later it was rediscovered by other groups \cite{ engbrant2012supersymmetric, Wellman}. Supersymmetric quantum mechanics was developed by Witten in 1982 \cite{witten1981dynamical}, as a toy model to test the breaking of supersymmetry. In general, supersymmetry plays a main role in the modern understanding of theoretical physics, in particular, quantum field theories, gravity theories, and string theories \cite{dumitrescu2013topics}.} 

\par{The path integral formulation of quantum mechanics offers an alternative point of view on quantum field theory \cite{zinn2010path}. In 1948, the path integral method was first developed by Richard Feynman \cite{freed1999five}. The path integral formulation of quantum mechanics leads to a new and deeper understanding of the relation between quantum and classical mechanics. As well, it provides new tools to study quantum mechanics in the semi-classical limit \cite{brack2003semiclassical}.}

\par{ This dissertation is structured as follows: in chapter \ref{ch2}, we start by introducing the algebra of anticommuting (Grassmann) variables. Next, we study the problem of a harmonic oscillator with fermionic as well as bosonic fields using the usual quantum mechanical operators method. Then, we consider supersymmetric classical mechanics and study generalized classical Poisson brackets to Dirac brackets and quantization rules in order to introduce the superalgebra. Furthermore, the formalism of supersymmetry in quantum mechanics is introduced. Subsequently, we give a general background in the concepts and methods of supersymmetric quantum mechanics. This followed by providing more specific study of the ground state of the supersymmetric harmonic oscillator.} 

\par{In chapter \ref{ch3}, we first give a brief review of the basic concepts of the path integral formulation of quantum mechanics. This is followed by introducing the concept of Euclidean path integral. To exemplify the use of the path integral method, we discuss the harmonic oscillator problem. Later the instanton method is described and applied to quantum mechanics of a double-well potential. Instanton effects are responsible for one of the most important quantum mechanical effects: tunneling through a potential barrier. Tunneling is discussed from the path integral point of view using the instantons method. Real-time path integral realization of instantons is considered in the last section of this chapter. Our conclusion is given in chapter \ref{ch4}. For completeness, some needed derivations are broadly outlined in the appendix.}

\par{ Finally, it needs to be pointed out that the purpose of this dissertation is to provide a clear and simple introduction of the topics of supersymmetry and path integral formulations in quantum mechanics. However, the dissertation is constructed to be comprehensible and understandable to someone with a decent background in quantum mechanics, but no previous knowledge of either supersymmetry nor path integrals.} 
\chapter{\textbf{Supersymmetric Quantum Mechanics}} \label{ch2}

\par{
Supersymmetry is a symmetry that relates bosons to fermions \cite{engbrant2012supersymmetric}. In particular, it is possible to introduce operators that change bosons which are commuting variables into fermions which are anticommuting variables and vice versa \cite{cooper1995supersymmetry}. Supersymmetric theories are then theories which are invariant under those transformations \cite{batzing2013lecture}. The supersymmetry algebra involves commutators as well as anticommutators \cite{krippendorf2010cambridge}. In this chapter, the main mathematical structure which involves supersymmetric quantum mechanics is derived starting with explaining the basic idea of supersymmetry and followed by introducing the necessary framework to make a supersymmetric theory.
} 

\section{Grassmann Numbers}

\par{The ideas in this section were introduced in \cite{wasay2010supersymmetric, wasay2016supersymmetric, morgan2003supersymmetric,seiberg2003noncommutative}. Supersymmetric theories necessarily include both bosons and fermions. Indeed we let to consider a system that includes both bosonic and fermionic degrees of freedom. The supersymmetry algebra is a mathematical formalism for describing the relation between bosons and fermions, we will discuss this later in Section \ref{Sec2.4}.}

\par{In quantum mechanics, bosons are integer spin particles which obey Bose-Einstein statistics. The bosonic operators satisfy the usual commutation relations
  \begin{align}
 &&&&&&&& [q_i,q_j]&= 0  & \Longleftrightarrow & & q_i q_j - q_j q_i = 0. & &&&&&&&&
  \end{align} 
  
  }
    
\par{
On the other hand, Fermions are half-integer spin particles characterized by Fermi–Dirac statistics and obey the Pauli exclusion principle. Fermions can be described by Grassmann variables, which are anticommuting objects. So it is important to describe the main characteristic properties of Grassmann number.
\begin{itemize} 
\item[(i)] Two Grassmann numbers, $\psi_1,\psi_2$ satisfy the following anticommutation relations
  \begin{align}
&&&&&&&&  \{\psi_i,\psi_j\} &= 0 & \Longleftrightarrow & & \psi_i \psi_j + \psi_j \psi_i = 0. & &&&&&&&&
  \end{align}
\item[(ii)] The above relations imply $\psi_i^2=0$.
\item[(iii)] Grassmann numbers commute with ordinary numbers. Therefore, a Grassmann number $\psi$ and an ordinary number of $q$ must satisfy the commutation relation
  \begin{align}
&&&&&&&& [\psi,q]&=0 & \Longleftrightarrow & & \psi q= q \psi. & &&&&&&&&
  \end{align}  
\item[(iv)] A Grassmann number, $\psi$, is called real if $\psi^{*}=\psi$ and imaginary if $\psi^{*} = - \psi$.  
  \end{itemize}
    
  }

\section{Harmonic Oscillator} \label{sec2.2}
\par{
In this section, we will describe the formalism of the one–dimensional harmonic oscillator which consists of a combination of bosonic and fermionic fields. There are many excellent reviews on this subject, we refer the reader to them for more and complete expositions \cite{Wellman,bagchi2000supersymmetry,hamdouni2005supersymmetry,nakahara2003geometry}. We begin by considering the Lagrangian formulation for the bosonic and fermionic harmonic oscillator
\begin{equation}\label{2.4}
   \mathit{L} = \frac{1}{2} \left( \dot{q}^2 + i \psi_{\alpha} \delta^{\alpha \beta} \dot{\psi}_{\beta} + \mathit{F}^2 \right) + \omega \left( i \psi_1 \psi_2 + q \mathit{F} \right),
  \end{equation}    
where $q(t)$ and $\mathit{F}(t)$ are real bosonic fields, $\psi_{\alpha}(t)$ (with $\alpha \in \{1,2\}$) are real fermionic fields, $\omega$ is a real parameter, $\delta^{\alpha \beta}$ is the Kronecker delta and we used the Einstein summation convention.
}
 
\par{
It is straightforward, using the classical equation of motion for $\mathit{F}+ \omega q =0$, to eliminate $\mathit{F}$ from Eq. (\ref{2.4}) and obtain the equivalent Lagrangian
\begin{equation} \label{2.5}
\mathit{L} = \frac{1}{2} \left( \dot{q}^2 + i \psi_{\alpha} \delta^{\alpha \beta} \dot{\psi}_{\beta} - \omega^2 q^2 \right) + i \omega \psi_1 \psi_2 .
\end{equation} 

} 
 
\par{
Then using the Legendre transformation from the variables $\{q,\dot{q},\psi,\dot{\psi}\}$ to $\{q,p,\psi,\pi\}$ we obtain
  \begin{equation}\label{2.6}
  \mathit{H}=\dot{q}p+\psi_{\alpha} \delta^{\alpha \beta} \pi^{\beta} - \mathit{L} = \frac{1}{2} \left( p^2 + \omega^2 q^2 \right) - i \omega \psi_1 \psi_2,
  \end{equation}
where $\pi^{\alpha}$ are the momenta conjugate to $\psi_{\alpha}$.}
\par{
Later in this chapter, we shall see that the general properties of supersymmetry have to be clear if the Hamiltonian is built as a larger Hamiltonian consisting of two components, one representing the bosonic field and the other representing the fermionic field. Leaving this aside for the moment, to derive the supersymmetric Hamiltonian let us define the bosonic valuable $\hat{q}$ and the bosonic momentum $\hat{p}$, respectively, as follows
\begin{align} \label{2.7}
&&&& \hat{q} &= q \mathbbmtt{I}_2, &&  \hat{p} = - i \hbar \mathbbmtt{I}_2 \dfrac{\partial}{\partial q}. &&&&
\end{align}
On the other hand, the fermionic variables $\hat{\psi}_1$, $\hat{\psi}_2$, and the fermionic momenta $\pi^{\alpha}$, respectively, have to be defined as
\begin{align} \label{2.8}
 &&&&&& \hat{\psi}_1 &= \sqrt{\frac{\hbar}{2}} \sigma_1, && \hat{\psi}_2 = \sqrt{\frac{\hbar}{2}} \sigma_2, &&&&&& \nonumber \\
&&&&&& \pi^{\alpha} &= \frac{\partial \mathit{L}}{\partial \dot{\psi}_{\alpha}}= - \frac{i}{2} \delta^{\alpha \beta} \psi_{\beta},  && \alpha, \beta \in \{1,2 \}. &&&&&&
  \end{align}
Hence, using Eqs. (\ref{2.7}, \ref{2.8}) we can rewrite the previous Hamiltonian (\ref{2.6}) in the following form           
  \begin{equation} \label{2.9}
  \mathit{\hat{H}} = \frac{1}{2} \left(- \hbar^2 \dfrac{d^2}{dq^2} + \omega^2 q^2 \right) \mathbbmtt{I}_2 + \left( \frac{1}{2} \hbar \omega \right) \sigma_3,
  \end{equation}
where $\mathbbmtt{I}_2$ is the $2 \times 2$ identity matrix and $\sigma_j$ are the Pauli matrices, that is
\begin{equation}
\mathbbmtt{I}_2 = \left( \begin{matrix} 1 & 0 \\ 0 & 1 \end{matrix} \right), \quad
\sigma_1 = \left( \begin{matrix} 0 & 1 \\ 1 & 0 \end{matrix} \right), \quad
\sigma_2 = \left( \begin{matrix} 0 & -i \\ i & 0 \end{matrix} \right), \quad
\sigma_3 = \left( \begin{matrix} 1 & 0 \\ 0 & -1 \end{matrix} \right).
\end{equation}

}

\par{
We want to obtain solutions of the time independent Schr\"{o}dinger equation $\mathit{H}\Psi = \mathit{E} \Psi$, that is
  \begin{equation}
  \frac{1}{2} \left(- \hbar^2 \dfrac{d^2}{dq^2} + \omega^2 q^2 \right) \mathbbmtt{I}_2 + \left( \frac{1}{2} \hbar \omega \right) \sigma_3 = \mathit{E} \Psi,
  \end{equation}
with the wave function $\Psi(x)$ constraint to satisfy appropriate boundary conditions.
}

\par{
As we mentioned, the supersymmetric Hamiltonian can be written as a summation of two terms representing the bosonic and the fermionic components of the Hamiltonian
  \begin{equation}
  \mathit{H} = \mathit{H}_B (p,q)+\mathit{H}_F(\psi).
  \end{equation}
  
  }
  
\par{
To this end we define the lowering (annihilation) and raising (creation) operators for bosons, $\hat{a}^{\dagger}$ and $\hat{a}$, respectively they are 
  \begin{align} \label{2.13}
&&  \hat{a} &= \sqrt{\frac{1}{2 \hbar \omega}} \left( \omega \hat{q}+i\hat{p} \right), &
   \hat{a}^{\dagger} &= \sqrt{\frac{1}{2 \hbar \omega}} \left( \omega \hat{q}-i\hat{p} \right).&& 
  \end{align}
  
  }
  
\par{
Later in Section \ref{Sec2.4} we will show that the bosonic operators, $\hat{q}$ and $\hat{p}$, satisfy the usual canonical quantum commutation condition $[\hat{q},\hat{p}]= i\hbar$. Taking this fact into account, we can find that the bosonic ladder operators $\hat{a}$ and $\hat{a}^{\dagger}$ satisfy the following commutation relations
  \begin{align} \label{2.14}
 [\hat{a},\hat{a}^{\dagger}] &=1  \quad  \Longleftrightarrow  \quad 
  \hat{a} \hat{a}^{\dagger} - \hat{a}^{\dagger}\hat{a} =1, \nonumber \\ 
  [ \hat{a},\hat{a} ]  &= [\hat{a}^{\dagger},\hat{a}^{\dagger}]=0.
  \end{align}
It is straightforward to verify Eq. (\ref{2.14})  using Eq. (\ref{2.13}), where
\begin{align}
[\hat{a}, \hat{a}^{\dagger}] &= \hat{a} \hat{a}^{\dagger} - \hat{a}^{\dagger} \hat{a} \nonumber \\
&= \frac{1}{2 \hbar \omega} \Big( (\omega \hat{q}+i\hat{p})(\omega \hat{q}-i\hat{p}) \Big) - \frac{1}{2 \hbar \omega} \Big( (\omega \hat{q}-i\hat{p}) (\omega \hat{q}+i\hat{p}) \Big) \nonumber \\
&= \frac{1}{2 \hbar \omega} \Big( \left( \omega^2 \hat{q}^2 - i \omega \hat{q} \hat{p} + i \omega \hat{p} \hat{q} + \hat{p}^2 \right) - \left( \omega^2 \hat{q}^2 + i \omega \hat{q} \hat{p} - i \hat{p} \hat{q} + \hat{p}^2 \right) \Big) \nonumber \\
&= \frac{1}{2 \hbar \omega} \Big(- 2i \omega (  \hat{q} \hat{p} -  \hat{p} \hat{q} ) \Big) = \frac{1}{2 \hbar \omega} \Big( -2i \omega [\hat{q}, \hat{p}] \Big) = 1,
\end{align}
and
\begin{align}
[\hat{a},\hat{a}] &= \hat{a} \hat{a} - \hat{a} \hat{a} \nonumber \\
&= \frac{1}{2 \hbar \omega} \left( \omega \hat{q}+i\hat{p} \right)^2 - \frac{1}{2 \hbar \omega} \left( \omega \hat{q}+i\hat{p} \right)^2 = 0,
\end{align}
\begin{align}
[\hat{a}^{\dagger},\hat{a}^{\dagger}] &= \hat{a}^{\dagger} \hat{a}^{\dagger} - \hat{a}^{\dagger} \hat{a}^{\dagger} \nonumber \\
&= \frac{1}{2 \hbar \omega} \left( \omega \hat{q}-i\hat{p} \right)^2 - \frac{1}{2 \hbar \omega} \left( \omega \hat{q}-i\hat{p} \right)^2 = 0.
\end{align}

}  
  
\par{
However, on the other hand, the lowering (annihilation) and raising (creation) operators for fermions, $\hat{b}^{\dagger}$ and $\hat{b}$, respectively are defined as
  \begin{align} \label{a2.13}
 && \hat{b} &= \sqrt{\frac{1}{2 \hbar}} \left(\hat{\psi}_1 - i\hat{\psi}_2 \right), & 
  \hat{b}^{\dagger} &=\sqrt{\frac{1}{2 \hbar}} \left(\hat{\psi}_1 + i\hat{\psi}_2 \right).&&
  \end{align}
Also, as will be shown in Section \ref{Sec2.4}, the fermionic operators $\psi_{\alpha}$ and $\psi_{\beta}$ satisfy the canonical quantum anticommutation condition $\{\psi_{\alpha},\psi_{\beta}\}= \hbar \delta_{\alpha \beta}$ \footnote{The derivation of the fermionic anticommutators is a bit subtle and will be discussed in the next section.}. Therefore, it can be seen that the fermionic ladder operators $\hat{b}$ and $\hat{b}^{\dagger}$ satisfy the following anticommutation relations
\begin{align} \label{a2.18}
 &&&& \{ \hat{b},\hat{b}^{\dagger} \} &=1  \: \:  \: \, \qquad \qquad \quad \Longleftrightarrow  \quad  \hat{b} \hat{b}^{\dagger} + \hat{b}^{\dagger}\hat{b} =1, &&&&\nonumber \\ &&&&
  \{ \hat{b},\hat{b} \} &= \{ \hat{b}^{\dagger},\hat{b}^{\dagger} \}=0 \quad \Longleftrightarrow \quad \hat{b}^2 = \hat{b}^{{\dagger}^2}=0.&&&&
  \end{align}
  
  }
  
\par{
Once again we can verify Eq. (\ref{a2.18}) using Eq. (\ref{a2.13}), where
\begin{align} 
\{\hat{b}, \hat{b}^{\dagger}\} &= \hat{b} \hat{b}^{\dagger} + \hat{b}^{\dagger} \hat{b} \nonumber \\
&= \frac{1}{2 \hbar} \Big( (\hat{\psi}_1 - i\hat{\psi}_2 )(\hat{\psi}_1 + i\hat{\psi}_2 ) \Big) + \frac{1}{2 \hbar} \Big( (\hat{\psi}_1 + i\hat{\psi}_2 )(\hat{\psi}_1 - i\hat{\psi}_2 ) \Big) \nonumber \\
&= \frac{1}{2 \hbar} \Big( \left( \hat{\psi}_1^2 + i \hat{\psi}_1 \hat{\psi}_2 - i \hat{\psi}_2 \hat{\psi}_1 + \hat{\psi}_2^2 \right) - \left( \hat{\psi}_1^2 - i \hat{\psi}_1 \hat{\psi}_2 + i \hat{\psi}_2 \hat{\psi}_1 + \hat{\psi}_2^2 \right) \Big) \nonumber \\
&= \frac{1}{2 \hbar} \Big( \left(- i \hat{\psi}_2  \hat{\psi}_1 - i \hat{\psi}_2 \hat{\psi}_1 \right) - \left( - i \hat{\psi}_1 \hat{\psi}_2 - i \hat{\psi}_1 \hat{\psi}_2 \right) \Big) \nonumber \\
&= \frac{1}{2 \hbar} \left( -2 i (\hat{\psi}_1 \hat{\psi}_2 + \hat{\psi}_2 \hat{\psi}_1)  \right) = \frac{1}{2 \hbar} \left( -2 i \{ \hat{\psi}_1, \hat{\psi}_2 \} \right) = 1,
\end{align}
and
\begin{align}
\{\hat{b},\hat{b}\} &= \hat{b} \hat{b} + \hat{b} \hat{b} \nonumber \\
&= \frac{1}{2\hbar} \Big( ( \hat{\psi}_1 - i\hat{\psi}_2 ) (\hat{\psi}_1 - i\hat{\psi}_2 ) \Big) + \frac{1}{2\hbar} \Big( ( \hat{\psi}_1 - i\hat{\psi}_2 ) (\hat{\psi}_1 - i\hat{\psi}_2 ) \Big) \nonumber \\
& = \frac{1}{\hbar} \Big( \hat{\psi}_1^2 - i \hat{\psi}_1 \hat{\psi}_2 - i \hat{\psi}_2 \hat{\psi}_1 - \hat{\psi}_2^2  \Big) = \frac{1}{\hbar} \Big( - i \hat{\psi}_1 \hat{\psi}_2 + i \hat{\psi}_1 \hat{\psi}_2  \Big) =0,
\end{align}
\begin{align}
\{\hat{b}^{\dagger},\hat{b}^{\dagger}\} &= \hat{b}^{\dagger} \hat{b}^{\dagger} + \hat{b}^{\dagger} \hat{b}^{\dagger} \nonumber \\
&= \frac{1}{2 \hbar} \Big( ( \hat{\psi}_1 + i\hat{\psi}_2 ) (\hat{\psi}_1 + i\hat{\psi}_2 ) \Big) + \frac{1}{2\hbar} \Big( ( \hat{\psi}_1 + i\hat{\psi}_2 ) (\hat{\psi}_1 + i\hat{\psi}_2 ) \Big) \nonumber \\
& = \frac{1}{\hbar} \Big( \hat{\psi}_1^2 + i \hat{\psi}_1 \hat{\psi}_2 + i \hat{\psi}_2 \hat{\psi}_1 - \hat{\psi}_2^2  \Big) = \frac{1}{\hbar} \Big(  i \hat{\psi}_1 \hat{\psi}_2 - i \hat{\psi}_1 \hat{\psi}_2  \Big) =0.
\end{align}

}  
\par{
It is possible using Eqs. (\ref{2.13}) to write the bosonic position and momentum variables in terms of the bosonic ladder operators as follows
\begin{align} \label{c2.23}
&&&&&&&& \hat{a} + \hat{a}^\dagger &= \sqrt{\frac{2 \omega}{\hbar}} \hat{q} && &\Rightarrow & && \hat{q} = \sqrt{\frac{\hbar}{2 \omega}} (\hat{a}^\dagger+ \hat{a}), &&&&&&&& \nonumber \\
&&&&&&&& \hat{a} - \hat{a}^\dagger &= \sqrt{\frac{2}{\hbar \omega}} i \hat{p} && &\Rightarrow & && \hat{p} =i \sqrt{\frac{\hbar \omega}{2}} (\hat{a}^\dagger- \hat{a}). &&&&&&&&
\end{align}
Similarly using Eqs. (\ref{a2.13}), we can write the fermionic variables $\psi_1$, $\psi_2$ in terms of the fermionic ladder operators as follows
\begin{align} \label{c2.24}
&&&&&&&& \hat{b}^\dagger + \hat{b} &= \sqrt{\frac{2}{\hbar}} \hat{\psi}_1 && &\Rightarrow & && \hat{\psi}_1 = \sqrt{\frac{\hbar}{2}} (\hat{b} + \hat{b}^\dagger), &&&&&&&& \nonumber \\
&&&&&&&& \hat{b}^\dagger - \hat{b} &= i \sqrt{\frac{2}{\hbar}}  \hat{\psi}_2 && &\Rightarrow & && \hat{\psi}_2 =i \sqrt{\frac{\hbar}{2}} (\hat{b} - \hat{b}^\dagger). &&&&&&&&
\end{align}

}

\par{
Actually, we derived Eqs. (\ref{c2.23}, \ref{c2.24}) here, however, it will be used later.  At this stage, let us define the following two hermitian operators, the bosonic number operator $\hat{\mathit{N}}_{B}$, and the fermionic number operator $\hat{\mathit{N}}_{F}$
  \begin{align}\label{2.16}
  \mathit{\hat{N}}_B &\equiv \hat{a}^{\dagger} \hat{a} = \frac{1}    {2 \hbar \omega} \left(\omega^2 \hat{q}^2 - \hbar \omega + \hat{p}^2 \right),  \nonumber \\
   \mathit{\hat{N}}_F &\equiv \hat{b}^{\dagger} \hat{b} \:=
\frac{-i}{\hbar} \left(\hat{\psi}_1 \hat{\psi}_2 \right).  
  \end{align} 
Using Eq. (\ref{2.16}), we can rewrite the Hamiltonian (\ref{2.9}) in the operator formalism 
  \begin{equation} \label{2.17}
  \mathit{\hat{H}} = \hbar \omega \left( \mathit{\hat{N}}_B + \frac{1}{2} \right) + \hbar \omega \left( \mathit{\hat{N}}_F - \frac{1}{2} \right).
  \end{equation} 
This means that the energy eigenstates can be labeled with the eigenvalues of $\mathit{n}_B$ and $\mathit{n}_F$.}

\par{
The action of the ladder operators, $\hat{a}$, $\hat{a}^{\dagger}$, $\hat{b}$, $\hat{b}^{\dagger}$, upon an energy eigenstate $\vert n_B , n_F \rangle$ are listed bellow
  \begin{align} \label{2.18}
  && \hat{a} \vert n_B , n_F \rangle &= \sqrt{\mathit{n}_B} \vert    \mathit{n}_B -1, n_F \rangle, & &
  & \hat{a}^\dagger \vert \mathit{n}_B, n_F \rangle &= \sqrt{\mathit{n}_B+1} \vert \mathit{n}_B +1, n_F \rangle, && \nonumber \\
 && \hat{b} \vert n_B, \mathit{n}_F \rangle &= \sqrt{\mathit{n}_F} \vert n_B, \mathit{n}_F -1 \rangle, && 
  & \hat{b}^\dagger \vert n_B, \mathit{n}_F \rangle &= \sqrt{\mathit{n}_F+1} \vert n_B, \mathit{n}_F +1 \rangle. &&
  \end{align}
It is straightforward from Eq. (\ref{2.18}) that the associated bosonic number operator $\mathit{\hat{N}}_B$ and fermionic number operator $\mathit{\hat{N}}_F$ obey
  \begin{equation} \label{2.19}
  \mathit{\hat{N}}_B \vert \mathit{n}_B, n_F \rangle = \mathit{n}_B \vert \mathit{n}_B, n_F \rangle,  \qquad   \mathit{\hat{N}}_F \vert n_B, \mathit{n}_F \rangle = \mathit{n}_F \vert n_B, \mathit{n}_F \rangle.
  \end{equation}
  
  }
  
\par{
Under Eq. (\ref{2.19}) the energy spectrum of the Hamiltonian (\ref{2.17}) is
  \begin{equation} \label{2.20}
  \mathit{E}_n = \hbar \omega \left( \mathit{n}_B + \frac{1}{2} \right) + \hbar \omega \left( \mathit{n}_F - \frac{1}{2} \right).
  \end{equation}
The form of Eq. (\ref{2.20}) implies that the Hamiltonian $\mathit{H}$ is symmetric under the interchange of $\hat{a}$ and $\hat{a}^{\dagger}$, and antisymmetric under the interchange of $\hat{b}$ and $\hat{b}^{\dagger}$.
}

\par{
The most important observation comes from Eq. (\ref{2.20}) is that $\mathit{E_n}$ remains invariant under a simultaneous annihilation of one boson ($\mathit{n}_B \rightarrow \mathit{n}_B -1$) and creation of one fermion ($\mathit{n}_F \rightarrow \mathit{n}_F + 1$) or vice versa. This is one of the simplest examples of a symmetry called "supersymmetry" (SUSY) and the corresponding energy spectrum reads
  \begin{equation}
  \mathit{E}_n= \hbar \omega \left( \mathit{n}_B+\mathit{n}_F \right).
  \end{equation}
  
  }
\par{
Moreover, we still have two more comments before ending this section:
\begin{itemize}
\item[(i)]
Consider the fermionic vacuum state $\vert 0 \rangle$ is defined by
\begin{equation}
\hat{b} \vert 0 \rangle \equiv 0.
\end{equation}
Then we define a fermionic one-particle state $\vert 1 \rangle$ by
\begin{equation}
\vert 1 \rangle \equiv \hat{b}^{\dagger} \vert 0 \rangle .
\end{equation}
It is now easy using the anticommutation relations (\ref{a2.18}) to see that there no other states, since operating on $\vert 1 \rangle$ with $\hat{b}$ or $\hat{b}^{\dagger}$ gives either the already known state $\vert 0 \rangle$, or nothing
\begin{align}
\hat{b} \vert 1 \rangle &= \hat{b} \hat{b}^{\dagger} \vert 0 \rangle = (1-\hat{b}^{\dagger} \hat{b}) \vert 0 \rangle = \vert 0 \rangle - \hat{b}^{\dagger} \hat{b} \vert 0 \rangle =  \vert 0 \rangle - 0 = \vert 0 \rangle, \nonumber \\
\hat{b}^{\dagger} \vert 1 \rangle &= \hat{b}^{\dagger} \hat{b}^{\dagger} \vert 0 \rangle = 0 \vert 0 \rangle =0.
\end{align}
So the subspace spanned by $\vert 0 \rangle$ and $\vert 1 \rangle$ is closed under the action of $\hat{b}$ and $\hat{b}^{\dagger}$, and therefore it is closed under the action of any product of these operators. We can show this directly using the two facts that the operator $\hat{b} \hat{b}^{\dagger}$ can be expressed as the linear combination $1 - \hat{b}^{\dagger} \hat{b}$ and $\hat{b}^2=\hat{b}^{{\dagger}^2}=0$. So any product of three or more $\hat{b}$, $\hat{b}^{\dagger}$ can be shortened by using either $\hat{b}^2 = 0$, or $\hat{b}^{{\dagger}^2} = 0$. For example $ \hat{b}^{\dagger} \hat{b} \hat{b}^{\dagger} = \hat{b}^{\dagger} - \hat{b} \hat{b}^{{\dagger}^2} = \hat{b}^{\dagger}$ and $\hat{b} \hat{b}^{\dagger} \hat{b} = \hat{b} - \hat{b}^{\dagger} \hat{b}^2 = \hat{b}$. So it seems to be clear that all products of $\hat{b}$, and $\hat{b}^{\dagger}$ can always be reduced  to linear combinations of the following four operators $1$, $\hat{b}$, $\hat{b}^{\dagger}$, and $\hat{b}^{\dagger} \hat{b}$. Based on this argument, there are only two possible fermionic eigenstate and hence two possible eigenvalues of the fermionic number operator, they are $\mathit{n}_F=\{0,1\}$.
\item[(ii)] The ground eigenstate has a vanishing energy eigenvalue (when $\mathit{n}_B= \mathit{n}_F=0$). In this case, the ground eigenstate is not degenerate and we say that supersymmetry is unbroken. This zero energy eigenvalue arises because of the cancellation between the bosonic and fermionic contributions to the supersymmetric ground-state energy. Since the ground-state energy for the bosonic and fermionic oscillators   have the values  $\frac{1}{2} \hbar \omega$ and $-\frac{1}{2} \hbar \omega$, respectively.
\end{itemize}

}

\section{The Constrained Hamiltonian Formalism}
\par{
Before discussing the supersymmetry algebra in the next section, we have to discuss the constrained Hamiltonian formalism in this section. This formalism is needed to get the correct fermionic anticommutators due to the complication of constraints. Good discussions of the constrained Hamiltonian systems are \cite{greenwood2006advanced, hanson1976constrained, marathe1983constrained, henneaux1992quantization, wipf1994hamilton, tavakoli2014lecture}. For instance, let us recall the Hamiltonian (\ref{2.6})
\begin{equation} \label{a2.31}
\mathit{H} = \frac{1}{2} \left( p^2 + \omega^2 q^2 \right) - i \omega \psi_1 \psi_2.
\end{equation}
A constrained Hamiltonian is one in which the momenta and positions are related by some constraints. In general, $M$ constraints between the canonical variables can be written as
\begin{equation}
\phi_m(q,p,\psi,\pi)=0, \qquad   m=1, \dots , M.
\end{equation}
These are called primary constraints. Secondary constraints are additional constraints relating momenta and positions which can arise from the requirement that the primary constraints are time-independent, \ie $\dot{\phi}_m = 0$. Fortunately, we do not have to deal with such constraints in this case. In principle there can also be tertiary constraints arising from the equation of motion of the secondary constraints and so on.}

\par{
The Hamiltonian (\ref{a2.31}) is an example of constrained Hamiltonians. In the previous section, we defined the fermionic momenta $\pi^\alpha$ in Eq. (\ref{2.8}) using the fermionic equation of motion follows from the usual Euler-Lagrange equations with the Lagrangian (\ref{2.5}) which is equivalent to this Hamiltonian. In fact, this definition gives us the relation between the fermionic momenta $\pi^\alpha$ and the canonical coordinates $\psi^\alpha$,
\begin{equation} \label{b2.34}
\pi^{\alpha} = \frac{\partial \mathit{L}}{\partial \dot{\psi}_{\alpha}}= - \frac{i}{2} \delta^{\alpha \beta} \psi_{\beta},  \qquad \alpha, \beta \in \{1,2 \}.
\end{equation}
In light of this equation, we have the primary constraint
\begin{equation} \label{b2.35}
\phi^\alpha = \pi^\alpha+ \frac{i}{2} \delta^{\alpha \beta} \psi_\beta = 0.
\end{equation}
Such constraints mean that the way we write the Hamiltonian is ambiguous since we can exchange position and momentum variables which changes the equations of motion one gets. For example, using Eq. (\ref{b2.34}) one can rewrite the Hamiltonian (\ref{a2.31}) in terms of the fermionic momenta as follows   
\begin{equation} \label{a2.36}
\mathit{H} = \frac{1}{2} \left( p^2 + \omega^2 q^2 \right) +4i \omega \pi^1 \pi^2.
\end{equation}
In the next subsection, we will briefly discuss the general properties of Poisson brackets and its relation with the Hamiltonian formalism in classical mechanics. Later in the same subsection, we will describe the problem caused by the constrained Hamiltonians. Next in Subsection \ref{SubSec.2.3.3}, we will show how to deal with this problem using the constraints.
}

\subsection{Poisson Brackets and First Class Constraints } \label{SubSec2.3.1}
\par{
The Poisson bracket of two arbitrary functions $\mathit{F}(\hat{q},\hat{p}, \hat{\psi},\hat{\pi})$ and $\mathit{G}(\hat{q},\hat{p}, \hat{\psi},\hat{\pi})$ is defined as
\begin{equation}
\{\mathit{F},\mathit{G}\}_P = \left( \dfrac{\partial \mathit{F}}{\partial q_i} \dfrac{\partial \mathit{G}}{\partial p^i} - \dfrac{\partial \mathit{F}}{\partial p^i}\dfrac{\partial \mathit{G}}{\partial q_i} \right)
 (-)^{\epsilon_\mathit{F}} \left( \dfrac{\partial \mathit{F}}{\partial \psi_\alpha} \dfrac{\partial \mathit{G}}{\partial \pi^\alpha} - \dfrac{\partial \mathit{F}}{\partial \pi^\alpha}\dfrac{\partial \mathit{G}}{\partial \psi_\alpha} \right),
\end{equation}
where $\epsilon_\mathit{F}$ is $0$ if $\mathit{F}$ is Grassmann even and $1$ if $\mathit{F}$ is Grassmann odd. Actually, in our argument here, we only consider the case of the odd Grassmann variables. In general, the Poisson brackets have the following properties:
\begin{itemize}
\item[(i)] A constraint $\mathit{F}$ is said to be a first-class constraint if its Poisson  bracket with all the other constraints vanish
\begin{align} 
\{\phi^\mathit{A}, \mathit{F} \}_P =0,  
\end{align} 
where $\mathit{A}$ runs over all the constraints. If a constraint is not a first-class, it is a second class.
\item[(ii)] Poisson bracket of two functions $\mathit{F}$ and $\mathit{G}$ is antisymmetric
\begin{equation}
\{\mathit{F},\mathit{G}\}_P = - \{ \mathit{G}, \mathit{F} \}_P.
\end{equation}
\item[(iii)]  For any three functions $\mathit{A}$, $\mathit{B}$, and $\mathit{C}$, Poisson bracket is linear in both entries
\begin{equation}
\{ \mathit{A},\mathit{B}+\mathit{C}\}_P= \{\mathit{A},\mathit{B}\}_P + \{\mathit{A},\mathit{C}\}_P. 
\end{equation}
\item[(iv)] Poisson bracket for any three functions satisfies the Leibniz rule
\begin{equation}
\{ \mathit{A},\mathit{B} \mathit{C}\}_P = \{ \mathit{A},\mathit{B}\}_P \mathit{C} + (-1)^{\epsilon_\mathit{A} \epsilon_\mathit{B}} \mathit{B} \{ \mathit{A},\mathit{C}\}_P, 
\end{equation}
where $\epsilon_\mathit{F}$ is the Grassmann parity of $\mathit{F}$ which is $0$ if $\mathit{F}$ is Grassmann even and $1$ if $\mathit{F}$ is Grassmann odd.
\item[(v)] Poisson bracket for any three functions obeys the Jacobi identity
\begin{equation}
\{ \mathit{A}, \{\mathit{B},\mathit{C}\}_P \}_P + \{\mathit{B},\{\mathit{C},\mathit{A}\}_P\}_P + \{ \mathit{C}, \{ \mathit{A},\mathit{B}\}_P\}_P =0.
\end{equation}
\item[(vi)] The time rate of change of any arbitrary function $\mathit{F}(q_i,p^i,\psi_\alpha,\pi^\alpha)$, which has no explicit time dependence, is given by its Poisson bracket with the Hamiltonian
\begin{equation}
\dot{\mathit{F}} = \{ \mathit{F}, \mathit{H} \}_P.
\end{equation}
\end{itemize}

}

\par{
In the Hamiltonian formalism of classical mechanics, the Hamilton's equations of motion have equivalent expressions in terms of the Poisson bracket. Using the last property (vi) of Poisson brackets, one can easily see that 
\begin{align} \label{b2.44}
&&&&&&&& \dot{q}_i &= \frac{\partial \mathit{H}}{\partial p^i} = \{ q_i, \mathit{H} \}_P, &&&& &\dot{p}^i &=- \frac{\partial \mathit{H}}{\partial q_i} = \{ q^i, \mathit{H} \}_P, &&&&&&&& \nonumber \\
&&&&&&&& \dot{\psi}_\alpha &= - \frac{\partial \mathit{H}}{\partial \pi^\alpha} = \{ \psi_\alpha, \mathit{H} \}_P, &&&& &\dot{\pi}^\alpha &=- \frac{\partial \mathit{H}}{\partial \psi_\alpha} = \{ \pi^\alpha, \mathit{H} \}_P. &&&&&&&&
\end{align}
However, the problem with the constrained system is that the equations of motion derived using Eqs. (\ref{b2.44}) not always consistence with those follow from the Lagrangian formalism. For example, if we consider the Hamiltonian (\ref{a2.31}), the equation of motion for $\psi_\alpha$ follows from the expected Poisson bracket equation have to be
\begin{equation} \label{b2.45}
\dot{\psi}_\alpha= \{ \psi_\alpha, \mathit{H} \}_P = - \frac{\partial \mathit{H}}{\partial \pi^\alpha} =0.
\end{equation}
Unfortunately, this equation is inconsistent with the equation of motion follows from the Lagrangian formalism. Even worst, the equation of motion for $\pi^\alpha$ follows from the expected Poisson bracket equation is also inconsistent with the equation derived from the Lagrangian formalism. For example, using the Poisson bracket we have
\begin{equation} \label{b2.46}
\dot{\pi}^1=\{ \pi^1,\mathit{H} \}_P = - \frac{\partial \mathit{H}}{\psi_1} = i \omega \psi_2.
\end{equation}
But if we use Eq. (\ref{b2.34}) to find an expression to the equation of motion for $\dot{\psi}_1$ and then substitute using Eq. (\ref{b2.46}), that gives us 
\begin{equation}
\dot{\psi}_1 = 2i\dot{\pi}^1=-2\omega\psi_2.
\end{equation}
This equation is different from the Lagrangian equations of motion and inconsistent with (\ref{b2.45}) (unless everything is trivial which is also not great). Moreover, using the Poisson bracket with the Hamiltonian (\ref{a2.36}), we can find the following set of equations of motion for the fermionic positions and momenta
\begin{align}
\dot{\psi}_1 &= \{ \psi_1, \mathit{H} \}_P = - \frac{\partial \mathit{H}}{\partial \pi^1}= - 4i\omega \pi^2= - 2 \omega \psi_1, \nonumber \\
\dot{\psi}_2 &= \{ \psi_2, \mathit{H} \}_P = - \frac{\partial \mathit{H}}{\partial \pi^2}= - 4i\omega \pi^1= - 2 \omega \psi_2, \nonumber \\
\dot{\pi}^\alpha &= \{ \pi^\alpha, \mathit{H} \}_P = - \frac{\partial \mathit{H}}{\partial \psi_\alpha}= 0.
\end{align}
These equations also are slightly different from the Lagrangian equations of motion and inconsistent with the equations follows from the Hamiltonian (\ref{a2.31}).
}

\par{
In brief, there is inconsistency between the equations of motion derived from the Hamiltonian formalism and the equations derived from the Lagrangian formalism. The reason for this inconsistency is that we have a constrained Hamiltonian. This problem will be solved in the next subsection by imposing the constraints onto the Hamiltonian.

}

\subsection{The Constraints and the Equations of Motion} \label{SubSec.2.3.3}
\par{
In the Lagrangian formalism, one can impose the constraints from the beginning by introducing Lagrange multipliers which enforce the constraints. So Hamilton's equations of motion, in the presence of constraints, can be derived from the extended action principle
\begin{equation} \label{a2.39}
\mathit{S} = \int \left( \dot{q}_i p^i + \dot{\psi}_{\alpha} \pi^{\alpha} - \mathit{H}-\phi^A \lambda_A \right) dt,
\end{equation}
where $\phi^A$ are our constraints, $\lambda_A$ are Lagrange multipliers and $\mathit{A}$ runs over the constraints. Using the extended we can write out the following two expressions of the extended Lagrangian and the extended Hamiltonian, respectively
\begin{align}
\mathit{L} &= \dot{q}_i p^i + \dot{\psi}_{\alpha} \pi^{\alpha} - \mathit{H} - \phi^A \lambda_A, \\ \label{a2.41}
\mathit{H} &= \dot{q}_i p^i + \dot{\psi}_{\alpha} \pi^{\alpha} - \phi^A \lambda_A - \mathit{L}.
\end{align}
Indeed, if the Lagrangian is independent of one coordinate, say $q_i$, then we call this coordinate an ignorable coordinate. However, we still need to solve the Lagrangian for this coordinate to find the corresponding equation of motion. Since the momentum corresponding to this coordinate, may still enter the Lagrangian and affect the evolution of other coordinates. Therefore using the Euler-Lagrange equation, we have
\begin{equation}
\dfrac{d}{dt} \dfrac{\partial \mathit{L}}{\partial \dot{q}_i} = \dfrac{\partial \mathit{L}}{\partial q_i} = 0.
\end{equation}
Thus, in general if we define the generalized momentum as $p^i = \dfrac{\partial \mathit{L}}{\partial \dot{q}_i}$, then the momentum  $p^i$ is conserved in the case that the Lagrangian $\mathit{L}$ is independent of the coordinate $q_i$, \ie $\dfrac{d p^i}{dt}=0$. This means that if the Lagrangian $\mathit{L}$ is independent of an coordinate $q^i$, it must be also independent of its corresponding momentum  $p^i$, \ie $\dfrac{\partial \mathit{L}}{\partial p^i} =0$. }

\par{
Furthermore, by extremising the Hamiltonian (\ref{a2.41}), we obtain the following generalized Hamilton's equations of motion
\begin{align}\label{b2.44}
&&&&&&&& \dot{q}_i &=  \dfrac{\partial \mathit{H}}{\partial p^i } + \dfrac{\partial \phi^A}{\partial p^i} \lambda_A & = \{q_i  , \mathit{H} + \phi^A \lambda_A \}_P,&&&&&&&& \nonumber \\
&&&&&&&& \dot{p}^i &= - \dfrac{\partial \mathit{H}}{\partial q_i} - \dfrac{\partial \phi^A}{\partial q_i} \lambda_A & = \{ p^i , \mathit{H} + \phi^A \lambda_A \}_P, &&&&&&&& \nonumber \\
&&&&&&&& \dot{\psi}_{\alpha} &= - \dfrac{\partial \mathit{H}}{\partial \pi^{\alpha}} - \dfrac{\partial \phi^A}{\partial \pi^{\alpha}} \lambda_A & = \{\psi_{\alpha}  , \mathit{H} + \phi^A \lambda_A \}_P, &&&&&&&& \nonumber \\
 &&&&&&&& \dot{\pi}^{\alpha} &= - \dfrac{\partial \mathit{H}}{\partial  \psi_{\alpha}} - \dfrac{\partial \phi^A}{\partial \psi_{\alpha}} \lambda_A & = \{\pi^{\alpha}  , \mathit{H} + \phi^A \lambda_A \}_P, &&&&&&&& \nonumber \\
&&&&&&&& \phi^A &= 0. &&&&&&&&
\end{align}
Eqs. (\ref{b2.44}) are derived in detail in Appendix \ref{AppendixA.1}. For example, using Poisson brackets we can show that Eqs. (\ref{b2.44}) imply for any arbitrary function $\mathit{F}= \mathit{F}(q_i,p^i,\psi_{\alpha},\pi^{\alpha})$, we can write $\dot{\mathit{F}}= \{ \mathit{F}, \mathit{H}+\phi^A \lambda_A \}_P$:
\begin{align} \label{b2.54}
\dot{\mathit{F}} &= \dfrac{\partial \mathit{F} }{\partial q_i} \dfrac{\partial q_i}{\partial t} + \dfrac{\partial \mathit{F} }{\partial p^i} \dfrac{\partial p^i}{\partial t} + \dfrac{\partial \mathit{F} }{\partial \psi_\alpha} \dfrac{\partial \psi_\alpha}{\partial t} + \dfrac{\partial \mathit{F} }{\partial \pi^\alpha} \dfrac{\partial \pi^\alpha}{\partial t} \nonumber \\
&= \dfrac{\partial \mathit{F} }{\partial q_i} \dot{q_i}+ \dfrac{\partial \mathit{F} }{\partial p^i} \dot{p^i} + \dfrac{\partial \mathit{F} }{\partial \psi_\alpha} \dot{\psi_\alpha} + \dfrac{\partial \mathrm{F} }{\partial \pi^\alpha} \dot{ \pi^\alpha} \nonumber \\ 
&= \dfrac{\partial \mathit{F} }{\partial q_i} \Big(  \dfrac{\partial \mathit{H}}{\partial p^i}+ \dfrac{\partial \phi^A}{\partial p^i} \lambda_A \Big) + \dfrac{\partial \mathit{F} }{\partial p^i} \Big(- \dfrac{\partial \mathit{H}}{\partial q^i} - \dfrac{\partial \phi^A}{\partial q^i} \lambda_A \Big) \nonumber \\
&{\quad}+ \dfrac{\partial \mathit{F} }{\partial \psi_\alpha} \Big(- \dfrac{\partial \mathit{H}}{\partial \pi^\alpha} - \dfrac{\partial \phi^A}{\partial \pi^\alpha} \lambda_A \Big) + \dfrac{\partial \mathit{F} }{\partial \pi^\alpha} \Big(- \dfrac{\partial \mathit{H}}{\partial \psi^\alpha} - \dfrac{\partial \phi^A}{\partial \psi_\alpha} \lambda_A \Big) \nonumber \\
&= \Big( \dfrac{\partial \mathrm{F} }{\partial q_i} \dfrac{\partial (\mathrm{H}+ \phi^A \lambda_A)}{\partial p^i}- \dfrac{\partial \mathrm{F} }{\partial p^i} \dfrac{\partial (\mathrm{H}+ \phi^A \lambda_A)}{\partial q^i} \Big) \nonumber \\
& \quad - \Big( \dfrac{\partial \mathrm{F} }{\partial \psi_\alpha} \dfrac{\partial (\mathrm{H}+ \phi^A \lambda_A)}{\partial \pi^\alpha} + \dfrac{\partial (\mathrm{H}+ \phi^A \lambda_A)}{\partial \psi^\alpha} \dfrac{\partial \mathrm{F} }{\partial \pi^\alpha}  \Big) \nonumber \\ 
& \equiv \{\mathrm{F}, \mathrm{H}+ \phi^A \lambda_A \}_P.
\end{align}

}
 
\par{
At this stage, we need to confirm that Eqs. (\ref{b2.44}) give us the correct equations of motion using the constrained Hamiltonian (\ref{a2.31}). Since $\alpha, \beta \in \{1,2\}$ in the primary constraint (\ref{b2.35}), then we have the two constraints
\begin{align} \label{aa2.57}
&&&&&&\phi^1 &= \pi^1 + \frac{i}{2} \psi_1, && \phi^2 = \pi^2 + \frac{i}{2} \psi_2. &&&&&&
\end{align}

}
\par{
By using Eqs. (\ref{b2.44}) and taking into count the Hamiltonian (\ref{a2.36}), and the constraints (\ref{aa2.57}), we find that
\begin{align} \label{aa2.59}
\dot{\psi}_1 &= - \dfrac{\partial \mathit{H}}{\partial \pi^1} - \dfrac{\partial \phi^1}{\pi^1} \lambda_1 = -  \lambda_1, \nonumber \\
\dot{\psi}_2 &= - \dfrac{\partial \mathit{H}}{\partial \pi^2} - \dfrac{\partial \phi^2}{\pi^2} \lambda_2 = - \lambda_2. 
\end{align}
On the other hand, using Eqs. (\ref{b2.44}) and considering the Hamiltonian (\ref{a2.31}), and the constraints (\ref{aa2.57}), we find that
\begin{align} \label{aa2.60}
\dot{\pi}^1 &= - \dfrac{\partial \mathit{H}}{\partial \psi_1} - \dfrac{\partial \phi^1}{\psi_1} \lambda_1 = i \omega \psi_2 - \frac{i}{2} \lambda_1, \nonumber \\
\dot{\pi}^2 &= - \dfrac{\partial \mathit{H}}{\partial \psi_2} - \dfrac{\partial \phi^2}{\psi_2} \lambda_2 = -i \omega \psi_1 - \frac{i}{2} \lambda_2.
\end{align}
Furthermore, using Eq. (\ref{b2.34}) we get
\begin{align} \label{aa2.62}
&&&&&&&&\pi^1 &= - \frac{i}{2} \psi_1 && \Rightarrow && \dot{\pi^1} = - \frac{i}{2} \dot{\psi}_1, &&&&&&&&  \nonumber \\
&&&&&&&&\pi^2 &= - \frac{i}{2} \psi_2 && \Rightarrow && \dot{\pi^2} = - \frac{i}{2} \dot{\psi}_2. &&&&&&&&
\end{align}
Eqs. (\ref{aa2.59}, \ref{aa2.60}, \ref{aa2.62}) imply the following set of equations of motion for the fermionic coordinates $\psi_\alpha$, 
\begin{align} \label{aa2.63}
&&&&&& \dot{\psi}_1 &= - \omega \psi_2, && \dot{\psi}_2 =  \omega \psi_1. &&&&&&
\end{align}
Fortunately, these equations are the same equations of motion that follow from the Lagrangian formalism.

}
\par{
For now let us use the Poisson bracket, the Hamiltonian (\ref{a2.31}), and the constraints (\ref{aa2.57}), to check that
\begin{align} \label{aa2.64}
& \dot{\phi}^1  = \{ \phi^1, \mathit{H}+\phi^1 \lambda_1   \}_P \nonumber \\
& = \resizebox{.95 \textwidth}{!} 
{$ \left( \dfrac{\partial \phi^1}{\partial q_i} \dfrac{\partial (\mathit{H}+\phi^1 \lambda_1) }{\partial p^i} - \dfrac{\partial \phi^1}{\partial p^i} \dfrac{\partial (\mathit{H}+\phi^1 \lambda_1)}{\partial q_i} \right) -  \left( \dfrac{\partial \phi^1}{\partial \psi_\alpha} \dfrac{\partial (\mathrm{H}+\phi^1 \lambda_1) }{\partial \pi^\alpha} + \dfrac{\partial \phi^1}{\partial \pi^\alpha} \dfrac{\partial (\mathit{H}+\phi^1 \lambda_1)}{\partial \psi_\alpha} \right)$} \nonumber \\
& =  0+0 - \Big(\frac{i}{2}\Big) \left(\lambda_1\right) - \left(1\right) \Big(-i\omega \psi_2+\frac{i}{2} \lambda_1 \Big) = i \left(- \lambda_1+\omega \psi_2 \right)=0.
\end{align}
The reason that the last step is zero comes from Eqs. (\ref{aa2.59}, \ref{aa2.63}) since we have $\lambda_1= - \dot{\psi}_1 = \omega \psi_2$. Similarly $\dot{\phi}^2$ can be calculated using Poisson bracket so follows
\begin{align} \label{aa2.65}
 \dot{\phi}^2 &= \{ \phi^2, \mathit{H}+\phi^2 \lambda_2 \}_P \nonumber \\
& = \resizebox{.94 \textwidth}{!} 
{$ \left( \dfrac{\partial \phi^2}{\partial q_i} \dfrac{\partial (\mathit{H}+\phi^2 \lambda_2) }{\partial p^i} - \dfrac{\partial \phi^2}{\partial p^i} \dfrac{\partial (\mathit{H}+\phi^2 \lambda_2)}{\partial q_i} \right) - \left( \dfrac{\partial \phi^2}{\partial \psi_\alpha} \dfrac{\partial (\mathit{H}+\phi^2 \lambda_2) }{\partial \pi^\alpha} + \dfrac{\partial \phi^2}{\partial \pi^\alpha} \dfrac{\partial (\mathit{H}+\phi^2 \lambda_2)}{\partial \psi_\alpha} \right)$} \nonumber \\
&=  0 + 0 - \Big( \frac{i}{2} \Big) \left( \lambda_2 \right) - \left(2\right) \Big( i\omega \psi_1+\frac{i}{2} \lambda_2 \Big) = i \left(- \lambda_2-\omega \psi_1 \right)=0.
\end{align}
Also, the last step is zero because from Eqs. (\ref{aa2.59}, \ref{aa2.63}) we have $\lambda_2= - \dot{\psi}_2 = - \omega \psi_1$. Subsequently, Eqs. (\ref{aa2.64}, \ref{aa2.65}) are zero if we impose the equations of motion (\ref{aa2.63}). In other words we can say that the constraints are consist with the equations of motion.}

\subsection{Dirac Brackets and Second Class Constraints }
\par{
A Constraint $\mathit{F}$ is said to be a second class constraint if it has non-zero Poisson brackets and therefore it requires special treatment. Given a set of second class constraints, we can define a matrix
\begin{equation}
\{ \phi^{\mathit{A}},\phi^{\mathit{B}} \}_P = \mathit{C}^{AB},
\end{equation}
where we define the inverse of $\mathit{C}^{\mathit{A} \mathit{B}},$ as $\mathit{C}_{\mathit{A} \mathit{B}}$, such that \label{b2.47}
\begin{equation}
\mathit{C}^{\mathit{A} \mathit{B}} \mathit{C}_{\mathit{B} \mathit{C}} = \delta^{\mathit{A}}_{\mathit{C}}.
\end{equation}
The Dirac bracket provides a modification to the Poisson brackets to ensure that the second class constants vanish
\begin{equation}\label{a2.47}
\{ \phi^A, \mathit{F} \}_D =0.
\end{equation}
The Dirac bracket of two arbitrary functions $F(q,p)$ and $G(q,p)$ is defined as
\begin{equation} \label{a2.48}
\{ F , G \}_D = \{ F,G \}_P - \{ F, \phi^A\}_P \mathit{C}_{AB} \{ \phi^B , G\}_P. 
\end{equation}

}
\par{
Dirac bracket has the same properties of Poisson bracket, which we have listed in the previous Subsection \ref{SubSec2.3.1}. In addition, we have to notice that
\begin{itemize} 
\item[(i)] If $\dot{\phi}^A=0$, then the Dirac bracket of any constraint with the extended Hamiltonian is equivalent to its Poisson bracket
\begin{equation}
\{ \mathit{F}, \mathit{H} + \phi^A \lambda_A \}_D = \{ \mathit{F}, \mathit{H} + \phi^A \lambda_A \}_P.
\end{equation}
\item[(ii)] In some cases, we can return to the original Hamiltonian (\ref{a2.31}), forget about the constraints
and the momenta $\pi^{\beta}$ at the cost of replacing Poisson brackets with Dirac brackets. For example, if we consider the case $\dot{\phi}^A =0$, then we have 
\begin{align}
\dot{\mathit{F}} & = \{ \mathit{F}, \mathit{H}+\phi^A \lambda_A \}_D  = \{ \mathit{F}, \mathit{H} \}_D + \{ \mathit{F},\phi^A \}_D \lambda_A = \{ \mathit{F}, \mathit{H} \}_D,
\end{align} 
where we used Eq. (\ref{a2.47}); $\{\phi^A, \mathit{F} \}_D = 0$.
\item[(iii)] If $\{\phi^A,p\}_P=0$, then using Eq. (\ref{a2.48}) we can find that
\begin{align}
\{ q,p\}_D & = \{ q,p\}_P - \{ q, \phi^A \}_P \mathit{C}_{AB} \{ \phi^B, p\}  \nonumber \\
& =  \{ q,p\}_P = \dfrac{\partial q}{\partial q} \dfrac{\partial p}{\partial p} - \dfrac{\partial q}{\partial p} \dfrac{\partial p}{\partial q} = 1.
\end{align}
\end{itemize}

}

\par{
We are now in a position to calculate $\{ \psi_{\alpha} , \psi_{\beta} \}_D$ for the Hamiltonian (\ref{a2.31}) with the constraints (\ref{b2.35}). That $\{ \psi_{\alpha} , \psi_{\beta} \}_D$ involves a primary constraint $\phi^\beta$ follows from the following relation which obtains from the Lagrangian formalism,
\begin{equation} \label{b2.65}
\phi^\beta = \pi^\beta + \frac{i}{2} \delta^{\alpha \beta} \psi_\alpha.
\end{equation}
Using the definition of Dirac bracket we  get 
\begin{equation} \label{b2.64}
\{ \psi_{\alpha} , \psi_{\beta} \}_D = \{ \psi_{\alpha} , \psi_{\beta} \}_P - \{ \psi_{\alpha} , \phi^{\alpha} \}_P \mathit{C}_{\alpha \beta} \{ \phi^{\beta}, \psi_{\beta} \}_P.
\end{equation}
Then, since the variables $\psi_{\alpha}$ and $\psi_{\beta}$ are independent, their Poisson bracket vanishes \ie
\begin{equation} \label{b2.66}
\{ \psi_{\alpha} , \psi_{\beta} \}_P = \{ \psi_{\beta} , \psi^{\alpha} \}_P = 0.
\end{equation}
Furthermore, using Eq. (\ref{b2.54}) we find that
\begin{align} \label{b2.67}
\{ \psi_{\alpha}, \phi^{\alpha} \}_P &= \left( \frac{\partial \psi_{\alpha}}{\partial q_i } \frac{\partial \phi^{\alpha}}{\partial p^i} - \frac{\partial \psi_{\alpha}}{\partial p^i} \frac{\partial \phi^{\alpha}}{\partial q_i} \right) - \left( \frac{\partial \psi_{\alpha}}{\partial \psi_\alpha } \frac{\partial \phi^{\alpha}}{\partial \pi^\alpha} + \frac{\partial \psi_{\alpha}}{\partial \pi^\alpha} \frac{\partial \phi^{\alpha}}{\partial \psi_\alpha} \right)   \nonumber \\
&=-1.
\end{align}
Similarly, using Eq. (\ref{b2.65}) we find that
\begin{align} \label{b2.68}
\{ \psi_{\beta}, \phi^{\beta} \}_P &= \left( \frac{\partial \psi_{\beta}}{\partial q_i } \frac{\partial \phi^{\beta}}{\partial p^i} - \frac{\partial \psi_{\beta}}{\partial p^i} \frac{\partial \phi^{\beta}}{\partial q_i} \right) 
- \left( \frac{\partial \psi_{\beta}}{\partial \psi_\beta } \frac{\partial \phi^{\beta}}{\partial \pi^\beta} + \frac{\partial \psi_{\beta}}{\partial \pi^\beta} \frac{\partial \phi^{\beta}}{\partial \psi_\beta} \right)   \nonumber \\
&=-1.
\end{align}
In addition, using the constraints definitions (\ref{b2.54}, \ref{b2.65}) we can define a matrix
\begin{align}
\mathit{C}^{ \alpha \beta} &= \{ \phi^\alpha, \phi^\beta\}_P \nonumber \\
&= \resizebox{.90 \textwidth}{!} {$\left( \frac{\partial \phi^{\alpha}}{\partial q_i } \frac{\partial \phi^{\beta}}{\partial p^i} - \frac{\partial \phi^{\alpha}}{\partial p^i} \frac{\partial \phi^{\beta}}{\partial q_i} \right)
- \left( \frac{\partial \phi^{\alpha}}{\partial \psi_\alpha } \frac{\partial \phi^{\beta}}{\partial \pi^\alpha} + \frac{\partial \phi^{\alpha}}{\partial \pi^\alpha} \frac{\partial \phi^{\beta}}{\partial \psi_\alpha} \right)  
- \left( \frac{\partial \phi^{\alpha}}{\partial \psi_\beta } \frac{\partial \phi^{\beta}}{\partial \pi^\beta} + \frac{\partial \phi^{\alpha}}{\partial \pi^\beta} \frac{\partial \phi^{\beta}}{\partial \psi_\beta} \right)$} \nonumber \\
&= -i \delta^{\alpha \beta}.
\end{align}
Moreover, from the definition (\ref{b2.47}) we have
\begin{equation} \label{b2.70}
\mathit{C}_{\alpha \beta} = -\mathit{C}^{\alpha \beta} = i \delta^{\alpha \beta}.
\end{equation} 
As a result of substituting with Eqs. (\ref{b2.66}, \ref{b2.67}, \ref{b2.68}, \ref{b2.70}) into Eq. (\ref{b2.64}), the Dirac bracket of the two variables $\psi_{\alpha}$ and $\psi_{\beta}$ gives us
\begin{equation}
\{ \psi_{\alpha} , \psi_{\beta} \}_D = i \delta^{\alpha \beta}.
\end{equation}

}

\subsection{Dirac Bracket and the Equations of Motion} 
\par{ 
In this subsection, we show that the Dirac brackets give the correct equations of motion for the Hamiltonian (\ref{b2.34}) using the expression
\begin{equation}
\dot{\psi}_{\alpha} = \{ \psi_{\alpha} , \mathit{H} \}_D,
\end{equation}
We check that
\begin{align}
\dot{\psi}_1 & = \{ \psi_1, \mathit{H} \}_D = \{ \psi_1, - i\omega \psi_1 \psi_2 \}_D = -i\omega \{ \psi_1,\psi_1 \psi_2 \}_D \nonumber \\
&= -i\omega \left( \{ \psi_1,\psi_1 \}_D \psi_2 + \psi_1 \{ \psi_1,\psi_2 \}_D \right) = -i\omega \left( -i \delta_{\alpha \beta} \psi_2 + 0 \right) \nonumber \\
&= - \omega \delta_{\alpha \beta} \psi_2 = - \omega \psi_2 ,
\end{align}
and similarly
\begin{align}
\dot{\psi}_2 &= \{ \psi_2, \mathit{H} \}_D = \{ \psi_2, - i\omega \psi_1 \psi_2 \}_D = -i\omega \{ \psi_2,\psi_1 \psi_2 \}_D \nonumber \\ 
&= -i\omega \left( \{ \psi_2,\psi_1 \}_D \psi_2 + \psi_1 \{ \psi_2,\psi_2 \}_D \right) = -i\omega \left( 0 + \psi_1 (-i \delta_{\alpha \beta}) \right) \nonumber \\
&= - \omega \psi_1 \delta_{\alpha \beta}  =  \omega \delta_{\alpha \beta} \psi_1  = \omega \psi_1.  
\end{align}

}
\par{
These equations are the same equations of motion which we have obtained in Subsection \ref{SubSec.2.3.3}  using the constrained Hamilton's equations of motion \ref{b2.44}.}
 
\subsection{The Dirac Bracket Superalgebra} \label{SubSec2.3.5}
\par{
In this subsection, we investigate the supersymmetry of our Hamiltonian (\ref{a2.31}). For this purpose we need to introduce some operators $\mathit{Q}_i$ are known as supercharges, which basically can be obtained from N\"{o}ether's theorem arising from a symmetry of the Lagrangian which exchange bosons and fermions. In the next section, we will look at the action of the supercharge operators. For instance, the supercharge operators in the case of the system under discussion here, can be written as follows
\begin{align}\label{q2.71}
\mathit{Q}_\alpha  = p \psi_\alpha + \omega q \epsilon_{\alpha \beta} \delta^{\beta \gamma} \psi_\gamma \quad \Rightarrow \quad 
\begin{split} 
\mathit{Q}_1  &= p \psi_1 + \omega q \psi_2  \\
\mathit{Q}_2  &= p \psi_2 - \omega q \psi_1,
\end{split}
\end{align}
where again we suppose that $\{\alpha,\beta\}$ can only take the values $\{1,2\}$, and $\epsilon_{\alpha \beta}$ is the two index Levi-Civita symbol which defined as  
\begin{equation}
\epsilon_{\alpha \beta }= \begin{cases} \; \; +1 \quad \texttt{if} \: (\alpha ,\beta) \: \texttt{is} \: (1,2), \\ \; \; -1 \quad \texttt{if} \: (\alpha ,\beta) \:\texttt{is} \: (2,1), \\ \; \; 0 \quad \: \: \: \, \texttt{if} \: \alpha=\beta. \end{cases}
\end{equation}
The system to be supersymmetric, the supercharges (\ref{q2.71}) together with the Hamiltonian (\ref{b2.34}) and the constraints (\ref{a2.31}) must satisfy the classical Dirac bracket super-algebra, which defined by
\begin{align} \label{aa2.73}
\{ \mathit{Q}_{\alpha}, \mathit{Q}_{\beta} \}_D & = - 2 i \delta_{\alpha \beta} \mathit{H}, \nonumber \\
\{ \mathit{Q}_{\alpha}, \mathit{H} \}_D &=0.
\end{align}  

}
\par{
Now we need to check that our system satisfies Eq. (\ref{aa2.73}). To do that, suppose $\{\alpha , \beta\}$ have the values $\{1,2\}$. Then, using the definition of Poisson and Dirac brackets we find that,(see Appendix \ref{AppendixA.2} for more details)
\begin{align}\label{aa2.74}
\{ \mathit{Q}_1, \mathit{Q}_1 \}_D &= -i \left( p^2+\omega^2 q^2 \right), && \{ \mathit{Q}_1, \mathit{Q}_2 \}_D = \omega \left( \psi_1^2+\psi_2^2 \right), \nonumber \\ 
\{ \mathit{Q}_2, \mathit{Q}_1 \}_D &= - \omega \left( \psi_1^2+\psi_2^2 \right), && \{ \mathit{Q}_2, \mathit{Q}_2 \}_D = -i \left( p^2+\omega^2 q^2 \right).
\end{align}
Substituting using Eqs. (\ref{aa2.74}), we find 
\begin{align} \label{aa2.75}
\{Q_\alpha, Q_\beta \}_D & = \{Q_1, Q_1 \}_D + \{Q_1, Q_2 \}_D + \{Q_2, Q_1 \}_D + \{Q_2, Q_2 \}_D \nonumber \\
& = - i ( p^2 + \omega^2 q^2) + \omega (\psi_1^2 + \psi_2^2)  - \omega (\psi_1^2 + \psi_2^2) - i ( p^2 + \omega^2 q^2) \nonumber \\
&= - 2i ( p^2 + \omega^2 q^2) \equiv -2 i \delta_{\alpha \beta} \mathit{H}.
\end{align}
Furthermore, using the following relations (see Appendix \ref{AppendixA.2})
\begin{align}\label{aa2.76}
\{ \mathit{Q}_1, \mathit{H} \}_P & = \omega p \psi_2 - \omega^2 q \psi_1, && \{ \mathit{Q}_2, \mathit{H} \}_D = - \omega p \psi_2 + \omega^2 q \psi_1, \nonumber \\ 
\{ \phi_1, \mathit{H} \}_P &= i \omega \psi_2, && \{ \phi_2, \mathit{H} \}_P = i \omega \psi_2.
\end{align}
we find 
\begin{align} \label{aa2.77}
\{Q_\alpha, \mathit{H}\}_D & = \{Q_1, \mathit{H}\}_D + \{Q_2, \mathit{H}\}_D \nonumber \\
& = \{Q_1, \mathit{H}\}_P - \{ Q_1, \phi^A \}_P C_{AB} \{ \phi^B, \mathrm{H} \}_P \nonumber \\
& \quad + \{Q_2, \mathit{H}\}_P - \{ Q_2, \phi^A \}_P C_{AB} \{ \phi^B, \mathit{H} \}_P  \nonumber \\
& = \{Q_1, \mathit{H}\}_P - \{ Q_1, \phi^1 \}_P C_{11} \{ \phi^1, \mathit{H} \}_P - \{ Q_1, \phi^2 \}_P C_{22} \{ \phi^2, \mathit{H} \}_P \nonumber \\
& \quad + \{Q_2, \mathit{H}\}_P - \{ Q_2, \phi^1 \}_P C_{11} \{ \phi^1, \mathit{H} \}_P - \{ Q_2, \phi^2 \}_P C_{22} \{ \phi^2, \mathit{H} \}_P \nonumber \\
& =   \omega P \psi_2 - \omega^2 q \psi_1 - \omega P \psi_2 + \omega^2 q \psi_1 - \omega P \psi_1 - \omega^2 q \psi_2 + \omega^2 q \psi_2 + \omega P \psi_1  \nonumber \\
& = 0.
\end{align}

}
\par{
It is clear from Eqs. (\ref{aa2.75}, \ref{aa2.77}), that the supercharges (\ref{q2.71}) together with the Hamiltonian (\ref{b2.34}) and the constraints (\ref{b2.35}) satisfy the two conditions (\ref{aa2.73}) of the classical Dirac bracket superalgebra.} 
 
\section{The Supersymmetry Algebra} \label{Sec2.4}
\par{
The material in this section elaborated in detail in \cite{kugo1983supersymmetry,drees1996introduction, bilal2001introduction,combescure2004n, bougie2012supersymmetric,labelle2010supersymmetry}. The supersymmetry algebra encodes a symmetry describing a relation between bosons and fermions. In general, the supersymmetry is constructed by introducing supersymmetric transformations which are generated by the supercharge $\mathit{Q}_i$ operators. Where the role of the supercharges $\mathit{Q}_i$ is to convert a fermionic degree of freedom into a bosonic degree of freedom and vice versa, \ie
\begin{equation}
  \mathit{Q} \vert \texttt{fermionic} \rangle = \vert \texttt{bosonic} \rangle , \qquad \mathit{Q} \vert \texttt{bosonic} \rangle= \vert \texttt{fermionic} \rangle.
\end{equation}
   
So far, we have restricted ourselves to study classical systems. In Subsection \ref{SubSec2.3.5} we have investigated the supersymmetry of the classical harmonic oscillator. Now we are going to quantize the theory to study the supersymmetry of quantum systems. In a quantum mechanical supersymmetric system, the supercharges $\mathit{Q}_i$ together with the Hamiltonian $\mathit{H}$ form a so-called superalgebra. The recipe for quantizing the Hamiltonian in a situation where we have second class constraints is to replace Dirac brackets with either commutator brackets for bosonic variables or anticommutator brackets for fermionic variables multiplied with the factor $i \hbar$, so we have
\begin{align} \label{d2.90}
&&&&&&&& \{ q,p\}_D &= 1 & \Rightarrow & \qquad &[ \hat{q}, \hat{p}] &= i \hbar, &&&&&&&& \nonumber \\
&&&&&&&& \{\psi_{\alpha}, \psi_{\beta} \}_D &= -i \delta_{\alpha \beta} & \Rightarrow & \qquad &\{ \hat{\psi}_{\alpha}, \hat{\psi}_{\beta} \} &= \hbar \delta_{\alpha \beta}. &&&&&&&&
\end{align}
Taking this into account, we can see that the superalgebra for N-dimensional quantum system is characterized by
\begin{align} \label{2.24}
 &&&&&&       [ \hat{\mathit{Q}}_i, \hat{\mathit{H}}] &= 0, & i &= 1 \cdots \mathit{N}, &&&&&& \nonumber \\  &&&&&&      
\{ \hat{\mathit{Q}}_i, \hat{\mathit{Q}}_j \} &= \hbar \delta_{ij} \hat{\mathit{H}}, & i,j &= 1     \dots \mathit{N}. &&&&&&
\end{align}
This will be elaborated further in the next section using the example of the supersymmetrical harmonic oscillator. One more notation before ending this section, the supercharges $\mathit{Q}_i$ are hermitian \ie $\mathit{Q_i^{\dagger}}=\mathit{Q_i}$, and this implies that
  \begin{equation}
  \{ \mathit{\hat{Q}}_i,\mathit{\hat{Q}}_j \}=  \{ \mathit{\hat{Q}}_i^{\dagger},\mathit{\hat{Q}}_j^{\dagger} \}.
  \end{equation}
  
  }

\section{Supersymmetric Harmonic Oscillator} \label{Sec2.5}
\par{ 
Now we turn back to the quantum harmonic oscillator problem as a simple example to show the supersymmetric property of a quantum mechanical system. Using Eq. (\ref{2.16}), the harmonic oscillator Hamiltonian (\ref{2.17}) can be written in terms of the bosonic and fermionic laddering operators as follows   
\begin{equation} \label{c2.93}
\mathit{\hat{H}} = \hbar \omega ( \hat{a}^\dagger \hat{a} + \hat{b}^\dagger \hat{b} ).
\end{equation}
Moreover, the supercharge operators $\mathit{\hat{Q}}_1$, $\mathit{\hat{Q}}_2$ also can be written in terms of the bosonic and fermionic laddering operators as follows 
\begin{align} \label{c2.94}
\mathit{\hat{Q}}_1 &= p \psi_1 + \omega q \psi_2 \nonumber \\
&= \left( i \sqrt{\frac{\hbar \omega}{2}} (\hat{a}^\dagger - \hat{a}) \right) \left( \sqrt{\frac{\hbar}{2}} (\hat{b} + \hat{b}^\dagger) \right) + \omega \left( \sqrt{\frac{\hbar}{2 \omega}} (\hat{a})^\dagger + \hat{a} ) \right) \left( i \sqrt{\frac{\hbar}{2}} (\hat{b}-\hat{b}^\dagger) \right) \nonumber \\
&= i \hbar \sqrt{\omega} (\hat{a}^\dagger \hat{b}- \hat{a} \hat{b}^\dagger ),
\end{align}
and similarly,
\begin{align} \label{c2.95}
\mathit{\hat{Q}}_2 &= p \psi_2 - \omega q \psi_1 \nonumber \\
&= \left( i \sqrt{\frac{\hbar \omega}{2}} (\hat{a}^\dagger - \hat{a}) \right) \left( i \sqrt{\frac{\hbar}{2}} (\hat{b}-\hat{b}^\dagger) \right) - \omega \left( \sqrt{\frac{\hbar}{2 \omega}} (\hat{a})^\dagger + \hat{a} ) \right) \left( \sqrt{\frac{\hbar}{2}} (\hat{b} + \hat{b}^\dagger) \right) \nonumber \\
&= - \hbar \sqrt{\omega} (\hat{a}^\dagger \hat{b}+ \hat{a} \hat{b}^\dagger ).
\end{align}

}
\par{
Using Eqs. (\ref{c2.94}, \ref{c2.95}) one may define non-hermitian operators $\mathit{\hat{Q}}$ and $\mathit{\hat{Q}}^\dagger$ as
\begin{align} \label{c2.96}
\mathit{\hat{Q}} &= \frac{1}{2} ( \mathit{\hat{Q}}_1 -i \mathit{\hat{Q}}_2 ) \nonumber \\
&=  \frac{1}{2} \left( i \hbar \sqrt{\omega} (\hat{a}^\dagger \hat{b}- \hat{a} \hat{b}^\dagger ) + i\hbar \sqrt{\omega} (\hat{a}^\dagger \hat{b}+ \hat{a} \hat{b}^\dagger ) \right) \nonumber \\
&= i \hbar \sqrt{\omega} (\hat{a}^\dagger \hat{b}),
\end{align} 
and similarly,
\begin{align} \label{c2.97}
\mathit{\hat{Q}}^\dagger &= \frac{1}{2} ( \mathit{\hat{Q}}_1 + i \mathit{\hat{Q}}_2 ) \nonumber \\
&=  \frac{1}{2} \left( i \hbar \sqrt{\omega} (\hat{a}^\dagger \hat{b}- \hat{a} \hat{b}^\dagger ) - i\hbar \sqrt{\omega} (\hat{a}^\dagger \hat{b}+ \hat{a} \hat{b}^\dagger ) \right) \nonumber \\
&= - i \hbar \sqrt{\omega} (\hat{a} \hat{b}^\dagger).
\end{align} 
We can directly use Eq. (\ref{c2.93}, \ref{c2.96}, \ref{c2.97}) to check that the supercharge operators $\mathit{\hat{Q}}$ and $\mathit{\hat{Q}}^\dagger$ with the Hamiltonian $\mathit{H}$ satisfy the quantum superalgebra
\begin{align} \label{c2.98}
&&&&&&&&&&&&&&&& \{ \mathit{\hat{Q}}, \mathit{\hat{Q}} \} & =& \{ \mathit{\hat{Q}}^\dagger , \mathit{\hat{Q}}^\dagger \} &= 0, &&&&&&&&&&&&&&&& \nonumber \\
&&&&&&&&&&&&&&&& \{ \mathit{\hat{Q}}, \mathit{\hat{Q}}^\dagger \} &=& \{ \mathit{\hat{Q}}^\dagger , \mathit{\hat{Q}} \} &= \mathit{H},&&&&&&&&&&&&&&&& \nonumber \\
&&&&&&&&&&&&&&&& [ \mathit{H}, \mathit{\hat{Q}} ] &= &[ \mathit{H}, \mathit{\hat{Q}}^\dagger ] &= 0. &&&&&&&&&&&&&&&&
\end{align} 

}

\par{
Now, it is easy using Eqs. (\ref{c2.96}, \ref{c2.97}) to see the action of the operators $\mathit{\hat{Q}}, \mathit{\hat{Q}}^{\dagger}$ on the energy eigenstates:
\begin{align}
\mathit{\hat{Q}} \vert \mathit{n}_{B}, \mathit{n}_{F} \rangle & \sim  \vert \mathit{n}_{B}-1, \mathit{n}_{F}+1 \rangle, \nonumber  \\
\mathit{\hat{Q}}^{\dagger} \vert \mathit{n}_{B}, \mathit{n}_{F} \rangle & \sim  \vert \mathit{n}_{B}+1, \mathit{n}_{F}-1 \rangle.
\end{align}
Remember that there are only two possible fermionic states $\mathit{n}_{F}=\{0,1\}$, so the effect of the operators $\mathit{\hat{Q}}$ and $\mathit{\hat{Q}}^{\dagger}$ can be written as follows  
\begin{align}
\mathit{\hat{Q}} \vert \mathit{n}_{B}, 0 \rangle & \sim \vert \mathit{n}_{B}-1, 1 \rangle, \nonumber  \\
\mathit{\hat{Q}}^{\dagger} \vert \mathit{n}_{B}, 1 \rangle & \sim  \vert \mathit{n}_{B}+1, 0 \rangle.
\end{align}
Also, from Eq. (\ref{c2.98})  the operators $\mathit{\hat{Q}}$ and $\mathit{\hat{Q}}^\dagger$ commute with the Hamiltonian $\mathit{H}$, for that reason we can see that
\begin{equation}
\mathit{H} (\mathit{\hat{Q}} \vert n_B, n_F \rangle)= \mathit{\hat{Q}} (\mathit{H} \vert n_B, n_F \rangle)= (E_B + E_F) (\mathit{\hat{Q}} \vert n_B, n_F \rangle),
\end{equation} 
this means that the whole energy of the system remains unchanged by the action of the supercharge operators.
}

\par
\begin{figure} [h!] 
\centering
\begin{tikzpicture} [scale=1.5]
  \draw[gray,very thick] (0,0)--(5,0);
  \draw[gray,very thick,->] (0,0) -- (0,4);

  \draw[ultra thick,blue] (1,0.025)--(2,0.025);
  \draw[ultra thick,blue] (1,1)--(2,1);
  \draw[ultra thick,blue] (1,2)--(2,2);
  \draw[ultra thick,blue] (1,3)--(2,3);

  \draw[black] (1,.22) node[above,left] {${\scriptscriptstyle \mathit{E}_0^{(1)}}$};
  \draw[black] (1,1.1) node[above,left] {${\scriptscriptstyle \mathit{E}_1^{(1)}}$};
  \draw[black] (1,2.1) node[above,left] {${\scriptscriptstyle \mathit{E}_2^{(1)}}$};
  \draw[black] (1,3.1) node[above,left] {${\scriptscriptstyle \mathit{E}_3^{(1)}}$};
  
  \draw[ultra thick,red] (3,1)--(4,1);
  \draw[ultra thick,red] (3,2)--(4,2);
  \draw[ultra thick,red] (3,3)--(4,3);
  
  \draw[black] (4,1.1) node[above,right] {${\scriptscriptstyle \mathit{E}_0^{(2)}}$};
  \draw[black] (4,2.1) node[above,right] {${\scriptscriptstyle \mathit{E}_1^{(2)}}$};
  \draw[black] (4,3.1) node[above,right] {${\scriptscriptstyle \mathit{E}_2^{(2)}}$};
  
  \draw[->] (1.5,3.05) to [bend left] (3.5,3.05);
  \draw[->] (3.5,2.95)  to [bend left] (1.5,2.95);
  \draw[black] (-.1,3.75) node[left] {$\scriptstyle \mathit{E}$};
  \draw[black] (0,0.1) node[above,left] {${\scriptscriptstyle \mathit{E}=0}$};
  \draw[black] (1.5,-.3) node[] {${\scriptstyle \mathit{N}_F = 0 }$};
  \draw[black] (3.5,-.3) node[] {${\scriptstyle \mathit{N}_F = 1 }$};
  \draw[black] (2.5,3.5) node[] {${\scriptscriptstyle \mathit{Q}}$};
  \draw[black] (2.5,2.4) node[] {${\scriptscriptstyle \mathit{Q}^{\dagger}}$};
\end{tikzpicture}
\caption{Energy spectrum for supersymmetric quantum mechanics. $\mathit{Q}$ and $\mathit{Q}^{\dagger}$ exchange bosons and fermions without affecting the energy.}
\label{f2.1} 
\end{figure}
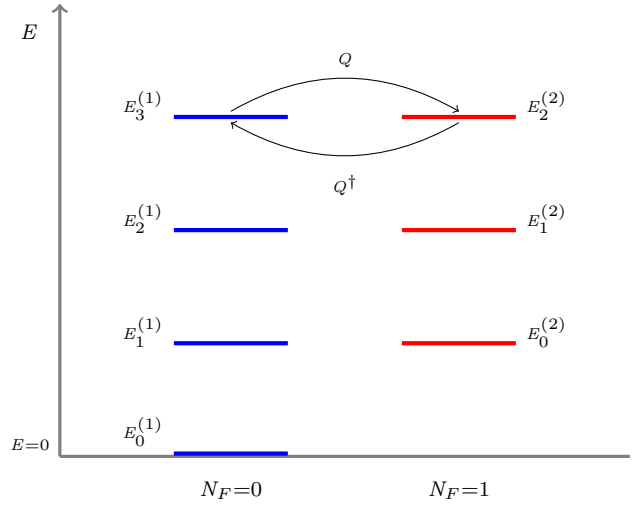

{
In short, the supercharge operator $\mathit{\hat{Q}}$ acts to change one boson to one fermion leaving the total energy of the system invariant. Conversely, $\mathit{\hat{Q}}^\dagger$ changes a fermion into a boson leaving the energy unchanged. This is illustrated in Fig. \ref{f2.1}.} 
 
\section{Supersymmetric Quantum Mechanics} \label{Sec2.6}

\par{
In this section, we will study the general formalism of one–dimensional supersymmetric quantum mechanics. The ideas in this section were discussed in \cite{witten1981dynamical, witten1982constraints, rodrigues2002quantum, ioffe2006susy, gudmundsson2014supersymmetric, berman2002supersymmetric}. We begin by considering a Hamiltonian of the form
\begin{equation} \label{2.26}
\mathit{\hat{H}}= \frac{1}{2} \left( \hat{p}^2 + V(\hat{x}) \right) \mathbbmtt{I}_2 + \frac{1}{2} \hbar \mathit{B}(\hat{x}) \sigma_3,
\end{equation}
where $V(\hat{x})$ is the potential, and $\mathit{B}$ is a magnetic field. If those potential and magnetic field can be written in terms of some function $\mathit{W}(x)$, as
\begin{equation} \label{2.27}
V(x)= \left( \dfrac{d \mathit{W}(x)}{dx} \right)^2 \equiv {\mathit{W}'}^2 \qquad \& \qquad \mathit{B}(x)= \left( \dfrac{d^2 \mathit{W}(x)}{dx^2} \right) \equiv \mathit{W}'',
\end{equation}
the Hamiltonian is supersymmetric and we shall refer to the function $\mathit{W}(x)$ as a superpotential. Using Eq. (\ref{2.27}) we can rewrite the Hamiltonian (\ref{2.26}) as
\begin{equation} \label{2.28}
\mathit{\hat{H}}= \frac{1}{2} \left( \hat{p}^2 + {\mathit{W}'}^2 \right) \mathbbmtt{I}_2 + \frac{1}{2} \hbar \sigma_3 \mathit{W}''. 
  \end{equation}
  
  }
\par{  
At this stage, we define the following two hermitian supercharge operators
  \begin{align} \label{2.29}
  \mathit{\hat{Q}}_1 &= \frac{1}{2} \left( \sigma_1 \hat{p}+\sigma_2 \mathit{W}'(\hat{x}) \right), \nonumber \\
  \mathit{\hat{Q}}_2 &= \frac{1}{2} \left( \sigma_2 \hat{p} - \sigma_1 \mathit{W}'(\hat{x}) \right).
  \end{align}  
The Hamiltonian (\ref{2.28}) together with the supercharges (\ref{2.29}) constitute a superalgebra. It is a simple matter of algebra to verify the conditions for the super-algebra given by Eq. (\ref{2.24}). It may be useful before verifying these conditions to check the following commutation relation
\begin{align}
[ \hat{p}, f(x)] \: g &= [ -i\hbar \frac{\partial }{\partial x}, f(x)] \: g = -i \hbar \left( \frac{\partial }{\partial x} (fg) + f \frac{\partial }{\partial x} (g) \right) \nonumber \\ 
&= -i \hbar \left( f \cdot \frac{\partial g }{\partial x} + \frac{\partial f }{\partial x} \cdot g - f \cdot \frac{\partial g }{\partial x}  \right) = -i\hbar f'(x)\: g.
\end{align}
Removing $g$ gives us the commutator of the momentum $\hat{p}$ with an arbitrary function of the position coordinate $x$:
\begin{equation} \label{d2.107}
[ \hat{p}, f(x)] = -i\hbar f'(x).
\end{equation}
Using this relation we find the following
\begin{align} \label{d2.108}
[\hat{p}, {\mathit{W}'}^2]&= \hat{p} {\mathit{W}'}^2 - {\mathit{W}'}^2 \hat{p} = - 2i\hbar \mathit{W}' \mathit{W}'', \nonumber \\
[ \mathit{W}', \hat{p}^2 ] &= \mathit{W}' \hat{p}^2- \hat{p}^2 \mathit{W}' = i \hbar \{ \mathit{W}'', \hat{p}\} , \nonumber \\
\{ \hat{p}, \mathit{W}''\} &= \hat{p} \mathit{W}'' + \mathit{W}'' \hat{p}.
\end{align}
Now we can use the outcome of Eqs. (\ref{d2.108}) to check the outcome of the commutator of the supercharge $\mathit{\hat{Q}}_1$ with the Hamiltonian $\mathit{\hat{H}}$:
\begin{align}  \label{2.30}
[\mathit{\hat{Q}}_1, \mathit{\hat{H}}] & = \mathit{\hat{Q}}_1 \mathit{\hat{H}} - \mathrm{H} \mathit{\hat{Q}}_1 \nonumber \\ 
& = \frac{1}{4} \left( \sigma_1 \hat{p} + \sigma_2 \mathit{W}' \right) \cdot \left( \left( \hat{p}^2 + {\mathit{W}'}^2\right) \mathbbmtt{I}_2 + \hbar \mathit{W}'' \sigma_3 \right) \nonumber \\  
& \quad - \frac{1}{4} \left( \left(\hat{p}^2 + {\mathit{W}'}^2\right) \mathbbmtt{I}_2 + \hbar \mathit{W}'' \sigma_3 \right) \cdot \left(\sigma_1 \hat{p} + \sigma_2 \mathit{W}'\right) \nonumber \\
& = \resizebox{.85 \textwidth}{!} {$\frac{1}{4}  \left( \sigma_1 \hat{p}^3 + \sigma_1 \hat{p} {\mathit{W}'}^2 + \sigma_1 \sigma_3 \hbar \hat{p} \mathit{W}''  + \sigma_2 \mathit{W}' \hat{p}^2 + \sigma_2 \mathit{W}' {\mathit{W}'}^2 +\sigma_2 \sigma_3 \hbar \mathit{W}'  \mathit{W}''\right) $}\nonumber \\
 & \quad - \resizebox{.85 \textwidth}{!} {$ \frac{1}{4} \left( \sigma_1 \hat{p}^3 +\sigma_2 \hat{p}^2 \mathit{W}' + \sigma_1 {\mathit{W}'}^2 \hat{p} +\sigma_2  {\mathit{W}'}^2 \mathit{W}'+ \sigma_3 \sigma_1 \hbar \mathit{W}'' \hat{p} +\sigma_3 \sigma_2 \hbar \mathit{W}'' \mathit{W}'\right) $} \nonumber \\ 
& =  \frac{1}{4} \left( \sigma_1 [\hat{p}, {\mathit{W}'}^2] - \sigma_2 i \hbar \{\hat{p}, \mathit{W}'' \} + \sigma_2 [\mathit{W}', \hat{p}] + 2 \sigma_1 i \hbar \mathit{W}' \mathit{W}''  \right) \nonumber \\
& = 0.
    \end{align}
Similarly, one can confirm that
\begin{equation} \label{2.31}
[\mathit{\hat{Q}}_2, \mathit{\hat{H}}] = \mathit{\hat{Q}}_2 \mathit{\hat{H}} - \mathit{\hat{H}} \mathit{\hat{Q}}_2 = 0. 
\end{equation}
Furthermore, we can find that
\begin{align} \label{2.32}
\{ \hat{Q}_1, \hat{Q}_1 \} & = \hat{Q}_1 \hat{Q}_1 + \hat{Q}_1 \hat{Q}_1 = 2 \hat{Q}_1^2 \nonumber \\
& = \frac{1}{2} \left( \sigma_1 \hat{p} + \sigma_2 \mathit{W}' \right)^2 \nonumber \\ 
& =  \frac{1}{2} \left(\sigma_1^2 \hat{p}^2 + \sigma_2^2 {\mathit{W}'}^2 
+\sigma_1  \sigma_2 \hat{p} \mathit{W}' + \sigma_2 \sigma_1 \mathit{W}' \hat{p} \right) \nonumber \\
&= \frac{1}{2} \left( ( \hat{p}^2 +  {\mathit{W}'}^2 ) \mathbbmtt{I}_2 + i \sigma_3  ( \hat{p} \mathit{W}' - \mathit{W}' \hat{p}) \right) \nonumber \\ 
&= \frac{1}{2} \left( ( \hat{p}^2 + {\mathit{W}'}^2 )\mathbbmtt{I}_2 + \hbar \sigma_3 \mathit{W}'' \right)  \equiv \mathrm{H},
\end{align}
and  similarly, we have
\begin{equation} \label{2.34}
\{ \mathit{\hat{Q}}_2^{\dagger}, \mathit{\hat{Q}}_2\} =\hat{\mathrm{H}}. 
\end{equation}
Note that, to do the previous calculations we used the Pauli matrices properties: $\sigma_1^2=\sigma_2^2=\sigma_3^2= \mathbbmtt{I}_2$ and $\sigma_1 \sigma_2=-\sigma_2 \sigma_1= i \sigma_3$.
It is clear from Eqs. (\ref{2.30},\ref{2.31}, \ref{2.32}, and \ref{2.34}) that the Hamiltonian (\ref{2.28}) together with the supercharges (\ref{2.29}) satisfy the conditions for the superalgebra given by Eq. (\ref{2.24}).

}

\par{
Indeed if we define the quantities $\mathit{Q}$ and $\mathit{Q}^{\dagger}$   as
\begingroup \allowdisplaybreaks
\begin{align}
\mathit{\hat{Q}} &= \mathit{\hat{Q}}_1 -i \mathit{\hat{Q}}_2 \nonumber \\
&= \frac{1}{2} \left( \sigma_1 \hat{p} + \sigma_2 \mathit{W}' \right) - \frac{1}{2} i \left( \sigma_2 \hat{p} - \sigma_1 \mathit{W}' \right) \nonumber \\
&= \frac{1}{2} \left( (\sigma_1 - i \sigma_2) \hat{p} + (\sigma_2+i \sigma_1) \mathit{W}'\right)  \nonumber \\
&= \frac{1}{2} (\sigma_1 - i \sigma_2) \left( \hat{p} + i \mathit{W}' \right) \nonumber \\
&= \sigma_{-} \left( \hat{p} +i \mathit{W}' \right), 
\end{align} \endgroup
and
\begin{equation}
\hat{Q}^\dagger = \hat{Q}_1 +i \hat{Q}_2 = \sigma_{+} \left( \hat{p} -i \mathit{W}' \right), 
\end{equation}
where $\sigma_{-}$ and $\sigma_{+}$ respectively are defined as
\begin{align}
\sigma_{-} &= \frac{1}{2} (\sigma_1 - i \sigma_2) = \left( \begin{matrix}0&0\\1&0 \end{matrix} \right), \nonumber \\
\sigma_{+} &= \frac{1}{2} (\sigma_1 + i \sigma_2) = \left( \begin{matrix}0&1\\0&0 \end{matrix} \right),
\end{align}
it is straightforward to check that the Hamiltonian $\mathit{\hat{H}}$ can be expressed as
  \begin{equation} \label{f2.116}
  \mathit{\hat{H}} = \{ \mathit{\hat{Q}}, \mathit{\hat{Q}}^{\dagger} \},
  \end{equation}  

and it commutes with both the operators $\mathit{\hat{Q}}$ and $\mathit{\hat{Q}}^{\dagger}$
\begin{equation} \label{2.39}
[\mathit{\hat{Q}}, \mathit{\hat{H}}]=0 \quad \& \quad [\mathit{\hat{Q}}^{\dagger}, \mathit{\hat{H}}]=0.
\end{equation}
Furthermore, one also finds
\begin{equation}
\{\mathit{\hat{Q}}, \mathit{\hat{Q}}\}=0 \quad \& \quad \{\mathit{\hat{Q}}^{\dagger}, \mathit{\hat{Q}}^{\dagger}\}=0.
\end{equation}

}
\par{
As we have seen in the case of the supersymmetric Harmonic oscillator, a supersymmetric system with two independent supercharges, with the possible exception of the energy of the ground eigenstate, all the energy levels are split into two eigenstates with either $n_F=0$ or $n_F=1$. For a spinor quantum mechanical system, this implies that the excited energy eigenstates comes in degenerate spin-up/spin-down pairs $\vert E_n, \uparrow \rangle$/$\vert E_n, \downarrow \rangle$. This degenerate spin-up/spin-down pairs related to the acts of the supercharge operators $\mathit{\hat{Q}}$ and $\mathit{\hat{Q}}^{\dagger}$. Where the supercharge operators $\mathit{\hat{Q}}$ convert the degenerate spin-up state to the degenerate spin-down state without making any change in the energy eigenvalue of the states
\begin{equation}
\hat{Q} \vert E_n , \uparrow \rangle \sim \left( \begin{matrix} 0 &0 \\ 1&0 \end{matrix} \right) \left( \begin{matrix} \uparrow \\ 0  \end{matrix} \right) = \left( \begin{matrix} 0 \\ \downarrow  \end{matrix} \right).
\end{equation}  
While the supercharge operators $\mathit{\hat{Q}}^{\dagger}$ convert the degenerate spin-down state to the degenerate spin-up state without making any change in the energy eigenvalue of the states
\begin{equation}
\hat{Q}^\dagger \vert E_n , \downarrow \rangle \sim \left( \begin{matrix} 0 &1 \\ 0&0 \end{matrix} \right) \left( \begin{matrix} 0 \\ \downarrow  \end{matrix} \right) = \left( \begin{matrix} \uparrow \\ 0  \end{matrix} \right).
\end{equation}

}   
  
\par{  
To sum up, the supersymmetric transformations occur due to the supercharge operators $\mathit{\hat{Q}}$ and $\mathit{\hat{Q}}^\dagger$. In this case, it causes transforms between the energy eigenstates spin-up/spin-down which have the same energy eigenvalues.}

\section{Supersymmetric Ground State} \label{Sec2.7}

\par{
So far, we have investigated all the required information to describe a supersymmetric quantum mechanical system. In this section, we will study the supersymmetric quantum mechanical ground state. The argument in this section follows \cite{argyres1996introduction, argyres2001introduction, guilarte2006n, junker2012supersymmetric, salomonson1982fermionic}. Let us now write the expression of the supersymmetric Hamiltonian (\ref{2.28}) as the summation of two separate terms; a Hamiltonian $\mathit{\hat{H}}_{+}$ and a Hamiltonian $\mathit{\hat{H}}_{-}$, then we have
\begin{equation}
\mathit{\hat{H}}_{\pm} = \frac{1}{2} \left( \hat{p}^2 + {\mathit{W}'}^2 \right) \pm \frac{1}{2} \hbar \mathit{W}''.
\end{equation}     
In the eigenbasis of $\sigma_3$, the supersymmetric Hamiltonian is diagonal:
\begin{equation} \label{f2.122}
\mathit{\hat{H}} \equiv \left( \begin{matrix}  \mathit{\hat{H}}_{+} & 0 \\ 0 & \mathit{\hat{H}}_{-}  \end{matrix} \right) = \left( \begin{matrix}  \mathit{A}^{\dagger} \mathit{A} & 0 \\ 0 & \mathit{A} \mathit{A}^{\dagger}  \end{matrix} \right), 
\end{equation} 
and the supercharge operators $\mathit{\hat{Q}}$ and $\mathit{\hat{Q}}^{\dagger}$ can be written as
\begin{equation} \label{f2.123}
\mathit{\hat{Q}} = \left( \begin{matrix}  0 & 0 \\ \mathit{A} & 0 \end{matrix} \right), \qquad 
\mathit{\hat{Q}}^{\dagger} = \left( \begin{matrix}  0 & \mathit{A}^{\dagger} \\ 0 & 0  \end{matrix} \right),
\end{equation}
where
\begin{align} \label{f2.124}
\mathit{A} &= \hat{p} + i \mathit{W}', \qquad
\mathit{A}^{\dagger} = \hat{p} - i \mathit{W}'.
\end{align}

}
\par{
According to Eq. (\ref{f2.116}), we can write the supersymmetric Hamiltonian $\mathit{\hat{H}}$ in terms of the supercharges operators $\mathit{\hat{Q}}$ and $\mathit{\hat{Q}}^{\dagger}$ as follows
\begin{align} \label{f2.125}
\mathit{\hat{H}} &= \mathit{\hat{Q}} \mathit{\hat{Q}}^{\dagger} + \mathit{\hat{Q}}^{\dagger} \mathit{\hat{Q}}  
= 2 {\mathit{\hat{Q}}}^2 = 2 \mathit{\hat{Q}}^{\dagger^2}.
\end{align}
In the last step, we used the fact that the supercharges are hermitian operators. We see from Eq. (\ref{f2.125}) that the Hamiltonian $\mathit{\hat{H}}$ can be written in terms of the squares of the supercharge operators. For this reason, the energy of any eigenstate of this Hamiltonian must be positive or zero. Let us now consider that $\vert \Psi_0 \rangle$ is the ground state of the supersymmetric Hamiltonian $\mathit{\hat{H}}$. Based on Eq. (\ref{f2.125}) it can be seen that the ground state can have a zero energy only if it satisfies the following two conditions
\begin{align} \label{f2.126}
&&&&\mathit{E}_0 &= \langle \Psi_0 \vert \mathit{\hat{H}} \vert \Psi_0 \rangle = \langle \Psi_0 \vert {\mathit{\hat{Q}}}^2 \vert \Psi_0 \rangle = 0  &\Longrightarrow& \quad \mathit{\hat{Q}} \vert \Psi_0 \rangle =0,&&&& \nonumber \\
&&&&\mathit{E}_0 &= \langle \Psi_0 \vert \mathit{\hat{H}} \vert \Psi_0 \rangle = \langle \Psi_0 \vert \mathit{\hat{Q}}^{\dagger^2} \vert \Psi_0 \rangle = 0 &\Longrightarrow& \quad \mathit{\hat{Q}}^{\dagger} \vert \Psi_0 \rangle =0. &&&&
\end{align}
Therefore, if there exists a state which is annihilated by each of the supercharge operators $\mathit{\hat{Q}}$ and $\mathit{\hat{Q}}^{\dagger}$ which means it is invariant under the supersymmetry transformations, such a state is automatically the zero-energy ground state. However, on the other hand, any state is not invariant under the supersymmetry transformations has a positive energy. Thus, if there is a supersymmetric state, it is the zero-energy ground state and it is said that the system on that case has good supersymmetry or the supersymmetry is unbroken.}  
\par{
Moreover, since the supersymmetry algebra (\ref{2.39}) implies that the supercharge operators $\mathit{\hat{Q}}$ and $\mathit{\hat{Q}}^{\dagger}$ commute with the supersymmetric  Hamiltonian $\mathit{\hat{H}}$, so all the eigenstates of $\mathit{\hat{H}}$ are doubly degenerate. For that reason,
it will be convenient to write the supersymmetric ground state of the system in terms of two components, $\psi_{0}^{\pm}$, as follows
\begin{equation} \label{f2.127}
\Psi_{0}= \left( \begin{matrix}  \psi_{0}^{+} \\  \psi_{0}^{-} \end{matrix} \right). 
\end{equation} 
One can now solve eigenvalue problem $\mathit{\hat{Q}} \vert \psi_0 \rangle =0$ to find the zero energy ground state wave function. Substitute using Eqs. (\ref{f2.123},  \ref{f2.127}) we get,
\begin{align}
\mathit{\hat{Q}} \vert \psi_{0} \rangle =
\left( \begin{matrix}  0 & 0 \\ \mathit{A} & 0 \end{matrix} \right) \left( \begin{matrix}  \psi_{0}^{+}  \\ \psi_{0}^{-} \end{matrix} \right) =0,
\end{align} 
then, substitute using Eq. (\ref{f2.124}), our problem reduces to solve the first-order differential equation,
\begin{equation}
\mathit{A} \vert \psi_{0}^{+} \rangle  =\left( \hat{p} + i \mathit{W}' \right) \vert \psi_{0}^{+} \rangle = 0 \quad \Longrightarrow \quad
- i\hbar \dfrac{\partial \psi_{0}^{+}}{\partial x} + i \mathit{W}' \psi_{0}^{+} = 0.
\end{equation}
It is simple and straightforward to solve the previous differential equation as follows
\begin{equation}
\dfrac{d \psi_{0}^{+}}{\psi_{0}^{+}} = \dfrac{\mathit{W}'}{\hbar} dx \quad \Longrightarrow \quad
\ln \psi_{0}^{+} = \frac{\mathit{W}'}{\hbar} + \ln A .
\end{equation}
Thus,
\begin{equation}
\psi_{0}^{+} = A e^{\frac{\mathit{W}}{\hbar}},
\end{equation}
and similarly,
\begin{equation}
\psi_{0}^{+} = B e^{- \frac{\mathit{W}}{\hbar}}.
\end{equation}
Now the general form of the ground state wave function may be expressed as
\begin{equation} \label{f2.133}
\Psi_0 = \left( \begin{matrix} A e^{\frac{\mathit{W}}{\hbar}} \\ B e^{- \frac{\mathit{W}}{\hbar}}. \end{matrix} \right).
\end{equation}
 
}
\par{ 
In fact, it has no physical meaning to find the ground state wave function $\psi_0$ if there are no normalizable solutions of such form. So we need now to normalize the ground state wave function (\ref{f2.133}) by determining the values of the two constants $A$ and $B$. The ground state wave function $\psi_0(x)$ to be normalizable, it must vanish at the positive and the negative infinite $x$-values. In order to satisfy this condition, we have to set $\vert \mathit{W}(x) \vert \rightarrow \infty$ as $\vert x \vert \rightarrow \infty$. Then we have the following three possible cases for the supersymmetric ground state wave function:}
 
\begin{itemize}
\item[1.] 
\par{ The first case, when $\mathit{W}(x) \rightarrow + \infty$ as $x\rightarrow \pm \infty$. In this case, the superpotential  $\mathit{W}(x)$ is positive at the boundaries and has an even number of zeros. Fig. (\ref{F2.1}) describe the behavior of the potential $\mathit{V}(x)$ in this case.} 

\par{Since $\mathit{W}(x)$ is positive, we can not normalize the wave function $\psi_{0}^{+} = \mathit{A} e^{\frac{\mathit{W}(x)}{\hbar}}$, but we can choose $\mathit{A}=0$. However we can normalize the wave function $\varphi_{0}^{-} = B e^{-\frac{\mathit{W}(x)}{\hbar}}$ to find the value of the constant $B$.}

\par{The complete ground state wave function, in this case, may be expressed in the general form
\begin{equation}
\Psi_0 = \left( \begin{matrix} 0 \\ B e^{-\frac{\mathit{W}(x)}{\hbar}} \end{matrix} \right).
\end{equation} 

\begin{figure} [h!]
\centering
\begin{tikzpicture} [scale=1.5]
      \draw[very thick, gray,->] (-2,0) -- (2,0) node[right,black] {$x$};
      \draw[very thick, gray,->] (-.5,-1.15) -- (-.5,1.25) node[above,black] {$V(x)$};
      \draw[domain=-1.75:1.75,smooth,variable=\x,blue,ultra thick] plot ({\x},{.5*\x*\x*\x*\x-\x*\x-.5});
    \end{tikzpicture} 
\caption{$\mathit{W}(x)$ has an even number of zeros, where $\mathit{W}(x) \rightarrow + \infty$ as $x \rightarrow \pm \infty$.}
\label{F2.1} 
\end{figure}
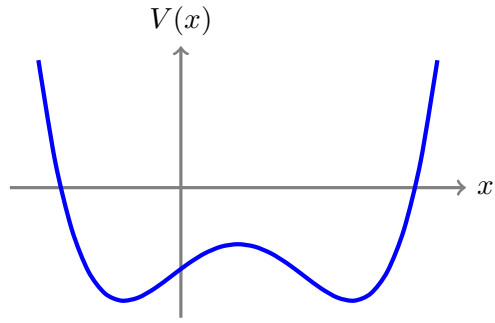 }

\item[2.] 
\par{The second case, when $\mathit{W}(x) \rightarrow - \infty$ as $x\rightarrow \pm \infty$. In this case the potential function $\mathit{W}(x)$ is negative at the boundaries and has an even number of zeros. The behavior of the potential $\mathit{V}(x)$ in this case is described in Fig. (\ref{F2.2}).} 

\par{Since $\mathit{W}(x)$ is negative, we can normalize the wave function $\Psi_0^a = A e^{\frac{\mathit{W}(x)}{\hbar}} \nonumber $ and find the value of the constant $A$, however we can not normalize the wave function $\Psi_0^b = B e^{-\frac{\mathit{W}(x)}{\hbar}}$, but we can choose $B=0$.}
\par{The complete ground state wave function, in this case, shall be expressed in the general form
\begin{equation}
\Psi_0 = \left( \begin{matrix} A e^{\frac{\mathit{W}(x)}{\hbar}} \\ 0 \end{matrix} \right).
\end{equation}

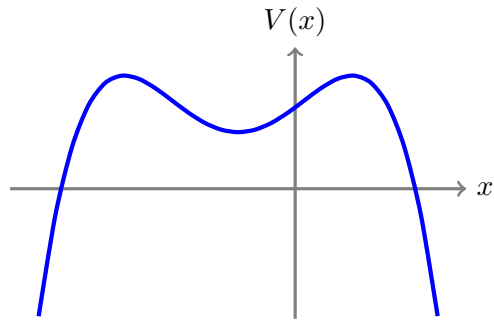
\begin{figure} [h!]
\centering
\begin{tikzpicture} [scale=1.5]
      \draw[very thick, gray,->] (-2,0) -- (2,0) node[right,thick,black] {$x$};
      \draw[very thick, gray,->] (.5,-1.15) -- (.5,1.25) node[above,black, thick] {$V(x)$};
      \draw[domain=-1.75:1.75,smooth,variable=\x,blue,ultra thick] plot ({\x},{-.5*\x*\x*\x*\x+\x*\x+.5});
    \end{tikzpicture}    
\caption{$\mathit{W}(x)$ has an even number of zeros, where $\mathit{W}(x) \rightarrow - \infty$ as $x \rightarrow \pm \infty$.} 
\label{F2.2} 
\end{figure} }

\item[3.] 
\par{The other two cases correspond to $\mathit{W}(x) \rightarrow \{+ \infty, - \infty \}$ as $x\rightarrow \{+ \infty, - \infty\}$, and  $\mathit{W}(x) \rightarrow \{- \infty, + \infty \}$ as $x\rightarrow \{+ \infty, - \infty\}$. In those two cases, the superpotential $\mathit{W}(x)$ is positive at one of the boundaries and negative at the other boundary and in the both two cases it has an odd  number of zeros. The behavior of the potential $\mathit{V}(x)$ in this case is described in Fig. (\ref{F2.3}).}
\par{In those two cases, both the two constants $A$ and $B$ are vanish and we can not normalize the wave function. Therefore we have
\begin{equation}
\Psi_0 = 0.
\end{equation}
Thus, since those two forms of the zero-energy ground state wave function can not be normalized, this means that they not exist.
\begin{figure} [h!]
\centering
\begin{subfigure}[b]{.4\textwidth}
\centering
\begin{tikzpicture} [scale=.9]
\draw[very thick, gray,->] (-2.8,.75) -- (2.8,.75) node[right,black] {$x$};
\draw[very thick, gray,->] (-.25,-2) -- (-.25,2.25) node[above,black] {$V(x)$};
\draw[domain=-2.595:2.595,smooth,variable=\x,blue,ultra thick] plot ({\x},{.1*\x*\x*\x*\x*\x-.708*\x*\x*\x+\x});\end{tikzpicture} 
\caption{$\mathit{W}(x) \rightarrow - \infty $ as $x \rightarrow  - \infty$, and $\mathit{W}(x) \rightarrow  \infty$ as $x \rightarrow   \infty$.}
\end{subfigure} 
\space
\begin{subfigure}[b]{.4\textwidth}
\centering
\begin{tikzpicture} [scale=.9]
\draw[very thick, gray,->] (-2.8,-.75) -- (2.8,-.75) node[right,black] {$x$};
\draw[very thick, gray,->] (-.25,-2) -- (-.25,2.25) node[above,black] {$V(x)$};
\draw[domain=-2.595:2.595,smooth,variable=\x,blue,ultra thick] plot ({\x},{-.1*\x*\x*\x*\x*\x+.708*\x*\x*\x-\x});
\end{tikzpicture} 
\caption{$\mathit{W}(x) \rightarrow  \infty$ as $x \rightarrow  - \infty$, and $\mathit{W}(x) \rightarrow  - \infty$ as $x \rightarrow   \infty$.}
\end{subfigure}
\caption{$\mathit{W}(x)$ has an odd number of zeros.}
\label{F2.3}   
\end{figure}
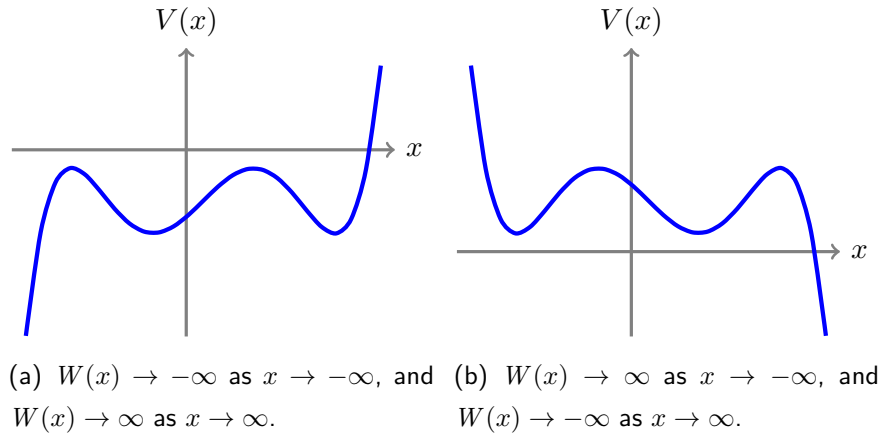  }
\end{itemize}
\par{
To conclude this section, the zero-energy ground state wave function can be normalized when the superpotential $\mathit{W}(x)$ has an even number of zeros. In this case, there exists a zero-energy ground state witch is the true vacuum state and the supersymmetry is unbroken.  However on the other hand, if the superpotential $\mathit{W}(x)$ has an odd number of zeros, the zero-energy ground state wave function can not be normalized. Then, we immediately realize that there is no zero-energy ground state in such case and the supersymmetry is spontaneously broken }

\section{Corrections to the Ground State Energy} \label{Sec.2.8}

\par{
So far, there is no unbroken supersymmetry that has been observed in nature and if nature is described by supersymmetry,  of course, it must be broken. As we have seen in the previous section, for supersymmetry to be broken, the ground state (the vacuum) must have non-zero energy. In this section, we show that the supersymmetry remains unbroken at any finite order of perturbation theory and therefore, only non-perturbative effects can be responsible for supersymmetry breaking. For this purpose, we consider the example of the supersymmetry harmonic oscillator which has a unique zero-energy ground state. Then using the conventional perturbation theory we prove that the corrections to the energy of the ground state vanish to any finite order in perturbation theory.  }   
\par{
In the next two subsections, based on \cite{goldstein1965classical,
griffiths2005introduction, book:17492, sakurai2011modern,
shankar2012principles},  we are going to compute both the first and the second corrections to the ground state energy of the supersymmetric harmonic oscillator. For instance, let us recall the general form of the supersymmetric quantum mechanics Hamiltonian, which is given by Eq. (\ref{2.28})   
  \begin{equation}\label{2.62}
\mathit{\hat{H}} = \frac{1}{2} (\hat{p}^2 + \mathit{W}'^2) \mathbbmtt{I}_2 + \frac{1}{2} \hbar \mathit{W}'' \sigma_3.
  \end{equation}  
Consider now the superpotential $\mathit{W}(x)$ is defined as
\begin{equation} \label{q2.138}
\mathit{W}(q)= \frac{1}{2} \omega \hat{q}^2 + \mathit{g} \hat{q}^3 \quad \Rightarrow \quad \begin{matrix}  \;\;\; \mathit{W}'  =  \omega \hat{q}+ 3 \mathit{g} \hat{q}^2 \\ \mathit{W}''  = \omega + 6 \mathit{g} \hat{q}, \end{matrix} 
\end{equation}
where $\mathit{g}$ is a perturbation. If $\mathit{g}=0$, the previous Hamiltonian reduces to the supersymmetric harmonic oscillator Hamiltonian,
\begin{equation}   
\mathit{\hat{H}}^{0} = \frac{1}{2} (\hat{p}^2 + \omega^2 q^2) \mathbbmtt{I}_2 + \frac{1}{2} \hbar \omega \sigma_3.
\end{equation} 
We have already solved this Hamiltonian and found that it has a unique zero energy ground state,
\begin{equation}
\mathit{E}_0^{(0)}=0,
\end{equation}
and we verify that this state is invariant under supersymmetry. Based on our argument in the previous section and after normalization, the unbroken supersymmetric ground state wave function can be written as follows
\begin{equation} \label{f2.142}
\Psi_0^{(0)}= \left( \begin{matrix} 0  \\ (\frac{\omega}{\pi \hbar})^{\frac{1}{4}} e^{- \frac{\omega q^2}{2 \hbar}} \end{matrix} \right).
\end{equation}

}
\par{
Now, when $\mathit{g}$ is small, the Hamiltonian (\ref{2.62}) can be written as
\begin{equation}
\mathit{\hat{H}}= \mathit{\hat{H}}^{0} + \mathit{\hat{H}}',
\end{equation}
where $\mathit{\hat{H}}^{0}$ is the Hamiltonian of the unperturbed system and $\mathit{\hat{H}}'$ a perturbation,
\begin{equation}
\mathit{\hat{H}}'= (\frac{9}{2} \mathit{g}^2 \hat{q}^4 + 3\omega \mathit{g} \hat{q}^3) \mathbbmtt{I}_2 + 3 \hbar \mathit{g} \hat{q} \sigma_3.
\end{equation}
Moreover, if we consider the matrix representation, the perturbative Hamiltonian $\mathit{\hat{H}}$ acting on the ground state wave function $\Psi_0^{(0)}$ yields,  
\begin{equation}
\mathit{\hat{H}}' \Psi_0^{(0)} = \left( \begin{matrix} \frac{9}{2} \mathit{g}^2 \hat{q}^4 + 3\omega \mathit{g} \hat{q}^3 + 3 \hbar \mathit{g} \hat{q} & 0 \\ 0 & \frac{9}{2} \mathit{g}^2 \hat{q}^4 + 3\omega \mathit{g} \hat{q}^3 - 3 \hbar \mathit{g} \hat{q} \end{matrix} \right) \left( \begin{matrix} 0  \\ (\frac{\omega}{\pi \hbar})^{\frac{1}{4}} e^{- \frac{\omega q^2}{2 \hbar}} \end{matrix} \right) =0.
\end{equation}
In view of this last equation, it is clear that for this combination the impact of the Hamiltonian $\mathit{\hat{H}}'$ on the wave function $\Psi_0^{(0)}$ is only due to the component,
\begin{equation}
\mathcal{\hat{H}}'= \frac{9}{2} \mathit{g}^2 \hat{q}^4 + 3\omega \mathit{g} \hat{q}^3 - 3 \hbar \mathit{g} \hat{q}.
\end{equation}
As a result, we see that it is enough to consider the Hamiltonian $\mathcal{\hat{H}}'$ to compute the first and second corrections to the ground state energy (see Appendix \ref{AppendixB.1}). }
\par{
Furthermore, we have to mention here that, for the supersymmetric harmonic oscillator, except for the ground state, all the energy levels are degenerate to two energy states. However, because we are just interested in computing the correction to the ground state energy, which is not degenerate, we can use the non-degenerate perturbation theory. For more explanation about this point see Appendix \ref{AppendixB.1}. }

\par{
In addition, it is useful to remember from quantum mechanics that the wave function of degree $n$ can be obtained by the following recursion formula,
  \begin{align} \label{2.60}
\hat{q} \psi_n &= \sqrt{\frac{(n+1)\hbar}{2 \omega}} \psi_{n+1} + \sqrt{\frac{n \hbar}{2 \omega}} \psi_{n-1}.
  \end{align}     
As well, it is important to recall the following relation,
\begin{equation}
\langle \psi_n \vert \psi_m \rangle = \delta_{nm}.
\end{equation}
We are ready now to move forward to the next two subsections and compute the first and the second corrections to the energy of the ground state. 
}  

\subsection{The First-Order Correction}
\par{
We know from the previous argument that the ground state of the supersymmetric Hamiltonian is non-degenerate. Therefore, we can use the time-independent non-degenerate perturbation theory to compute the first-order correction to the ground state energy of the supersymmetric harmonic oscillator as follows
\begin{align} \label{2.34}
\mathit{E}_0^{(1)} & = \langle \psi_0^{(0)} \vert \mathcal{\hat{H}}' \vert \psi_0^{(0)} \rangle \nonumber \\
& = \langle \psi_0^{(0)} \vert \frac{9}{2} \mathit{g}^2 \hat{q}^4 + 3\omega \mathit{g} \hat{q}^3 - 3 \hbar \mathit{g} \hat{q} \vert \psi_0^{(0)} \rangle  \nonumber  \\ 
& = \frac{9}{2} \mathit{g}^2 \langle \psi_0^{(0)} \vert \hat{q}^4 \vert \psi_0^{(0)} \rangle  + 3\omega \mathit{g} \langle \psi_0^{(0)} \vert \hat{q}^3\vert \psi_0^{(0)} \rangle  - 3 \hbar \mathit{g} \langle \psi_0^{(0)} \vert \hat{q} \vert \psi_0^{(0)} \rangle. 
\end{align}
The terms $\hat{q}^3 \psi_0$, and $\hat{q} \psi_0$ are linearly independent of $\psi_0$ so that both the second and the third terms of Eq. (\ref{2.34}) are zero. However
\begin{equation}
\hat{q}^4 \psi_0 = \frac{3\hbar^4}{2\omega^4}\psi_4 + \frac{7\hbar^4}{4\omega^4}\psi_2+\frac{9\hbar^4}{16\omega^4}\psi_0,
\end{equation}    
then, substituting into Eq. (\ref{2.34}) gives us,
\begin{align}\label{2.71}
\mathit{E}_0^{(1)} & = \frac{9}{2} \mathit{g}^2 \langle \psi_0^{(0)} \vert \hat{q}^4 \vert \psi_0^{(0)} \rangle = \frac{9}{2} g^2 \times \frac{3}{4} \frac{\hbar^2}{\omega^2} \langle \psi_0^{(0)} \vert \psi_0^{(0)} \rangle  \nonumber \\
&= \frac{27}{8} \frac{\hbar^2}{\omega^2} \mathit{g}^2 \simeq  \mathcal{O}(g^2).  
\end{align}
In the view of the last equation, the first-order correction to the ground state energy reduces to zero. Another method with more detailed to calculations of the first-order correction of the ground state energy is given in Appendix \ref{AppendixB.1}.
}
\subsection{The Second-Order Correction}

\par{
From the non-degenerate time-independent perturbation theory, the second-order correction to the energy is given by
\begin{equation}\label{2.72}
\mathit{E}_2^{(0)} = \sum_{m \neq n} \frac{\vert \langle \psi_n^{(0)} \vert \mathcal{\hat{H}}' \vert \psi_m^{(0)} \rangle \vert^2}{\mathit{E}_m^{(0)} - \mathit{E}_n^{(0)}}.
\end{equation}
Since the perturbation contains only terms of $q$, $q^3$, and $q^4$, the numerator of Eq. (\ref{2.72}) is zero for all $m$ values expects $m=1,3,4$. For more explanation about this point see Appendix \ref{AppendixB.2}. Therefore, the second-order correction to ground state energy of the supersymmetric harmonic oscillator is computed as follows
\begin{align}\label{2.73}
\mathit{E}_2^{(0)} & = \frac{\vert \langle \psi_0^{(0)} \vert \frac{9}{2} \mathit{g}^2 \hat{q}^4 + 3 \omega \mathit{g} \hat{q}^3 - 3 \hbar \mathit{g} \hat{q} \vert \psi_1^{(0)} \rangle \vert^2}{E_0^{(0)} - \mathit{E}_1^{(0)}} +  \frac{\vert \langle \psi_0^{(0)} \vert \frac{9}{2} \mathit{g}^2 \hat{q}^4 + 3 \omega \mathit{g} \hat{q}^3 - 3 \hbar \mathit{g} \hat{q}  \vert \psi_2^{(0)} \rangle \vert^2}{\mathit{E}_0^{(0)} - \mathbb{E}_2^{(0)}}
\nonumber \\
&{\quad} + \frac{\vert \langle \psi_0^{(0)} \vert \frac{9}{2} \mathit{g}^2 \hat{q}^4 + 3 \omega \mathit{g} \hat{q}^3 - 3 \hbar \mathit{g} \hat{q}  \vert \psi_3^{(0)} \rangle \vert^2}{\mathit{E}_0^{(0)} - \mathit{E}_3^{(0)}}+  \frac{\vert \langle \psi_0^{(0)} \vert \frac{9}{2} \mathit{g}^2 \hat{q}^4 + 3 \omega \mathit{g} \hat{q}^3 - 3 \hbar \mathit{g} \hat{q}  \vert \psi_4^0 \rangle \vert^2}{\mathit{E}_0^{(0)} - \mathit{E}_4^{(0)}}.
\end{align}
We can simplify the previous equation by calculating the numerator of each term using the recurrence relation for the harmonic oscillator energy eigenfunction given by Eq. (\ref{2.60}). We get
\begin{align} \label{2.74}
\hat{q} \psi_1 &= \sqrt{\frac{\hbar}{\omega}} \psi_2 + \sqrt{\frac{\hbar}{2 \omega}} \psi_0, \nonumber \\
\hat{q}^3 \psi_1 &= \sqrt{\frac{3 \hbar^3}{\omega^3}} \psi_4+ 3 \sqrt{\frac{\hbar^3}{\omega^3}} \psi_2+ \sqrt{\frac{9 \hbar^3}{8 \omega^3}} \psi_0, \nonumber \\
\hat{q}^3 \psi_3 &= \sqrt{\frac{15 \hbar^3}{\omega^3}} \psi_6 + \sqrt{\frac{72 \hbar^3}{\omega^3}} \psi_4+ \sqrt{\frac{243 \hbar^3}{8 \omega^3}} \psi_2 + \sqrt{\frac{3 \hbar^3}{4 \omega^3}} \psi_0, \nonumber \\
\hat{q}^4 \psi_2 &= \sqrt{\frac{45\hbar^4}{2\omega^4}} \psi_6 + \sqrt{\frac{147\hbar^4}{\omega^4}} \psi_4 + 12.5 \sqrt{\frac{\hbar^4}{\omega^4}} \psi_2 + \sqrt{\frac{9\hbar^4}{2\omega^4}} \psi_0,  \nonumber \\
\hat{q}^4 \psi_4 &= \sqrt{\frac{105 \hbar^4}{\omega^4}} \psi_8 + \sqrt{\frac{945\hbar^4}{4\omega^4}} \psi_6 + \sqrt{\frac{2725\hbar^4}{16\omega^4}} \psi_4 + \sqrt{\frac{81\hbar^4}{2\omega^4}} \psi_2 + \sqrt{\frac{3\hbar^4}{2\omega^4}} \psi_0.
\end{align} 

}
 
\par{  
In Eq. (\ref{2.74}), we only list the terms which contain $\psi_0$. The other terms do not affect our calculus, since all of it gives us zero, (see Appendix \ref{AppendixB.1}). Using Eq. (\ref{2.74}) to calculate Eq. (\ref{2.73}), we get
\begin{align} \label{2.75}
\mathit{E}_2^{(0)} & = - \frac{\mathit{g}^2}{\hbar \omega} \left( 3\omega \times \frac{9 \hbar^3}{\omega^3} - 3 \hbar \times \sqrt{\frac{\hbar}{2\omega}} \right)^2
- \frac{81 \mathit{g}^4}{8\hbar \omega} \left( \sqrt{\frac{9\hbar^4}{2 \omega^4}} \right)^2 \nonumber \\
&{\quad}-\frac{g^2}{3 \hbar \omega} \left( 3\omega \times \sqrt{\frac{3\hbar^3}{4\omega^3}} \right)^2 - \frac{\mathit{g}^4}{4\hbar \omega} \left( \frac{9}{2} \times \sqrt{\frac{3 \hbar^4}{2 \omega^4}} \right)^2 \nonumber \\
& = - \frac{9}{8} \frac{\hbar^2}{\omega^2} \mathit{g}^2 - \frac{729}{16} \frac{\hbar^4}{\omega^4} - \frac{9}{4} \frac{\hbar^2}{\omega^2} \mathit{g}^2 - \frac{243}{32} \frac{\hbar^4}{\omega^4} \nonumber \\
& = - \frac{27}{8} \frac{\hbar^2}{\omega^2} \mathit{g}^2  - \frac{1701}{32} \frac{\hbar^3}{\omega^5} \mathit{g}^4.
\end{align}   
\par{For more details, another method to compute the second-order correction of the ground state energy is given in Appendix \ref{AppendixB.2}.}

\par{Finally, with regard to Eqs. (\ref{2.71} and \ref{2.75}) we realize that up to the second-order the energy corrections to the ground state energy vanishes. This is concluded as
\begin{align}
\mathcal{O}(g)&=0, \nonumber \\
\mathcal{O}(g^2)& = \frac{27}{8}\frac{\hbar^2}{\omega^2} - \frac{27}{8}\frac{\hbar^2}{\omega^2}=0.
\end{align}  

}

\section{Properties of SUSY Quantum Mechanics} \label{Sec.2.9}
\par{
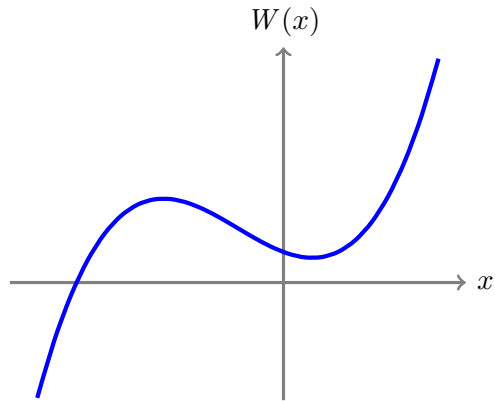
\begin{figure} [h!]
\centering
\begin{tikzpicture} [scale=1.2]
      \draw[very thick, gray,->] (-2.5,-0.6) -- (2.5,-0.6) node[right,black] {$x$};
      \draw[very thick, gray,->] (.5,-1.9) -- (.5,2) node[above,black] {$\mathit{W}(x)$};
      \draw[domain=-2.2:2.2,smooth,variable=\x,blue,ultra thick] plot ({\x},{.3*\x*\x*\x-.6*\x});
    \end{tikzpicture} 
\caption{The Superpotential of the Harmonic Oscillator Ground State.}
\label{F2.4} 
\end{figure}
A supersymmetric quantum mechanics system is said to have unbroken supersymmetry if it has a zero-energy ground state, which is $\mathit{E}_0=0$.  While if the system has a positive ground state energy, $\mathit{E}_0>0$, it is said to have a broken supersymmetry \cite{engbrant2012supersymmetric}. For example, in Section \ref{Sec2.5}, we have studied the Hamiltonian of the supersymmetric harmonic oscillator and showed that it obeys the superalgebra. Then, we calculated the ground state energy for that system and found evidence that it vanishes $E_0=0$, and consequently, this supersymmetric harmonic oscillator has a good supersymmetry or the supersymmetry for that system is unbroken.}
\par{
In the previous section, we applied a small perturbation $\mathit{g}$ to the supersymmetric harmonic oscillator and calculated the effect on the ground state energy. We found that the perturbation does not affect the energy of the ground state at second order in perturbation theory, but this result can be expanded to any finite order. This means that the supersymmetry breaking does not occur because of perturbation, and it must be due to the non-perturbative effects.}

\par{Again, let us consider the same potential Eq. (\ref{q2.138}), which we have used before in Sec. \ref{Sec.2.8},
\begin{equation}
W(q)= \frac{1}{2} \omega \hat{q}^2 + g \hat{q}^3.
\end{equation}  
As we discussed in Sec. \ref{Sec2.7}, the wave function of the ground state should has three zeros, $\Psi_0 \rightarrow +\infty$ as $x \rightarrow +\infty$, and $\Psi_0 \rightarrow -\infty$ as $x \rightarrow -\infty$. Fig. (\ref{F2.4}), shows an arbitrary diagram how the wave function of the ground state should look like, and Eq. (\ref{2.78}) gives us the non-normalized form of the ground state wave function, 
\begin{equation}\label{2.78}
\Psi_0 = \left( \begin{matrix} A e^{-\frac{W}{\hbar}} \\ B e^{-\frac{W}{\hbar}} \end{matrix} \right).
\end{equation}

}
\par{      
Acutely, Eq. (\ref{2.78}) could not be normalized, and the only way to solve the Schr\"{o}dinger equation for the ground state is taking $\Psi_0=0$. This means that the wave function of the ground state is not exited, even the perturbation theory told us that it is exited and has no energy correction.}

\par{
Thus, the perturbation technique gives us incorrect results for both the wave function as well as the energy spectrum and fails to give an explanation to the supersymmetry breaking.
}

\chapter{\textbf{Path Integrals and Instantons}} \label{ch3}

\par{In the previous chapter, we have reviewed supersymmetry in quantum mechanics. As discussed, so far there has been no unbroken supersymmetry observed in nature, and if supersymmetry exists, definitely it must be broken \cite{Wellman}. In fact, supersymmetry may be broken spontaneously at any order of perturbation theory, or dynamically due to non-perturbative effects \cite{witten1981dynamical}. Instantons are a non-perturbative effect in quantum mechanics, which can not be seen in perturbation theory \cite{marino2015instantons}. Instantons can lead to the dynamical breaking of supersymmetry, because it may connect bosonic states to fermionic states \cite{shifman1999instantons}.}

\par{In this chapter, based on \cite{rajaraman1982solitons, van1983instantons, zinn2002quantum} we will give a brief review of the path integral formulation of quantum mechanics, and in particular, we will study in detail Euclidean instantons. As a simple example, we will consider the quantum mechanics of the harmonic oscillator. Then, we will encounter the problem of the symmetric double-well potential and discuss quantum tunneling due to the instanton effects. Finally, we will end this chapter by explaining the quantum tunneling phenomena using the real-time path integral formalism.}

\section{Path Integral Formulation of Quantum Mechanics} \label{Sec3.1}

\par{The path integral formulation of quantum mechanics is discussed in many texts, see for instance \cite{coleman1988aspects, feynman2005quantum, rattazzi2009path, blaunotes}. To start, let us consider a spinless particle moving in a one-dimensional potential $V(x)$. If we take the mass of the particle, $m=1$ with $\hbar =1$, the Lagrangian and the Hamiltonian of the system can be written as follows
\begin{align}
\mathit{L} &= \frac{1}{2} \left( \frac{dx}{dt} \right)^2 - V(x), \nonumber \\
\mathit{H} &=  \frac{1}{2} p^2 + V(x).
\end{align}

}

\par{Suppose that the particle is initially at position $x_i$ at time $-t_0/2$ and ends up at position $x_f$ at time $t_0/2$. The path integral approach to quantum mechanics, to calculate the amplitude of such a process, was developed by Feynman. In the Feynman path integral, the amplitude of an event is equal to the sum over all paths joining the points $(x_i; -t_0/2)$ and $(x_f; t_0/2)$.  The contribution of each path weighted by the factor $e^{i\mathit{S}[x]}$, where $\mathit{S}[x]$ is the action:
\begin{equation}\label{3.2}
\mathit{S}[x] = \int_{-t_0/2}^{t_0/2} dt \left[ \frac{1}{2} \left( \frac{dx}{dt} \right)^2 - V(x) \right].
\end{equation}

}
\par{
Formally we write the path integral formula for the transition amplitude as 
\begin{align} \label{3.3}
\mathit{T}(x_f,t_0/2 ; x_i,-t_0/2) &\equiv \langle x_f \vert e^{-i \mathit{H}t_0} \vert x_i \rangle = \mathcal{N} \int_{x_i}^{x_f} \mathcal{D}[x] e^{i\mathit{S}[x]}.
\end{align}  
In this formula, $\mathcal{N}$ is the normalization factor, we shall return to discuss it later. $\mathcal{D}[x]$ donates the path integral integration overall functions $x(t)$.}
\par{
It is hard to compute the path integral (\ref{3.3}) directly in the real-time framework because of the imaginary power of the exponential term. To avoid this problem, let us make a Wick rotation \cite{muller2006introduction} from real-time framework  into Euclidean-time framework by $t \rightarrow - i \tau$, and consequently we write $i \mathit{S} \rightarrow - \mathit{S}_E$. Applying this rotation on the action (\ref{3.2}), simply we find the Euclidean action
\begin{equation} \label{3.4}
\mathit{S}_{E}[x] = \int_{-{\tau=}_0/2}^{{\tau}_0/2} d \tau \left[ \frac{1}{2} \left( \frac{dx}{d \tau} \right)^2 + V(x) \right],
\end{equation}  
which is, in fact, corresponding to the Hamiltonian $\mathit{H}$ of the system. Thus, the Euclidean-time path integral formula for the transition amplitude becomes 
\begin{equation} \label{3.5}
\langle x_f \vert e^{- \mathit{H}{\tau}_0} \vert x_i \rangle = \mathcal{N} \int_{x_i}^{x_f} \mathcal{D}[x] e^{- \mathit{S}_E}.
\end{equation}
Before going through the process of calculating this path integral, it is important here to explain how the path integral gives us the properties of the ground state. The transition amplitude, which is given by Eq. (\ref{3.5}), can be also expressed in terms of energy eigenstates as follows
\begin{equation}
\langle x_f \vert e^{- \mathit{H}{\tau}_0} \vert x_i \rangle = \sum_n e^{- \mathit{E}_n \tau_0} \langle x_f \vert n \rangle \langle n \vert x_i \rangle,
\end{equation}
where $\vert n \rangle$ is the energy eigenstate $\mathit{H} \vert n \rangle = \mathit{E}_n \vert n \rangle $. In the limit $\tau \rightarrow \infty$, the sum over the intermediate states is dominated by the ground state;
\begin{equation} \label{z3.7}
\langle x_f \vert e^{- \mathit{H}{\tau}_0} \vert x_i \rangle \simeq e^{- \mathit{E}_0 \tau_0} \psi_0(x_f) \psi_{0}^{*}(x_i).
\end{equation}
Comparing the result of the path integral (\ref{3.5}) with Eq. (\ref{z3.7}), leads us to obtain the energy $\mathit{E}_0$ and the wave function $\psi_0(x)$ of the ground state.}

\par{
We now turn to calculate the path integral (\ref{3.5}). Let us consider the configuration $x(\tau)$, an arbitrary function satisfies the boundary conditions. Then we can write $x(\tau)$ as the sum of a quantum fluctuation $\delta x(\tau)$, around a classical solution, $\mathit{X}(\tau)$, that is
\begin{align} \label{3.6}
x(\tau) &= \mathit{X}(\tau) + \delta x(\tau) \nonumber \\
&= \mathit{X}(\tau) + \sum_n \mathit{c}_n x_n (\tau).
\end{align}
Where $\mathit{X}(\tau)$ satisfies the boundary conditions,
\begin{equation}
\mathit{X}(- \tau_0 /2) = x_i, \quad \texttt{and} \quad \mathit{X}( \tau_0 /2) = x_f.
\end{equation}
Note that the fluctuation $\delta x(\tau)$ can be written as, $\delta x(\tau)=\sum_n \mathit{c}_n x_n (\tau)$, where the $x_n(\tau)$ are some complete set of orthonormal function that vanish at the boundaries,
\begin{align} \label{3.8}
& \int_{- {\tau}_{0} / {2} }^{{\tau}_{0} / {2} } d \tau x_n(\tau)  x_m({\tau}) = \delta_{nm}, \nonumber \\
& x_n(- {\tau}_0/2) = x_n( {\tau}_0/2) =0.
\end{align} 
To sum over all paths, we consider all possible fluctuations, so the path integral measure $\mathcal{D}[x]$ can be taken to be, 
\begin{equation} \label{3.9}
\mathcal{D}[x] = \prod_n \frac{d \mathit{c}_n}{\sqrt{2 \pi}}.
\end{equation} 
where the factor $\sqrt{2 \pi}$ is taken for convenience.}

\par{
Furthermore, consider the action of the classical solution $\mathit{S}_0 = \mathit{S}[\mathit{X}]$. Then, the expansion of the Euclidean action around the classical action $\mathit{S}_0$ can be computed using a steepest-descent approximation as follows

\begin{align}\label{3.10}
\mathit{S}_{\mathit{E}} [\mathit{X} + \delta x] = \mathit{S}_0 & + \int_{- {\tau}_0/2}^{{\tau}_0/2} d \tau \left[ - \dfrac{d^2 x}{d {\tau}^2} + V'(x) \right] \delta x \nonumber \\ & + \frac{1}{2} \int_{- {\tau}_0/2}^{{\tau}_0/2} d \tau \delta x \left[ - \dfrac{d^2}{d {\tau}^2} \delta x + V''(x) \delta x \right] + \mathcal{O}(\delta x^3),
\end{align}
where 
\begin{equation}
V'(x) \equiv \dfrac{d V(x)}{dx}, \quad \texttt{and} \quad  V'' \equiv \dfrac{d^2 V(x)}{dx^2}.
\end{equation}
The second term of Eq. (\ref{3.10}) which is linear in $\delta x$, vanishes due to the classical equation of motion,
\begin{equation}\label{3.12}
\dfrac{d^2 x}{d {\tau}^2} =  V'(x).
\end{equation}
Note that since the Euclidean formulation is considered, the sign of the potential is effectively reversed. Also, we assume the last term of Eq. (\ref{3.10}) is very small and can be ignored. Therefore, we can rewrite Eq. (\ref{3.10}) as follows
\begin{equation}\label{3.13}
\mathit{S}_{\mathit{E}} \left[ \mathit{X} + \delta x \right] = \mathit{S}_0   + \frac{1}{2} \int_{- {\tau}_0/2}^{{\tau}_0/2} d \tau \delta x \mathit{\widehat{W}} \delta x ,
\end{equation}
where $\mathit{\widehat{W}}$ a Hermitian operator is called the wave operator and defined as 
\begin{equation}
\mathit{\widehat{W}} = - \dfrac{d^2}{d {\tau}^2} + V''(x). 
\end{equation}
In order to make further progress, we consider a complete set of eigenfunctions and eigenvalues of the Hermitian operator $\mathit{\widehat{W}}$
\begin{equation}
\mathit{\widehat{W}} x_n(\tau) = \epsilon_n x_n(\tau).
\end{equation}
Since $\mathit{\widehat{W}}$ is a Hermitian operator, we can choose these functions as the orthonormalized system which must satisfy the boundary conditions (\ref{3.8}). Then the Euclidean action (\ref{3.13}) can simply be written in the diagonal form
\begin{equation}\label{3.16}
\mathit{S}_E= \mathit{S}_0   + \frac{1}{2}  \sum_n \epsilon_n \mathit{c}^2_n.
\end{equation}
Substituting Eqs. (\ref{3.9}, \ref{3.16}), the transition amplitude ,(\ref{3.5}), becomes
\begin{align} 
\langle x_f \vert e^{- \mathit{H}{\tau}_0} \vert x_i \rangle &= \mathcal{N} \int \prod_n \frac{d \mathit{c}_n}{\sqrt{2 \pi}} e^{- \mathit{S}_0 - \frac{1}{2}  \epsilon_n \mathit{c}^2_n} \nonumber \\
&= \mathcal{N} e^{- \mathit{S}_0} \prod_n \int \frac{dc_n}{\sqrt{2\pi}} e^{- \frac{1}{2}  \epsilon_n \mathit{c}^2_n}.
\end{align}
This integration is a simple Gaussian integral
\begin{equation}
\int \frac{dc_n}{\sqrt{2\pi}} e^{- \frac{1}{2}  \epsilon_n \mathit{c}^2_n} = \epsilon_n^{- \frac{1}{2}}.
\end{equation} 
With regard to this approximation, the Euclideanized transition amplitude reads
\begin{equation} \label{3.19}
\langle x_f \vert e^{i \mathit{H}{\tau}_0} \vert x_i \rangle = \mathcal{N} e^{- \mathit{S}_0} \prod_n \epsilon_n^{- \frac{1}{2}}.
\end{equation} 
Moreover, since $\prod_n \epsilon_n = \det \mathit{\widehat{W}}$, we can write
\begin{equation} \label{3.20}
\prod_n \epsilon_n^{- \frac{1}{2}} = \left[\det \left( - \dfrac{d^2}{d {\tau}^2} + V''(X(\tau))\right) \right]^{- \frac{1}{2}}.
\end{equation}
Before we end this section, it is important to remark three points. Firstly, in the view of the last equation, the calculation of the Euclideanized transition amplitude (\ref{3.19}) depends only on the eigenvalues $\epsilon_n$ and no knowledge about the behavior of the wave functions is required. Secondly, in the previous argument, we have assumed that all the eigenvalues are positive $\epsilon_n > 0$, but that is not the case in general. Thirdly, the normalization factor $\mathcal{N}$ has not been fixed yet.}

\section{Path Integral for Harmonic Oscillator} \label{Sec3.2}
\par{
Having introduced the general machinery of path integrations we now turn to the quantum mechanics of harmonic oscillator as a simple application of the path integral method. Good dissections of this topic can be found in \cite{perepelitsapath, mackenzie2000path,flip2010tanedo}. Consider a one-dimensional harmonic oscillator for a particle of mass, $m=1$, placed in the potential
\begin{equation} \label{3.21}
V(x) = \frac{1}{2} \omega^2 x^2.
\end{equation}
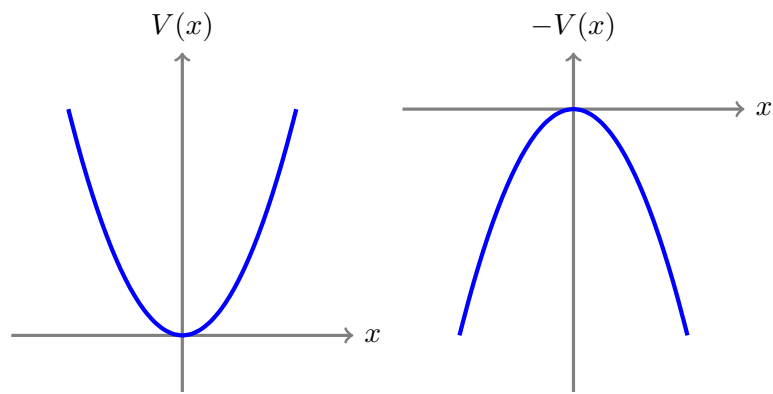
\begin{figure} [h!]
\centering
\begin{tikzpicture} [scale=1.5]
      \draw[very thick, gray,->] (-1.5,0) -- (1.5,0) node[right,black] {$x$};
      \draw[very thick, gray,->] (0,-0.5) -- (0,2.5) node[above,black] {$V(x)$};
      \draw[domain=-1:1,smooth,variable=\x,blue,ultra thick] plot ({\x},{2*\x*\x});
    \end{tikzpicture}
\begin{tikzpicture} [scale=1.5]
      \draw[very thick, gray,->] (-1.5,0) -- (1.5,0) node[right,black] {$x$};
      \draw[very thick, gray,->] (0,-2.5) -- (0,.5) node[above,black] {$-V(x)$};
      \draw[domain=-1:1,smooth,variable=\x,blue,ultra thick] plot ({\x},{-2*\x*\x});
    \end{tikzpicture}    
\caption{The potential for the quantum mechanical harmonic oscillator problem in both the real and the Euclidean time.}
\label{F3.1} 
\end{figure} 

}
\par{
An example of a harmonic oscillator potential is shown in Fig. \ref{F3.1}. Using Eq. (\ref{3.12}) and Eq. (\ref{3.21}) we get the equation of motion for the Euclidean harmonic oscillator, that is 
\begin{equation} \label{3.22}
\dfrac{d^2 x}{d {\tau}^2} = V'(x)=  \omega^2 x.
\end{equation}
Obviously, there is a minimum of $x=0$, and we have 
\begin{equation} \label{3.23}
V''(x=0) =  \omega^2.
\end{equation}
Now the general solution of Eq. (\ref{3.22}) takes the form
\begin{equation}
x(\tau)= \mathit{A} \cosh (\omega \tau) + \mathit{B} \sinh (\omega \tau),
\end{equation}
where $\mathit{A}$ and $\mathit{B}$ are arbitrary constants. When the boundary conditions are set, $\mathit{X}(- \tau_0/2)=\mathit{X}(\tau_0/2)=0$, there is only one path with such properties which is the trivial solution $\mathit{X}(\tau)=0$ and consequently  $\mathit{\dot{X}}(\tau)=0$. This implies that the classical action of the harmonic oscillator vanishes based on our choices of the boundary conditions, \ie
\begin{equation} \label{3.24}
\mathit{S}_0 =0.
\end{equation}

 }  
\par{
Taking into account Eqs. (\ref{3.23}, \ref{3.24}), thus the Euclideanized transition amplitude (\ref{3.19}) becomes 
\begin{equation} \label{3.25}
\langle x_f \vert e^{- \mathit{H}{\tau}_0} \vert x_i \rangle = \mathcal{N} \left[\det \left(- \dfrac{d^2}{d {\tau}^2} + \omega^2 \right) \right]^{- \frac{1}{2}}
\end{equation} 
To solve this equation,  we still need to fix the normalization factor $\mathcal{N}$. Leaving that aside for the moment, our problem reduces to computing the determinant of the wave operator $\det \mathit{\widehat{W}}$. The wave operator in this case reads
\begin{equation}
\mathit{\widehat{W}} = - \dfrac{d^2}{d {\tau}^2} + \omega^2,
\end{equation}
So the eigenvalue equation of this wave operator becomes
\begin{equation} \label{3.27}
\mathit{\widehat{W}} x_n = \epsilon_n x_n \quad \Rightarrow \quad \dfrac{d^2 x_n}{d {\tau}^2} + \left( \epsilon_n - \omega^2 \right) x_n = 0.
\end{equation}
The boundary conditions $x_n(\pm \tau/2)=0$, allow us to solve Eq. (\ref{3.27}) for the eigenfunctions
\begin{equation}
x_n= A \sin \sqrt{\epsilon_n - \omega^2} \left( \tau + \frac{1}{2} \tau_0 \right), 
\end{equation}
 and the eigenvalues
 \begin{equation}
 \epsilon_n = \left( \frac{n \pi}{\tau_0} \right)^2 + \omega^2, \quad n= 1,2, \dots .
\end{equation}

}
\par{ 
In light of the knowledge of the eigenvalues $\epsilon_n$, we are able to determine Eq. (\ref{3.25}) as follows
\begin{align}\label{3.30}
\langle x_f \vert e^{- \mathit{H}{\tau}_0} \vert x_i \rangle &= \mathcal{N} \left[\det \left(- \dfrac{d^2}{d {\tau}^2} + \omega^2 \right) \right]^{- \frac{1}{2}} \nonumber \\
&= \mathcal{N} \prod_n \epsilon_n^{- \frac{1}{2}} = \mathcal{N} \prod_{n=1}^{\infty} \left[ \left( \frac{n\pi}{\tau_0} \right)^2 + \omega^2  \right]^{- \frac{1}{2}} \nonumber \\
&= \left[ \mathcal{N} \prod_{n=1}^{\infty} \left( \frac{n^2 {\pi}^2}{{\tau}_0^2}  \right)^{- \frac{1}{2}} \right] \times \left[ \prod_{n=1}^{\infty} \left( 1 + \frac{\omega^2 
\tau_0^2}{\pi^2 n^2} \right) \right]^{- \frac{1}{2}}.
\end{align}
In the last step, we have split the determinant into two factors, in order to ignore the immediate requirement of explicit determination of the normalization factor $\mathcal{N}$. The first factor corresponds to the motion of a free particle, which is when $\omega=0$. Since, if $\omega=0$, the potential (\ref{3.21}) becomes zero and consequently, the harmonic oscillator Hamiltonian reduces to the free particle Hamiltonian, $\mathit{H}= \frac{1}{2} \mathit{p}^2$. Therefore, it is straightforward to determine the result of the first factor of Eq. (\ref{3.30}) by inserting a complete set of momentum eigenstates $\vert \mathit{\hat{p}} \rangle$,
\begin{equation}
\langle x \vert p \rangle = \frac{1}{\sqrt{2 \pi}} e^{-px}.
\end{equation}
Thus, by Fourier transform, one finds
\begin{align} \label{3.31}
\mathcal{N} \prod_{n=1}^{\infty} \left( \frac{n^2 {\pi}^2}{{\tau}_0^2}  \right)^{- \frac{1}{2}} &= \langle x_f=0 \vert e^{- \mathit{H} \tau_0} \vert x_i=0 \rangle \nonumber \\
&= \int_\infty^\infty dp \langle x_f=0 \vert p \rangle e^{- \frac{1}{2} \hat{p}^2 \tau_0} \langle p \vert x_i=0 \rangle \nonumber \\
&= \dfrac{1}{2 \pi} \int_\infty^\infty dp e^{- \frac{1}{2} \hat{p}^2 \tau_0} = \frac{1}{\sqrt{2\pi \tau_0}}.
\end{align}

}
\par{
However, we have not computed an explicit expression for the normalization factor $\mathcal{N}$, but this is not needed since it is implicit in  Eq. (\ref{3.31}). For instance the second factor in Eq. (\ref{3.30}) can be calculated using the formula \cite{gradshteyn2014table} 
\begin{equation}
\prod_{n=1}^{\infty} \left( 1+ \frac{y^2}{n^2} \right) = \frac{\sinh \pi y} {\pi y},
\end{equation}
wherein our case $y= \omega \tau_0 / \pi$, therefore we get
\begin{align} \label{3.33}
\left[ \prod_{n=1}^{\infty} \left( 1 + \frac{ (\omega \tau_0 /\pi)^2}{ n^2} \right) \right]^{- \frac{1}{2}} &= \left( \frac{\sinh (\omega \tau_0)}{\omega \tau_0} \right)^{-\frac{1}{2}}. 
\end{align}
Substituting from Eqs. (\ref{3.31}, \ref{3.33}) into Eq. (\ref{3.30}), we get our final formula to the Euclideanized transition amplitude of the harmonic oscillator, that is 
\begin{align}
\langle x_f=0 \vert e^{- \mathit{H}{\tau}_0} \vert x_i=0 \rangle &=  \frac{1}{\sqrt{2\pi \tau_0}}  \left( \frac{\sinh (\omega \tau_0)}{\omega \tau_0} \right)^{-\frac{1}{2}} \nonumber \\
&= \left( \frac{\omega}{\pi} \right)^{\frac{1}{2}} \left( 2 \sinh \omega \tau_0 \right)^{- \frac{1}{2}}.
\end{align}
Moreover, using the exponential form of the $\sinh$ function and taking the limit of $\tau_0 \rightarrow \infty$, the transition amplitude can be written in the form
\begin{align} \label{q3.37}
\langle x_f=0 \vert e^{- \mathit{H}{\tau}_0} \vert x_i=0 \rangle &= \left( \frac{\omega}{\pi} \right)^{\frac{1}{2}} \left( e^{\omega \tau_0} - e^{-\omega \tau_0} \right)^{-\frac{1}{2}} \nonumber \\
&= \left( \frac{\omega}{\pi} \right)^{\frac{1}{2}} e^{- \omega \tau_0/2} \left( 1 - e^{-\omega \tau_0} \right)^{-\frac{1}{2}} \nonumber \\
&= \left( \frac{\omega}{\pi} \right)^{\frac{1}{2}} e^{- \omega \tau_0/2} \left( 1 + \frac{1}{2} e^{-2\omega \tau_0} + \dots \right).
\end{align}
This exponential form of the transition amplitude can lead us to calculate the energy eigenvalues since the transition amplitude can be also expanded in terms of energy eigenstates $\psi_n(x)=\langle x \vert n \rangle$, as follows
\begin{align} \label{q3.38}
\langle x_f=0 \vert e^{- \mathit{H}{\tau}_0} \vert x_i=0 \rangle &= \sum_n  \langle x_f=0 \vert n \rangle  e^{- \mathit{H}{\tau}_0} \langle n \vert x_i=0 \rangle \nonumber \\
&= \sum_n e^{- \mathit{E}_n \tau_0} \vert \psi_n(0) {\vert}^2.
\end{align}
We have here to note that, the $e^{-\mathit{E}_n \tau_0}$ expression arises only after the infinite sum, and the terms with odd $n$ do not contribute since for them $\psi_n(0)=0$. For instance, if we compare the last expression (\ref{q3.38}) with the exponential form of the transition amplitude (\ref{q3.37}), it is easy to see that the energy spectrum of all the even energy levels can be determined by the following formula
\begin{equation}
\mathit{E}_{2n} = \frac{\omega}{2} + 2n\omega.
\end{equation} 
This exactly agrees with the energy spectrum of the harmonic oscillator. In the case of the ground state, it has the following energy eigenvalue,
\begin{equation}
\mathit{E}_0 = \frac{\omega}{2}.
\end{equation} 
Also, comparing the two formulas (\ref{q3.37}) and (\ref{q3.38}) leads to the exact expression of the ground state wave function of the harmonic oscillator,
\begin{equation}
\psi_0(0)= \left( \frac{\omega}{\pi} \right)^{\frac{1}{4}}.
\end{equation}
This clearly shows how the Euclidean path integral in the large $\tau_0$ limit, teaches us about the properties of the ground state.
}

\section{Double-Well Potential and Instanton Calculation} \label{Sec3.3}
\par{
In this section, we will study a particle in a symmetric double-well potential, which is a quantum mechanical system with tunneling. Indeed, it is the simplest system for which we can apply the instanton method. Here the materials are heavily borrowed from \cite{novikov1999abc, kazamalectures, collinucci2005instantons, pareek2013quantum, schiappa2014resurgence}.}
\par{
For instance, consider the problem of one-dimensional motion of a particle, with mass $m=1$,  placed in a symmetric double-well potential, which may be written in the form
\begin{equation} \label{3.38}
 V(x) = \lambda \left( x^2 - a^2 \right)^2.
\end{equation} 

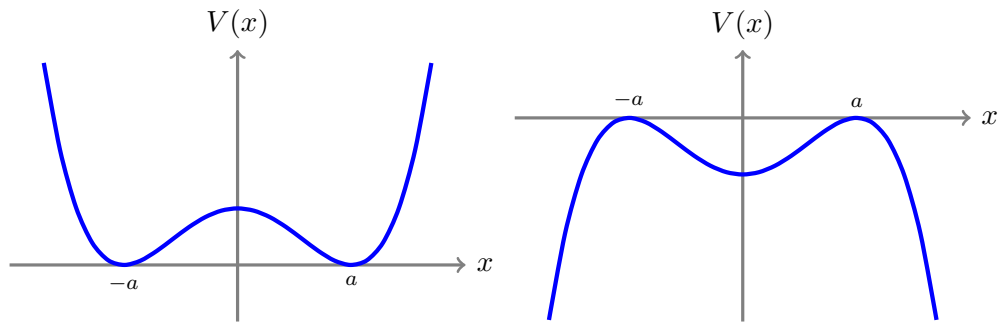
\begin{figure} [h!]
\centering
\begin{tikzpicture} [scale=1.5]
      \draw[very thick, gray,->] (-2,0) -- (2,0) node[right,black] {$x$};
      \draw[very thick, gray,->] (0,-0.5) -- (0,1.9) node[above,black] {$V(x)$};
      \draw[black, thick] (-1,0) node[below] {${\scriptstyle -a}$};
      \draw[black, thick] (1,0) node[below] {${\scriptstyle a}$};
      \draw[domain=-1.7:1.7,smooth,variable=\x,blue,ultra thick] plot ({\x},{.5*\x*\x*\x*\x-\x*\x+.5});
    \end{tikzpicture}
\begin{tikzpicture} [scale=1.5]
      \draw[very thick, gray,->] (-2,0) -- (2,0) node[right,thick,black] {$x$};
      \draw[very thick, gray,->] (0,-1.8) -- (0,.6) node[above,black, thick] {$V(x)$};
      \draw[black, thick] (-1,0) node[above] {${\scriptstyle -a}$};
      \draw[black, thick] (1,0) node[above] {${\scriptstyle a}$};
      \draw[domain=-1.7:1.7,smooth,variable=\x,blue,ultra thick] plot ({\x},{-.5*\x*\x*\x*\x+\x*\x-.5});
    \end{tikzpicture}    
\caption{The double-well potential and its flipped over Euclidean version.}
\label{F3.2}
\end{figure} 

}
\par{
Taking a closer look at this potential equation, we deduce  the following points:
\begin{itemize}
\item This potential has two classical minima located at $x =\pm a$. Those two minima are equivalent because of the reflection symmetry of the system.
\item When $a$ tends to infinity, the barrier width becomes infinite and consequently, the system decomposes into two separated and independent harmonic oscillator potentials with frequency $\omega = \mathit{V}(x=\pm a)= 8 a^2 \lambda$.
\item The parameter $\lambda$ is some constant controlling the depth of the potential wells, relate to $a$ and $\omega$ by
\begin{equation} \label{3.39}
\lambda = \frac{\omega^2}{8 a^2}.
\end{equation}
\item Eq. (\ref{3.38}) implies that the maximum height of the potential barrier at the center is 
\begin{equation}
V_{\textsf{max}}(0)= \frac{\omega^4}{64 \lambda}.
\end{equation}
\item Classically, a particle with energy less than the central potential barrier, is trapped in one of the wells and can never reach to the other well. But, quantum mechanics allows it to possibly tunnel through the barrier.
\item The potential barrier is high enough to consider the classical picture, if $\lambda \ll \omega^3$.
\item Quantum mechanical tunneling can be studied in perturbation theory. First we expand the potential (\ref{3.38}) around each minima. Then we have the expansion
\begin{equation}
V(x) = \frac{\omega^2}{2} (x-a)^2 + \mathcal{O}\left( (x-a)^3 \right).
\end{equation}
This is, to first order, the harmonic oscillator, which we have studied in the previous section. To lowest order, we can think of the system as two decoupled harmonic oscillators which means the ground state is should be doubly degenerate.
\item However, from the exact quantum mechanical solution of this problem, it is well-known that there is no degeneracy of the energy eigenvalues.     
\end{itemize}

}
\par{ The goal of this section is to solve this problem using the path integral method, more precisely via semi-classical trajectories in the Euclidean time called instantons. In particular, we are looking to use the path integral form (\ref{3.19}) to calculate the transmitted amplitude for the double-well potential problem
\begin{equation} \label{3.42}
\langle a \vert e^{- \mathit{H} \tau_0} \vert - a \rangle = \mathcal{N} e^{- \mathit{S}_0}  \left[ \det \left( - \dfrac{d^2}{d \tau^2} + V''\left(\mathit{X} \right) \right) \right]^{-\frac{1}{2}}.
\end{equation}
It will be a while before we can evaluate Eq. (\ref{3.42}). The process is explained step-by-step in the following subsections.}
  
\subsection{The Classical Solution}
\par{
We begin in this subsection, by finding the solution of the classical equation of motion. The Euclideanized equation of motion is given by
\begin{equation}
\dfrac{d^2 X}{d \tau^2} = \dfrac{d V(X)}{d X}, 
\end{equation}
where $X$ represents the classical solution for the equation of motion. To get the explicit solution of this equation, let us first multiply both sides by $\dot{X}\equiv \dfrac{d X}{d \tau}$, then we have
\begin{equation}
\dot{X} \ddot{X} = \dfrac{d X}{d \tau} \dfrac{d V(X)}{dX} = \dfrac{d V(X)}{d \tau },
\end{equation}
therefore, by integration we get
\begin{equation} \label{3.44}
\frac{1}{2} \dot{X}^2 = V(X) + c.
\end{equation}
We want a solution with finite action in the limit $\tau_0 \rightarrow \infty$. From the relation above,
\begin{equation}
\mathit{S[X]} = \int_{- \infty}^{\infty} d \tau (\dot{X}^2 - c).
\end{equation}
For this to be finite, we need $\dot{X}^2(\pm \infty)=c$. Then, using the boundary conditions $\dot{X}(\pm \infty)=0$, implies $c=0$. Thus, Eq. (\ref{3.44}) becomes
\begin{equation}
\frac{1}{2} \dot{X}^2 = V(X).
\end{equation}
Furthermore, putting in the explicit form of the double-well potential Eq. (\ref{3.38}), we have
\begin{equation}
\dot{X} = \pm \sqrt{2 \lambda} (a^2 - X^2) \quad \Rightarrow \quad
\dfrac{d X}{a^2 - X^2}  = \pm \sqrt{2 \lambda} (\tau - \tau_c).
\end{equation} 
This is a simple integration and may be easily solved by using the boundary conditions $X(\pm \infty)= \pm a$.  Then we find the following solution
\begin{equation}
X= \pm a \tanh \left( \sqrt{2 \lambda}a (\tau-\tau_c) \right).
\end{equation}
Using the relation (\ref{3.39}), this may be written as
\begin{equation} \label{3.50}
X= \pm a \tanh \left( \frac{\omega}{2} (\tau-\tau_c) \right).
\end{equation}

}
\par{
\begin{itemize}
\item The solutions with positive and negative signs are respectively called the instanon and
anti-instanton. See Fig. (\ref{F3.3}).
\item The instanton solution describes a particle moving from $x=-a$ at $- \tau_0/2$ to $x=a$ at $\tau_0/2$.
\item Similarly, the anti-instanton solution describes the inverse motion beginning at $x=a$ and ending at $x=-a$.
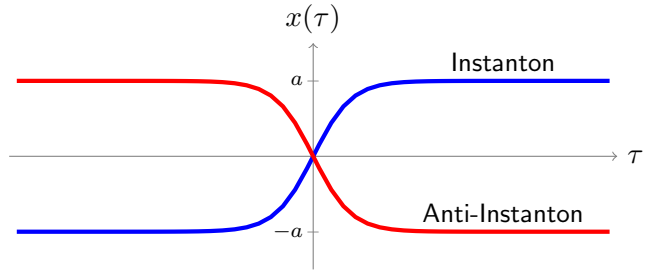
\begin{figure} [h!]
\centering
\begin{tikzpicture}[xscale=.5,yscale=1]
  \draw[->, gray] (0,-1.5) -- (0,1.5) node[above,black] {$x(\tau)$};
  \draw[->, gray] (-8,0)--(8,0) node[right,black] {$\tau$};
  \draw[domain=-7.8:7.8,samples=50,blue, ultra thick] plot(\x,{tanh(\x)});
  \draw[domain=-7.8:7.8,samples=50,red, ultra thick] plot(\x,{-1*tanh((\x))});
   \draw[black] (5,1) node[above] { \small Instanton};
   \draw[black] (5,-1) node[above] { \small Anti-Instanton};
   \draw[black] (0,1) node[left] {${\scriptstyle a}$};
   \draw[black] (0,-1) node[left] {${\scriptstyle -a}$};
   \draw[black, gray] (0,1) node[] {${\scriptscriptstyle -}$};
   \draw[black, gray] (0,-1) node[] {${\scriptscriptstyle -}$};   
\end{tikzpicture}
\caption{The instanton and the anti-instanton solutions in a double-well potential.}
\label{F3.3} 
\end{figure}
\end{itemize}

}

\subsection{The Action for the Instanton Solution} 
\par{
In this section, we calculate the instanton's action. Using the explicit expression for the instanton solution (\ref{3.50}) gives
\begin{equation} \label{3.51}
\dot{X} = \left( \frac{a \omega}{2} \right) \frac{1}{\cosh^2 \frac{\omega}{2} (\tau - \tau_c)}.
\end{equation}
Therefore, we are able to compute an explicit expression for the action by substituting Eq. (\ref{3.51}) into the following relation,
\begin{align}
S_0 & = S[X]= \int_{- \infty}^{\infty} d \tau \dot{X}^2 \nonumber \\
& = \frac{a^2 \omega^2}{4} \int_{-\infty}^{\infty} d \tau \frac{1}{\cosh^4 \frac{\omega}{2}(\tau-\tau_0)}.
\end{align}
Now let us take
\begin{equation}
u \equiv \frac{\omega}{2}(\tau-\tau_0) \quad \Rightarrow \quad d \tau= \frac{2}{\omega} du,
\end{equation}
then, we have
\begin{align} \label{3.54}
S_0 &= \frac{\omega^3}{16 \lambda} \int_{-\infty}^{\infty} d u \frac{1}{\cosh^4 u} \nonumber \\
& = \frac{\omega^3}{16 \lambda} \times \frac{4}{3} = \frac{\omega^3}{12 \lambda}.
\end{align} 
This gives us the power of the principle exponential factor in Eq. (\ref{3.42}), $e^{-\mathit{S}_0}$.}

\subsection{Calculating the Determinant}
\par{
Our next task is to calculate the determinant factor which appears in Eq. (\ref{3.42}). This is not a simple problem. For that reason, rather than calculating this determinant directly, we will calculate it in terms of the determinant of the harmonic oscillator, which already has been calculated in section (\ref{Sec3.2}). Therefore, our task reduces to calculating the ratio between the determinant of the double-well potential and the determinant of the harmonic oscillator potential. Thus we have   
\begin{equation} \label{3.55}
\left[ \det \left(- \dfrac{d^2}{d \tau^2} + V''(X)\right) \right]^{-\frac{1}{2}} = \det \left(- \dfrac{d^2}{d \tau^2} + \omega^2 \right)^{-\frac{1}{2}} \times \left\{  \frac{\det \left(- \dfrac{d^2}{d \tau^2} + V''(X)\right)}{\det \left(- \dfrac{d^2}{d \tau^2} + \omega^2 \right)} \right\}^{-\frac{1}{2}}.
\end{equation}
We have multiplied and divided by the determinant of the harmonic oscillator. That is the first factor of the right side of Eq. (\ref{3.55}). Hence, to calculate the left side of Eq. (\ref{3.55}), we merely need to compute the ratio of the second factor of the right side. In order to do that, we need the solution to the following eigenvalue problem,
\begin{equation}
- \dfrac{d^2}{d \tau^2} x_n + V''(X) x_n= \epsilon_n x_n.
\end{equation}

}
\par{
Using the explicit form of the classical solution (\ref{3.50}) and by a simple calculation,  we have the following expression for the second derivative of the potential (\ref{3.38}) (see Appendix \ref{AppendixC.1}),
\begin{equation}
V''(X)= \omega^2 \left( 1- \frac{3}{2 \cosh^2 \frac{\omega}{2}(\tau-\tau_c)} \right).
\end{equation}
Substituting the explicit expression of $V''(X)$, we get the eigenvalue equation,
\begin{equation} \label{3.58}
- \dfrac{d^2}{d \tau^2} x_n + \left( \omega^2 - \frac{3 \omega^2}{2 \cosh^2 \frac{\omega}{2} (\tau_2 - \tau_c)} \right) x_n = \epsilon_c x_n. 
\end{equation} 
This is formally identical to  the Schr\"{o}dinger equation with  a potential $U(\tau)= V''(X(\tau))$, see Fig (\ref{F3.4}). Taking $u\equiv \frac{\omega}{2} (\tau-\tau_c)$, and rearranging Eq. (\ref{3.58}), it can be written in the form
\begin{equation} \label{3.59}
\dfrac{d^2 x_n}{d \tau^2} - (\omega^2 - \epsilon_n)x_n + \frac{3 \omega^2}{2 \cosh^2 u} x_n = 0.
\end{equation}
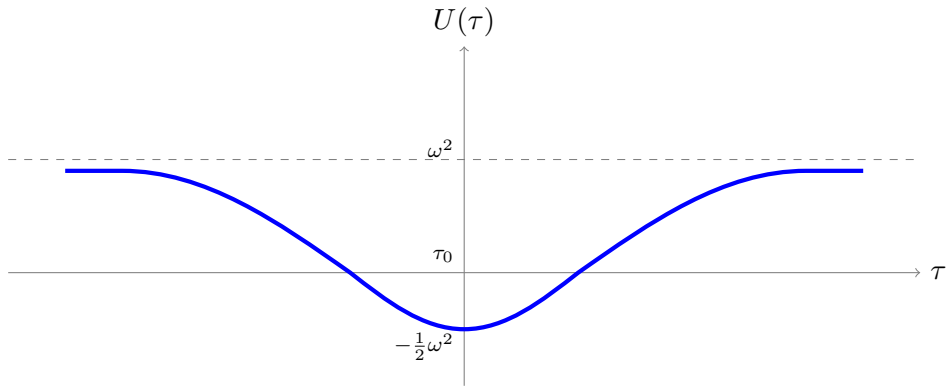
\begin{figure} [h!]
\centering
\begin{tikzpicture}[xscale=1.5,yscale=1.5]
  \draw[->, gray] (0,-1) -- (0,2) node[above,black] {$U(\tau)$};
  \draw[->, gray] (-4,0)--(4,0) node[right,black] {$\tau$};
  \draw[dashed, gray] (-4,1)--(4,1);
  \draw[blue, ultra thick]{(-3.5,.9) cos(-3,.9)  cos (-1,0) sin (0,-0.5) cos (1,0) sin (3,0.9) sin(3.5,0.9)};
   \draw[black] (0,0.15) node[above,left] {${\scriptstyle \tau_0}$};
   \draw[black] (0,1.1) node[above,left] {${\scriptstyle \omega^2}$};
   \draw[black] (0,-0.65) node[below,left] {${\scriptstyle - \frac{1}{2} \omega^2}$};
   \end{tikzpicture}
\caption{A sketch of the lowest wave function correspond to the solution of Schr\"{o}edinger equation with a potential $U(\tau)$.}
\label{F3.4} 
\end{figure}
}
\par{
Next, we have to simplify this equation in steps to be exactly solvable. Firstly, the instanton solution (\ref{3.50}) in terms of $u$ can be written as
\begin{equation}
X= \pm a \tanh u.
\end{equation}
If we make the change of variables
\begin{equation} \label{3.61}
\xi = \frac{X_+}{a} = \tanh u \quad \Rightarrow \quad 1-\xi^2 = 1- \tanh^2 u = \frac{1}{\cosh^2 u},
\end{equation} 
then we have
\begin{equation}
\dot{\xi}= \frac{\omega}{2} (1- \zeta^2) \quad \Rightarrow
\quad \dfrac{d}{d \tau} =\frac{\omega}{2} (1-\xi^2) \dfrac{d }{d \xi}.
\end{equation}
Furthermore,
\begin{align} \label{3.63}
\dfrac{d^2}{d \tau^2} &= \dfrac{d}{d \tau} \left[ \frac{\omega}{2} (1-\xi^2) \dfrac{d }{d \xi} \right] = \dot{\xi} \dfrac{d}{d \xi} \left[ \frac{\omega}{2} (1-\xi^2) \dfrac{d }{d \xi} \right] \nonumber \\
&= \frac{\omega^2}{4} (1-\xi^2) \dfrac{d}{d \xi} (1-\xi^2) \dfrac{d}{d \xi}.
\end{align}
Substituting with Eq. (\ref{3.61}, \ref{3.63}), into Eq. (\ref{3.59}), we get
\begin{equation}
\left( \frac{\omega^2}{4} (1-\xi^2) \dfrac{d}{d \xi} (1-\xi^2) \dfrac{d}{d \xi} \right) x_n + \left( \frac{3 \omega^2}{2} (1-\xi^2)+ \epsilon_n- \omega^2 \right) x_n =0.
\end{equation}
Then, dividing by the factor $\frac{\omega^2}{4} (1-\xi^2)$,  we get
\begin{equation} \label{3.65}
\dfrac{d}{d \xi} (1-\xi^2) \dfrac{d}{d \xi} x_n + \left( 6+ \frac{4(\epsilon_n - \omega^2)}{\omega^2(1-\xi^2)} \right) x_n =0
\end{equation}
If we take, 
\begin{equation} \label{3.a65}
a=6, \qquad \text{and} \qquad b=\frac{4(\epsilon_n - \omega^2)}{\omega^2},
\end{equation}
then, Eq. (\ref{3.65}) becomes
\begin{equation} \label{3.66}
\dfrac{d}{d \xi} (1-\xi^2) \dfrac{d}{d \xi} x_n + \left( a + \frac{b}{1-\xi^2} \right) x_n = 0.
\end{equation}
Actually, we can solve this kind of equations in terms of a hypergeometric function as follows
\begin{equation} \label{3.67}
x = \left(1-\xi^2\right)^c \chi,
\end{equation}
where $c$ is a constant, which will be chosen appropriately later, and for simplicity, we have omitted the subscript $n$.}
\par{
Moreover, substituting Eq. (\ref{3.67}) into Eq. (\ref{3.66}), and following this by a simple algebra (see Appendix \ref{AppendixC.2}), we get
\begin{align} \label{3.68}
(1-\xi^2) \dfrac{d^2 \chi}{d \xi^2} - (4c+2) \xi \dfrac{d \chi}{d \xi} + \left( a-2c-4c^2+ \frac{b+4c^2}{1-\xi^2} \right) \chi&=0.
\end{align}
In addition, we have to make another change of variables in the form
\begin{equation}\label{3.69}
z=\frac{1}{2} (1-\xi). 
\end{equation}
This implies that
\begin{align}\label{3.70}
(1-z)&= \frac{1}{2} (1-\xi) &  \Rightarrow  & \quad (1-\xi^2) = 4z(1-z)&&&& \nonumber \\
 \dfrac{dz}{d \xi} &= -\frac{1}{2} & \Rightarrow  & \quad \dfrac{d^2}{d \xi}= - \frac{1}{2} \dfrac{d}{d z} \quad \& \quad \dfrac{d^2}{d \xi^2}= \frac{1}{4} \dfrac{d^2}{d z}.&&&&&&&&&&
\end{align}
Thus, the corresponding asymptotic values of our variables are
\begin{align}
&&&&\tau & \rightarrow \infty & \Longleftrightarrow & & \xi  \rightarrow   &1 & \Longleftrightarrow & &z\rightarrow0& &&&& \nonumber \\ 
&&&&\tau & \rightarrow -\infty & \Longleftrightarrow & & \xi  \rightarrow  &-1 & \Longleftrightarrow & &z\rightarrow 1&. &&&&
\end{align}
Substituting Eqs. (\ref{3.69}, \ref{3.70}) into Eq. (\ref{3.68}), we get
\begin{align} \label{3.72}
z(z-1) \dfrac{d^2 \chi}{dz^2} +\frac{1}{2} (4c+2)(1-2z) \dfrac{d \chi}{dz} + \left( a+2c-4c^2+\frac{b+4c^2}{4z(z-1)} \right) \chi &=0 \nonumber \\
z(z-1) \dfrac{d^2 \chi}{dz^2} + \left( 2c+1- (4c+2)z \right) \dfrac{d \chi}{dz} + \left( a+2c-4c^2+\frac{b+4c^2}{4z(z-1)} \right) \chi &=0
\end{align} \label{3.74}
If we choose $c$ such that
\begin{equation}
b+4c^2=0,
\end{equation}
and then taking
\begin{equation} \label{3.75}
\gamma=2c+1, \quad \alpha+\beta=4c+1, \quad \alpha \beta = 4c^2 + 2c-a.
\end{equation}
 Eq. (\ref{3.72}) becomes the familiar hypergeometric equation
\begin{equation} \label{3.66}
z(1-z) \frac{d^2 \chi}{d z^2}+\left( \gamma -(\alpha + \beta +1)z \right) \dfrac{d \chi}{dz} - \alpha \beta \chi =0
\end{equation}
The solution is the hypergeometric function $F(\alpha,\beta,\gamma;z)$, which around $z=0$ has the expansion
\begin{equation} \label{3.77}
F(\alpha, \beta,\gamma;z)= 1+ \frac{\alpha \beta}{\gamma} \frac{z}{1 !} + \frac{\alpha (\alpha+1)(\beta+1)}{\gamma(\gamma+1)} \frac{z^2}{2!}+ \dots .
\end{equation}  
Remember from Eq. (\ref{3.a65}) that $a=6$, so the parameters (\ref{3.75}) can be simplified to
\begin{equation} \label{3.78}
\alpha= 2c-2, \qquad 
\beta=2c+3, \qquad \gamma=2c+1.
\end{equation} 
In terms of the hypergeometric function (\ref{3.77}), the solution (\ref{3.67}) of the eigenvalue equation (\ref{3.58}) can be rewritten as
\begin{equation} \label{3.79}
x = \left(1-\xi^2\right)^c F(\alpha, \beta,\gamma;z).
\end{equation} 

} 
\subsection{Continuous Spectrum} \label{Sec3.3.4}
\par{
It can be seen from the form of the double-well potential that there are discrete, as well as continuous, spectra.  The spectrum is continuous for $\epsilon \geq \omega^2$, and it can be parametrized by the real positive momentum
\begin{equation} \label{3.80}
p \equiv \sqrt{\epsilon -\omega^2}, \qquad \text{or} \qquad k \equiv \frac{p}{\omega}.
\end{equation} 
In addition, there is no barrier for $\epsilon > \omega^2$, and for that reason, we expect that there is no reflection as the particle moves from $\tau=- \infty$ to $\tau=\infty$. Consequently, all the scattering dynamics are contained in the knowledge of the phase shift $\delta_p$. If we set $\tau_c =0$, then the phase shift can be defined as
\begin{align}
x_p (\tau \rightarrow \infty) &= e^{ip\tau}, \nonumber \\
x_p (\tau \rightarrow -\infty)& = e^{i p \tau + i \delta p}.
\end{align}

}
\par{
We will see later that the knowledge of the phase shift is enough to compute the contribution to the determinant. Moreover, recall from Eq. (\ref{3.a65}) that $b=\frac{4 (\epsilon_n - \omega^2)}{\omega^2}$. By using the definition of the momentum (\ref{3.80}), and comparing the result we get with Eq. (\ref{3.74}), we have $b=4k^2=-4c^2$. This implies that we get $c^2=-k^2$. If we choose $c=-ik$ and substitute this value into Eq. (\ref{3.79}), then we get the solution which is asymptotic to $e^{ip\tau}$ as $\tau \rightarrow \infty$ (\ie $z \rightarrow 0$), that is  
\begin{equation} \label{3.82}
x=\left(1-\xi^2 \right)^{-ik} F(\alpha,\beta,\gamma;z).
\end{equation}
\begin{itemize}
\item Near $\tau \rightarrow \infty$ (\ie $z \rightarrow 0$), if we take $u= \frac{1}{2}$, then we have $2uk=\omega \tau k= p\tau$.  Therefore the first factor of Eq. (\ref{3.82}) becomes
\begin{equation} \label{3.83}
(1-\xi^2)^{-ik}  = (\cosh^2 u)^{iK} \simeq (4 e^{-2u})^{-ik}= 4^{-iK} e^{ip \tau}.
\end{equation}
\item However, near $\tau \rightarrow -\infty$ (\ie $z \rightarrow 1$), the first factor of Eq. (\ref{3.82}) becomes
\begin{equation} \label{3.84}
(1-\xi^2)^{-ik} \simeq (4 e^{2u})^{-iK}= 4^{-ik} e^{-ip \tau}.
\end{equation}
\end{itemize}
Actually, the reflected wave is represented only by Eq. (\ref{3.84}). On the other hand, the hypergeometric function can be rewritten as
\begin{equation}
F(\alpha,\beta,\gamma;z)= F_1(1-z) + F_2(1-z),
\end{equation} 
where $F_1(1-z)$ and $F_2(1-z)$ are defined respectively as
\begin{align}
F_1(1-z) &= \frac{\Gamma(\gamma) \Gamma(\gamma-\alpha-\beta)}{\Gamma(\gamma-\alpha) \Gamma(\gamma-\beta)} F(\alpha,\beta,\alpha+\beta+1-\gamma;1-z), \\
F_2(1-z) &= \frac{\Gamma(\gamma) \Gamma(\alpha+\beta-\gamma)}{\Gamma(\alpha)\Gamma(\beta)} (1-z)^{\gamma-\alpha-\beta} \times F(\gamma-\alpha,\gamma-\beta,\gamma+1-\alpha-\beta;1-z).
\end{align}
Since we have chosen $c=-ik$, we can rewrite the parameters $\alpha, \beta,\gamma$ in Eq. (\ref{3.78}) in terms of $k$ as follows
\begin{equation}
\alpha= -2ik-2 \quad \& \quad \beta = -2ik+3 \quad \& \quad \gamma=-2ik+1.
\end{equation}
It can be seen that $F_1(1-z)$ vanishes due to the dominant factor $\Gamma(\gamma-\beta)=\Gamma(-2)$, which diverges. Subsequently, near $\tau \rightarrow -\infty$ (\ie $z \rightarrow 1$), the complete  Eq. (\ref{3.82}) becomes
\begin{align}
(1-\xi^2)^{-ik}  F(1-z)&= (1-\xi^2)^{-ik}  F_2(1-z)\nonumber \\ &= \frac{\Gamma(-2ik+1)\Gamma(-2ik)}{\Gamma(-2ik+3)\Gamma(-2ik-2)} (1-z)^{2ik} (1-\xi)^{-ik}\left(1+\mathcal{O}(1-z)\right) \nonumber \\
&= \frac{(1+2ik)(1+ik)}{(1-2ik)(1-ik)} (1-\xi)^{ik} 4^{-2ik}\left(1+\mathcal{O}(1-z)\right) \nonumber \\ 
& \simeq \frac{(1+2ik)(1+ik)}{(1-2ik)(1-ik)} 4^{-ik}e^{ip\tau}.
\end{align}
Comparing this result with Eq. (\ref{3.83}) leads us to read off the phase shift as 
\begin{equation} \label{3.90}
e^{i\delta p}= \frac{(1+2ik)(1+ik)}{(1-2ik)(1-ik)}.
\end{equation}

}
\subsection{Discretization of the Spectrum}
\par{
In this subsection, we must regularize the continuous spectrum, before going ahead to calculate the contribution to the determinant ratio in the next subsection. As usual, the appropriate way is to put the system in a box of interval $- \frac{\tau_0}{2} < 0< \frac{\tau_0}{2}$. We also have to take the boundary conditions as
\begin{equation} \label{3.91}
x(- \tau_0/2) = x(\tau_0/2) =0.
\end{equation} 
Based on these boundary conditions, there will be a reflected wave which vanishes as $\tau_0 \rightarrow  \infty$. Since $x_p(-\tau)$ and $x_p(\tau)$ are independent solutions, the general solution must be of the form
\begin{equation}
x = A x_p (\tau)+ B x_p (-\tau).
\end{equation}
Then, using the boundary conditions (\ref{3.91}), we have
\begin{align}
A x_p (\tau_0/2) + B x_p (-\tau_0/2) &=0, \nonumber \\
A x_p (-\tau_0/2) + B x_p (\tau_0/2) &=0.
\end{align}
Solving those two equations gives us the non-trivial solution $A=\pm B$. This implies that we have
\begin{equation}
\frac{x_p(\tau_0/2)}{x_p(-\tau_0/2)} = \pm 1.
\end{equation} 
Using the asymptotic form of the solution, we find that $e^{ip\tau_0-i\delta_p}=\pm 1$. Hence the spectrum $\tilde{p}_n=$ which satisfies the boundary conditions (\ref{3.91}), is given by 
\begin{equation} 
\tilde{p}_n= \frac{\pi n+\delta_p}{\tau_0}, \qquad n= 0,1,2,\dots.
\end{equation}

}
\subsection{Contribution to the Determinant Ratio}
\par{
At this stage, we are able to calculate the contribution of these modes to the determinant ratio. Let us recall from the harmonic oscillator, the spectrum is
\begin{equation}
p_n= \frac{\pi n}{\tau_0}.
\end{equation} 
In the limit $\tau_0 \rightarrow \infty$, it can be seen that the contribution from a finite number of eigenvalues with $n \sim \mathcal{O}(1)$ cancel against the contribution from the harmonic oscillator states as they both become $\omega^2$. For instance, remember from Eq. (\ref{3.80}) that $p=\sqrt{\epsilon - \omega^2}$, and so $\epsilon = \omega^2 + p^2$. Thus, the ratio to be calculated can be taken as
\begin{equation} \label{3.97}
\mathit{R} \equiv \prod^{\infty}_{n=1} \frac{\omega^2+\tilde{p}^2_n}{\omega^2+p^2_n}.
\end{equation}
Let us define the difference $\bar{\Delta} p_n $ as follows
\begin{equation} \label{3.a98}
\bar{\Delta} p_n \equiv \tilde{p}_n - p_n = \frac{\pi n + \delta_p}{\tau_0} - \frac{\pi n}{\tau_0} = \frac{\delta_p}{\tau_0}.
\end{equation}
It is easy to see that the difference $\bar{\Delta} p_n$ tends to zero as $\tau_0 \rightarrow \infty$. Thus, we can expand in powers of $\bar{\Delta}p_n$ to the first order, therefore we have
\begin{equation}
\tilde{p}^2_n = p^2_n + 2 p_n \bar{\Delta}p_n + \dots .
\end{equation}
Hence, Eq. (\ref{3.97}) becomes
\begin{align} \label{3.98}
\mathit{R} &= \exp \left[ \sum \ln \frac{ \omega^2 + \tilde{p}^2_n}{\omega^2+p^2_n} \right] = \exp \left[ \sum \ln \frac{ \omega^2 + p^2_n + 2 p_n \bar{\Delta}p_n + \dots}{\omega^2+p^2_n} \right] \nonumber \\
& \simeq \exp \left[ \sum \ln \left(1+ \frac{2 p_n \bar{\Delta} p_n }{\omega^2 + p^2_n} \right) \right] \simeq \exp \left[ \sum \frac{2 p_n \bar{\Delta} p_n }{\omega^2 + p^2_n} \right].
\end{align}
We used the Taylor expansion $\ln(1+x)= x- \frac{x^2}{2}+ \frac{x^3}{3}- \dots$ to the first order to get the least approximation. Moreover, since at the same time the interval $\pi/\tau_0$ goes to zero, we can rewrite (\ref{3.a98}) as 
\begin{equation}
\tilde{\Delta}p_n= \frac{\delta_p}{\tau_0}=\frac{\delta_p}{\pi}\frac{\pi}{\tau_0}= \frac{\delta_p}{\pi}(p_{n+1}-p_n)=\frac{\delta_p}{\pi} \Delta p_n.
\end{equation}
In terms of this notation, we may rewrite Eq. (\ref{3.98}) in the following form,
\begin{align}
\mathit{R} & \simeq  \exp \left[ \frac{1}{\pi} \int_0^{\infty} \delta_p \frac{2 p}{\omega^2+p^2} dp \right] = \exp \left[ \frac{1}{\pi} \int_0^{\infty} \delta_p \dfrac{d}{dp} \ln \left( 1+ \frac{p^2}{\omega^2} \right) dp \right].
\end{align}
Using integration by parts and noticing that $\delta_p$ is zero at the boundaries, we get
\begin{align}
\mathit{R} & = \exp\left[ - \frac{1}{\pi} \int_0^{\infty} dp \frac{ d \delta_p}{dp} \ln \left(1+\frac{p^2}{\omega^2})\right) \right].
\end{align}
Remember from Eq. (\ref{3.80}) that $k= \frac{p}{\omega}$, therefore we can rewrite the previous expression in terms of $k$ instead of $p$, that is
\begin{align} \label{3.104}
\mathit{R} & = \exp\left[ - \frac{1}{\pi} \int_0^{\infty} dk \frac{d \delta_k}{dk} \ln(1+k^2) \right].
\end{align}
From the explicit expression (\ref{3.90}) for the phase shift, we easily get
\begin{equation} \label{3.105}
\frac{d \delta_k}{dk}= \frac{2}{1+k^2}+\frac{4}{1+4k^2}.
\end{equation}
In the next step, we have to use the following formula for integration
\begin{equation} \label{3.106}
\int_0^{\infty} dk \frac{\ln (1+k^2)}{1+a^2k^2}= \frac{\pi}{a} \ln\left(1+\frac{1}{a}\right).
\end{equation}
Subsequently, we shall substitute Eq. (\ref{3.105}) into Eq. (\ref{3.104}) and then use Eq. (\ref{3.106}) to find the final value for the determinant ratio
\begin{align}
\mathit{R} &= \exp \left[ - \frac{1}{\pi} \int_0^{\infty} dk \left( \frac{2}{1+k^2}+\frac{4}{1+4k^2}\right) \ln (1+k^2) \right] \nonumber \\
&= \exp \left[ - \frac{2}{\pi} \int_0^{\infty} dk \frac{\ln (1+k^2)}{1+k^2} - \frac{4}{\pi} \int_0^{\infty} dk \frac{\ln (1+k^2)}{1+4k^2} \right] \nonumber \\
&= \exp \left[ - \frac{2}{\pi} (\pi \ln 2)- \frac{4}{\pi} (\frac{\pi}{2} \ln \frac{3}{2}) \right] = \exp \left[ \ln \frac{1}{9} \right] = \frac{1}{9}.
\end{align}
This is exactly the ratio between the determinant of the double-well and the determinant of the harmonic oscillator.}

\subsection{Discrete Spectrum}
\par{
As we discussed in Subsection \ref{Sec3.3.4}, for the double-well potential there are discrete as well as continuous spectra. We have already calculated the contribution due to the continuous spectrum. In this subsection, we consider the discrete part of the spectrum in the case $\omega^2 - \epsilon_n >0$. Thus, we must set
\begin{equation} \label{3.108}
c=k \equiv \frac{\sqrt{\omega^2 - \epsilon_n}}{\omega} > 0.
\end{equation}  
Based on our argument in subsection \ref{Sec3.3.4}, the solution of Shr\"{o}dinger equation (\ref{3.59}) near $\tau_0 \rightarrow \infty$ (\ie $z\rightarrow 0$) is 
\begin{equation}
x_n = (1-\xi^2)^k \mathit{F}(\alpha, \beta, \gamma;z),
\end{equation}
where the hypergeometric function $\mathit{F}(\alpha, \beta, \gamma;z)$ is given by Eq. (\ref{3.77}). In order for the solution, in this case, to be finite, we have to set the parameters of the hypergeometric function to be
\begin{equation}
\alpha= 2k-2, \qquad \beta= 2k+3, \qquad \gamma=2k+1.
\end{equation}

On the other hand, near $\tau_0 \rightarrow -\infty$ (\ie $z\rightarrow 1$), for this solution to be finite, the series must terminate at finite terms. Since $\beta, \gamma, k >0$ and $\alpha 2k-2 >-2$, this occurs only when
\begin{equation} \label{3.111}
\alpha=2k-2=-n, \qquad n=0,1.
\end{equation} 
This implies that we have only two values for $k$, they are
\begin{equation} \label{3.112}
k=1 \qquad \& \qquad k= \frac{1}{2}.
\end{equation}
From the definition of $k$ Eq. (\ref{3.112}), it can be seen that $\epsilon_n = \omega^2 (1-k^2)$. Using Eq. (\ref{3.104}) we realize that there are two discrete energy levels (bound states), they are
\begin{equation}
\epsilon_0 = 0, \qquad \epsilon_1= \frac{3}{4} \omega^2.
\end{equation} 

}
\subsection{Treatment of the Zero Mode}
\par{
In the previous subsection, we found that there are two discrete modes of energy levels. Let us constraint ourselves with the one which corresponds to the zero eigenvalue $\epsilon_0=0$. Actually, we should avoid including the zero-mode into the determinant calculation. Inclusion of such a mode makes the determinant vanish.} 
\par{
As shown in Eqs. (\ref{3.108},\ref{3.111}, \ref{3.112}), for the zero mode we have $\alpha=0$ and $k=c=1$. Therefore, in this case the hypergeometric function  $\mathit{F}(\alpha,\beta,\gamma;z)=1$. The corresponding wave function can be obtained by substituting those values into Eq. (\ref{3.79}), then we have
\begin{equation}
x_0(\tau) \propto \frac{1}{\cosh^2 \frac{\omega}{2} (\tau - \tau_c)}.
\end{equation} 
As usual, we have to normalize the wave function. Suppose $\mathcal{N}$ is the normalization factor, then we write the zero-mode normalized wave function as follows
\begin{equation}
x_0(\tau) = \frac{\mathcal{N}}{\cosh^2 \frac{\omega}{2} (\tau - \tau_c)}.
\end{equation}   
The wave function must obey the quantum mechanical normalization condition 
\begin{equation}
\int_{-\infty}^{\infty} \vert x_0(\tau) \vert^{2}=1.
\end{equation}
Thus we have
\begin{equation}
1= \int_{-\infty}^{\infty} d \tau \frac{\mathcal{N}^2}{\cosh^4 \frac{\omega}{2} (\tau - \tau_c)} = \mathcal{N}^2 \frac{8}{3 \omega}.
\end{equation}
This implies that $\mathcal{N}= \sqrt{\frac{8}{3 \omega}}$. Hence the zero-mode normalized wave function is
\begin{equation}\label{3.118}
x_0(\tau) = \sqrt{\frac{8}{3 \omega}} \frac{1}{\cosh^2 \frac{\omega}{2} (\tau - \tau_c)}.
\end{equation}

}
\par{
More importantly, we can compute the correctly normalized zero mode in terms of the instanton solution and the classical action. Since $\partial X (\tau,\tau_c)/\partial \tau_c= - \partial X(\tau,\tau_c)/\partial \tau$, we may write
\begin{equation}
x_0(\tau) = \mathcal{N} \frac{\partial X}{\partial \tau}.
\end{equation} 
Once again the normalization condition reads
\begin{equation}
1= \int_{- \infty}^{\infty} d \tau \vert x_0(\tau) \vert^2 = \mathcal{N}^2 \int_{- \infty}^{\infty} d \tau \left( \frac{\partial X}{\partial \tau} \right)^2 = \mathcal{N}^2 \mathit{S}_0. 
\end{equation}
This implies that $\mathcal{N}= \mathit{S}_0^{- \frac{1}{2}}$. Therefore the normalized zero mode can be written as
\begin{equation} \label{3.121}
x_0 = \mathit{S}_0^{- \frac{1}{2}} \frac{\partial X}{\partial \tau}.
\end{equation} 
For instance, let us recall the explicit forms for the instanton solution and the classical action (\ref{3.50}, \ref{3.54}) respectively, then we have 
\begin{align} \label{3.122}
\mathit{S}_0 &= \frac{\omega^3}{12 \lambda}, \nonumber \\
\frac{\partial X}{\partial \tau} &= \frac{a \omega}{2} \frac{a}{\cosh^2 \frac{\omega}{2} (\tau - \tau_c) } = \sqrt{\frac{\omega^4}{32 \lambda}} \frac{1}{\cosh^2 \frac{\omega}{2} (\tau - \tau_c)}. 
\end{align}
Substituting from Eq. (\ref{3.122}) into Eq. (\ref{3.121}), we get
\begin{equation}
x_0(\tau) = \frac{\omega^3}{12 \lambda}  \sqrt{\frac{\omega^4}{32 \lambda}} \frac{1}{\cosh^2 \frac{\omega}{2} (\tau - \tau_c)} =  \sqrt{\frac{3 \omega}{8}} \frac{1}{\cosh^2 \frac{\omega}{2} (\tau - \tau_c)}.
\end{equation}
This is exactly identical to Eq. (\ref{3.118}). This means that we got the same result without solving the Schr\"{o}dinger equation.}
\subsection{Integration over the Zero Mode (Collective Coordinate)}
\par{
In this subsection, we turn back to our argument in section \ref{Sec3.1}. We have performed the formal Gaussian integration $c_n$ defined by Eq. (\ref{3.6})
\begin{equation} \label{3.124}
x(\tau) = \mathit{X}(\tau) + \sum_n \mathit{c}_n x_n (\tau),
\end{equation}
and obtained $\left( \det \widehat{\mathit{W}} \right)^{- \frac{1}{2}}$. However, integration with respect to the coefficient $c_0$ corresponding to the zero-mode is non-Gaussian. Thus, we must treat its integration separately. Actually, the integration over $dc_0$ is the same as the integration over $d\tau_0$. To determine this change of variables, we merely need the coefficient of proportionality (the Jacobian). If $c_0$ in Eq. (\ref{3.124}) changes by $\Delta c_0$, then the corresponding change of $x(\tau)$ is given by
\begin{equation} \label{3.125}
\Delta x(\tau) = x_0 (\tau) \Delta c_0.
\end{equation} 
On the other hand, the change of $x(\tau)$ corresponds to the change $\Delta \tau_c$ and is given by
\begin{equation} \label{3.126}
\Delta x(\tau) = \Delta X(\tau_c) \Delta \tau_c = - \frac{d X}{ d\tau} \Delta \tau_c= - \sqrt{\mathit{S}_0} x_0(\tau) \Delta \tau_c.
\end{equation}
Comparing Eqs. (\ref{3.125}, \ref{3.126}), we obtain the Jacobian factor \cite{zinnbarrier} which corresponds to the change of these two variables, that is
\begin{equation}
\mathit{J} = \Big{\vert} \frac{d c_0}{d \tau_0} \Big{\vert} = \sqrt{\mathit{S}_0}.
\end{equation}
This method of changing variables is called a collective coordinate. Using this trick we are able to find the zero-mode integration by integrating over $d \tau_c$ rather than integrating over $d c_0$.}
\par{
So far, we only dealt with the treatment of the zero mode. In fact, this is not yet the end of the story, we still have one last factor contributing to the calculation of the determinant. This factor is due to the second discrete level with $\epsilon_1 = \frac{3}{4} \omega^2$. In the limit $\tau_0 \rightarrow \infty$, the contribution of this level to the determinant ration is $\frac{3}{4}$.}  
\subsection{Complete One Instanton Contribution}
\par{
We have now arrived at the point at which we are able to compute the complete contribution of one instanton for large $\tau_0$ to the transition amplitude. Combining all the results obtained in the previous subsections, we get
\begin{align}
\langle -a \vert e^{-\mathit{H} \tau_0} \vert a  \rangle_{\textsf{one-inst}} & = \frac{1}{\sqrt{2 \pi \tau_0}} \left( \frac{\sinh \omega \tau_0}{\omega \tau_0} \right)^{-\frac{1}{2}} 
 \left( \frac{1}{\omega^2} \times \frac{3}{4} \times \frac{1}{9} \right)^{-\frac{1}{2}} e^{- \mathit{S}_0} \sqrt{\mathit{S}_0} \dfrac{d\tau_c}{\sqrt{2 \pi}} \nonumber \\
& =\left( \sqrt{\frac{\omega}{\pi}} e^{-\omega \tau/2} \right) \left( \sqrt{\frac{6}{\pi}} \sqrt{\mathit{S}_0} e^{- \mathit{S}_0} \right) \omega d \tau_c.
\end{align} 
The factor inside the first brackets represents the harmonic oscillator-like contribution. While the factor inside the second brackets, which can be denoted by $\rho$, is called the instanton density.}
 
\section{Dilute Gas of Instantons and Anti-Instantons}
\par{
Up to now, we have concentrated on the calculation of the one-instanton contribution to the transition amplitude at large $\tau_0$. This is not the whole story, we have to consider the possibility that the particle oscillates between the two tops of the inverted potential in such a large time interval, as shown in Fig. (\ref{F3.5}). This issue was discussed further in \cite{marino2015instantons, forkel2000primer,casahorran2002instantonic}.
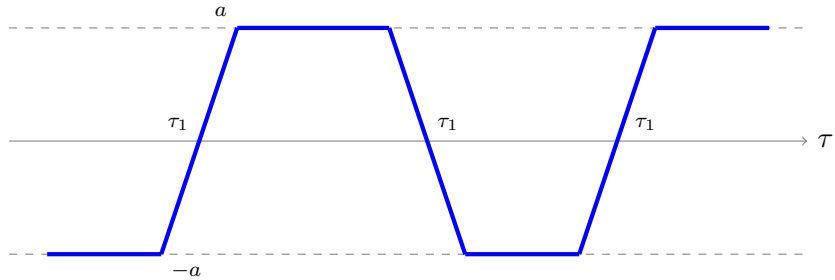
\begin{figure} [h!]
\centering
\begin{tikzpicture}[xscale=1,yscale=1.5]
  \draw[->, gray] (-.5,0)--(10,0) node[right,black] {$\tau$};
  \draw[dashed, gray] (-.5,1)--(10,1);
  \draw[dashed, gray] (-.5,-1)--(10,-1);
  
  \draw[blue, ultra thick] (0,-1)--(1.5,-1);
  \draw[blue, ultra thick] (1.5,-1)--(2.5,1);
  
  \draw[blue, ultra thick] (2.5,1)--(4.5,1);
  \draw[blue, ultra thick] (4.5,1)--(5.5,-1);
  
  \draw[blue, ultra thick] (5.5,-1)--(7,-1);
  \draw[blue, ultra thick] (7,-1)--(8,1);

  \draw[blue, ultra thick] (8,1)--(9.5,1);  
  
   \draw[black] (2,0.15) node[above,left] {${\scriptstyle \tau_1}$};
   \draw[black] (5,0.15) node[above,right] {${\scriptstyle \tau_1}$};
   \draw[black] (7.6,0.15) node[above,right] {${\scriptstyle \tau_1}$};
   
   \draw[black] (2.5,1.15) node[above,left] {${\scriptstyle a}$};
\draw[black] (1.5,-1.15) node[above,right] {${\scriptstyle -a}$};
   \end{tikzpicture}
\caption{The chain of $n$ well-separated instantons and anti-instantons .}
\label{F3.5} 
\end{figure}

}
\par{
In this condition, paths constructed from alternating instanton and anti-instanton $(IA)$ configurations. If the separation $\vert \tau_i - \tau_j\vert$ between their centers is much larger than the characteristic time scale of the system $\omega^{-1}$, then such paths constitute also classical solutions.}
\par{
For instance, we have to evaluate the contribution of such a configuration to the transition amplitude. Suppose we have $n$ such $IA$'s with centers $\tau_1, \tau_2, \dots, \tau_n$, whenever the points $\tau_i$ satisfy the condition
\begin{equation}
- \frac{\tau_0}{2} \leq \tau_1 \leq \tau_2 \leq \dots \leq \tau_n \leq \frac{\tau_n}{2}.
\end{equation} 
The corresponding action for such a configuration is $n\mathit{S}_0$, where $\mathit{S}_0$ is the one-instanton action.}
\par{
Finally, the contribution of $n$-$IA$ configuration to the transition amplitude can be written in the form 
\begin{align} \label{3.130}
\mathcal{A}_n & = \left( \sqrt{\frac{\omega}{\pi}} e^{-\omega \tau/2} \right) \rho^n I_n,
\end{align} 
where we define $I_n$ as follows
\begin{equation} \label{3.131}
I_n = \int_{-\tau_0/2}^{\tau_0/2} \omega d \tau_n \dots \int_{-\tau_0/2}^{\tau_3} \omega d \tau_2 \int_{-\tau_0/2}^{\tau_2} \omega d \tau_1.
\end{equation}
Moreover, if we define the variables 
\begin{align}
x_i  \equiv \omega \left( \tau_i +\frac{\tau_0}{2} \right), \qquad
x  \equiv \omega \left( \tau +\frac{\tau_0}{2} \right).
\end{align}
Then, we can rewrite the integration (\ref{3.131}) in the form
\begin{equation} \label{3.133}
I_n (\tau) \equiv J_n( x \equiv \omega \tau) = \int_{0}^{x} \omega d x_n \dots \int_{0}^{x_3} d x_2 \int_{0}^{x_2} d x_1,
\end{equation}
where
\begin{equation}
\dfrac{d J_n(x)}{d x} = J_{n-1}(x) \quad \Rightarrow \quad \dfrac{d^{n-1} J_n}{d x^{n-1}} = J_1(x) =x.
\end{equation}
But since $J_k(0)=0$ for all k, then the integration (\ref{3.133}) can be solved directly to obtain
\begin{equation} \label{3.135}
J_n(x) = \frac{x^n}{n!} \qquad \Rightarrow \qquad I_n = \frac{(\omega \tau_0)^n}{n!}.
\end{equation}
In order to compute the energy levels, we have to consider the amplitude $\langle a \vert e^{- \mathit{H}\tau_0} \vert a \rangle$  as well as the amplitude $\langle -a \vert e^{- \mathit{H}\tau_0} \vert a \rangle$. Another essential point, in the first case the amplitude is obtained by summing over the even number of $IA$. While to obtain the amplitude in the second case, we need to take the summation over the odd number of $IA$. Thus, substituting  Eq. (\ref{3.135}) into Eq. (\ref{3.130}), we get
\begin{align} \label{3.136}
\langle a\vert e^{-\mathit{H} \tau_0}\vert a\rangle &= \sqrt{\frac{\omega}{\pi}} e^{-\omega \tau_0/2} \sum_{n=0,2,\dots} \frac{(\omega \rho \tau_0 )^n}{n!}  = \sqrt{\frac{\omega}{\pi}} e^{-\omega \tau_0/2} \cosh (\omega \rho \tau_0 )\nonumber \\
&= \sqrt{\frac{\omega}{\pi}} e^{-\frac{\omega}{2} \tau_0} \left(\frac{ e^{\omega \rho \tau_0} + e^{-\omega \rho \tau_0}}{2} \right) = \sqrt{\frac{\omega}{4 \pi}} \left( e^{-(\frac{\omega}{2} - \omega \rho) \tau_0} + e^{-(\frac{\omega}{2}  + \omega \rho )\tau_0} \right), 
\end{align}
and
\begin{align} \label{3.137}
\langle -a \vert e^{-\mathit{H} \tau_0}\vert a \rangle &= \sqrt{\frac{\omega}{\pi}} e^{-\omega \tau_0/2} \sum_{n=1,3,\dots} \frac{(\omega \rho \tau_0 )^n}{n!} = \sqrt{\frac{\omega}{\pi}} e^{-\omega \tau_0/2} \sinh (\omega \rho \tau_0 )\nonumber \\
&= \sqrt{\frac{\omega}{\pi}} e^{-\frac{\omega}{2} \tau_0} \left(\frac{ e^{\omega \rho \tau_0} - e^{-\omega \rho \tau_0}}{2} \right) = \sqrt{\frac{\omega}{4 \pi}} \left( e^{-(\frac{\omega}{2} -\omega \rho) \tau_0} - e^{-(\frac{\omega}{2} \tau_0 -\omega \rho) \tau_0} \right) . 
\end{align}
Now using the  completeness relation of energy eigenstates, one should obtain
\begin{align} \label{3.138}
\langle a\vert e^{-\mathit{H} \tau_0}\vert a\rangle&= \sum_n e^{-E_n \tau_0} \vert \langle a \vert n\rangle \vert^2 , \\ \label{3.139}
\langle -a \vert e^{-\mathit{H} \tau_0}\vert a \rangle &= \sum_n e^{-E_n \tau_0}  \langle -a \vert n\rangle \langle n \vert a \rangle. 
\end{align}
By comparing the above completeness relations  Eqs. (\ref{3.136}, \ref{3.138}) with transition amplitude relations Eqs. (\ref{3.137} ,\ref{3.139}), the lowest two energy eigenvalues of the system can be immediately found as
\begin{align}
E_0 &= \frac{\omega}{2}- \omega \rho = \frac{\omega}{2} - \frac{\omega}{2} \sqrt{\frac{2\omega^3}{\pi \lambda}} e^{-\omega^3/12\lambda}, \\
E_1 &=\frac{\omega}{2}+ \omega \rho = \frac{\omega}{2} + \frac{\omega}{2} \sqrt{\frac{2\omega^3}{\pi \lambda}} e^{-\omega^3/12\lambda}.
\end{align}  
In addition, if the ground state of the system is donated by $\vert 0 \rangle$, then the ground state wave function can be written as  
\begin{equation}
\langle a \vert 0\rangle = \langle -a \vert 0 \rangle = \left( \frac{\omega}{4\pi} \right)^{-\frac{1}{4}}. 
\end{equation}
Thus, we see that the energy is split by non-perturbative tunneling contribution, and the symmetry of the ground state is indeed broken.}

\section{Real-Time Path Integral of Instantons} \label{Sec3.5}
\par{
As we have seen above, the tunneling phenomena are well-understood in the Euclidean-time path integral. However, it is not yet well understood in the real-time Feynman path integral formulation. Since physics actually happens in the real-time, so we look forward in this section, based on the argument of \cite{ yubo1999path, razavy2003quantum, shuryak2004qcd, song2008tunneling, kleinert2009path, rad2014feynman}, to understand such phenomena directly in the real-time path integral.} 

\subsection{The Problem with the Real-Time Formulation}
\par{
Consider the same problem of the symmetric double-well potential, which has been studied in the previous section in the case of Euclidean time. The real-time equation of motion for this problem is
\begin{equation} \label{3.143}
- \dfrac{d^2 x(t)}{d t^2}= \frac{\omega^2}{2 a^2} x(x^2-1).
\end{equation}
If we consider a real solution for this equation of motion, then the corresponding action will also be real. Since we are looking for a path integral of the weight $e^{i\mathit{S}[x]}$, that means the real-time instanton configuration will be complex. This is the main problem which meant instantons were never formulated in the real-time Feynman path integral. To be more definite, we have two essential inconveniences:
\begin{itemize}
\item[(i)] If that is the case, the field space should be complexified, which leads to an undesirable doubling of the degrees of freedom in the complexified path integral.
\item[(ii)] We have to avoid including pathological directions in the complex field space which may yield exponentially large contribution. 
\end{itemize} 

}    
\subsection{Pure Imaginary Solution}
\par{
For instance, we can use the Wick rotation to go back from Euclidean-time to real-time. Taking into account Eq. (\ref{3.50}), the solution of Eq. (\ref{3.143}) can be written in the form
\begin{equation} \label{3.144}
x(t)= \pm ia \tan \left( \frac{\omega}{2}(t-t_c) \right).
\end{equation}

}
\par{
 Unfortunately, this solution is not convenient for many reasons. Firstly, it is purely imaginary so it can not possibly give a contribution like $e^{-\mathit{S}[x]}$, and therefore it can not describe a path integral interpolating from $-a$ to $a$. Secondly, the solution is never located near the bottoms of the wells, so the description of the tunneling from $-a$ to $a$ is not clear. Third, the action is not well-defined since the solution has singularities  at $t=(2n+1)\pi/2,n\in \mathbb{Z}$.} 

\subsection{Complex Solution}
\par{
Based on \cite{cherman2014real, aoyama1995complex, levkov2005real, tanizaki2014real}, the complex-time formalism is developed in the framework of the path integral, to incorporate the imaginary-time formalism with the real-time formalism. To understand how this method works, suppose we start with real-time and then make a complex Wick rotation, which can be defined as
\begin{equation} \label{3.145}
t \rightarrow \tau e^{-i\alpha}.
\end{equation}    
Such rotation allows for the complexification of configuration space to deal with the inconveniences (i) and (ii). Making the complex Wick rotation  (\ref{3.145}) to the real-time equation of motion, we get the complex-time equation of motion, which is
\begin{equation}
- e^{2i\alpha} \dfrac{d^2 x(\tau)}{d \tau^2} = \frac{\omega^2}{2 a^2} x(x^2-a^2).
\end{equation} 
Indeed, the solution of such an equation of motion is
\begin{equation} \label{3.147}
x(\tau) = \pm a \tanh \left( \frac{\omega}{2} e^{-i (\alpha - \frac{\pi}{2})} \right).
\end{equation}
Moreover, the $\alpha$-dependent action can be written as
\begin{equation}
\mathit{S}_\alpha = \frac{\omega^2 e^{- i \alpha}}{2a} \int_{- \infty}^{ \infty} d\tau \left[ e^{2i \alpha} \dfrac{d^2 x}{d\tau^2} - (x^2-a^2)^2 \right]. 
\end{equation}
It can be seen that, for small $\alpha$, $\tau \approx t$ and $\mathit{S}_\alpha \approx \mathit{S}$. While, for $\alpha = \pi/2$, $\tau$ becomes the standard Euclidean time coordinate and $i\mathit{S}_{\pi/2}=- \mathit{S}_{\mathit{E}}$. }
\par{
Returning to the solution (\ref{3.147}), to make it a legitimate solution, we have to complexify the space of path histories from $x : \mathbb{R} \rightarrow \mathbb{R}$ to $x : \mathbb{R} e^{-i\alpha} \rightarrow \mathbb{C}$. A closer look at the solution itself shows that $\alpha$ acts like a crucial regulator which tells us whatever the solution represents Euclidean-time or real-time instantons.}
\begin{itemize}
\item[(i)] When $\alpha=\frac{\pi}{2}$, the complex instanton solution (\ref{3.147}) reduces to the pure real instanton solution (\ref{3.50}) see Fig. (\ref{F3.6}).
\item[(ii)] When $\alpha=0$, Eq. (\ref{3.147}) reduces to the pure imaginary solution (\ref{3.144}). The sketch for this situation is like the opposite of the first case  see Fig. (\ref{F3.7}). 
\item[(iii)] For $0 < \alpha \leq \frac{\pi}{2}$, consider the instanton starting at $x=-a$ at $\tau = - \infty$, it is supposed to whip around in the left half of complex-$x$ plane for $\tau < \tau_0$, then crosses to the right half of the plane at $\tau = \tau_0$, whips around $x=a$ for $\tau > \tau_0$, until it reaches $x=a$ and remains there  at $\tau=\infty$.
\begin{figure} [h!]
\centering
\includegraphics[scale=.5]{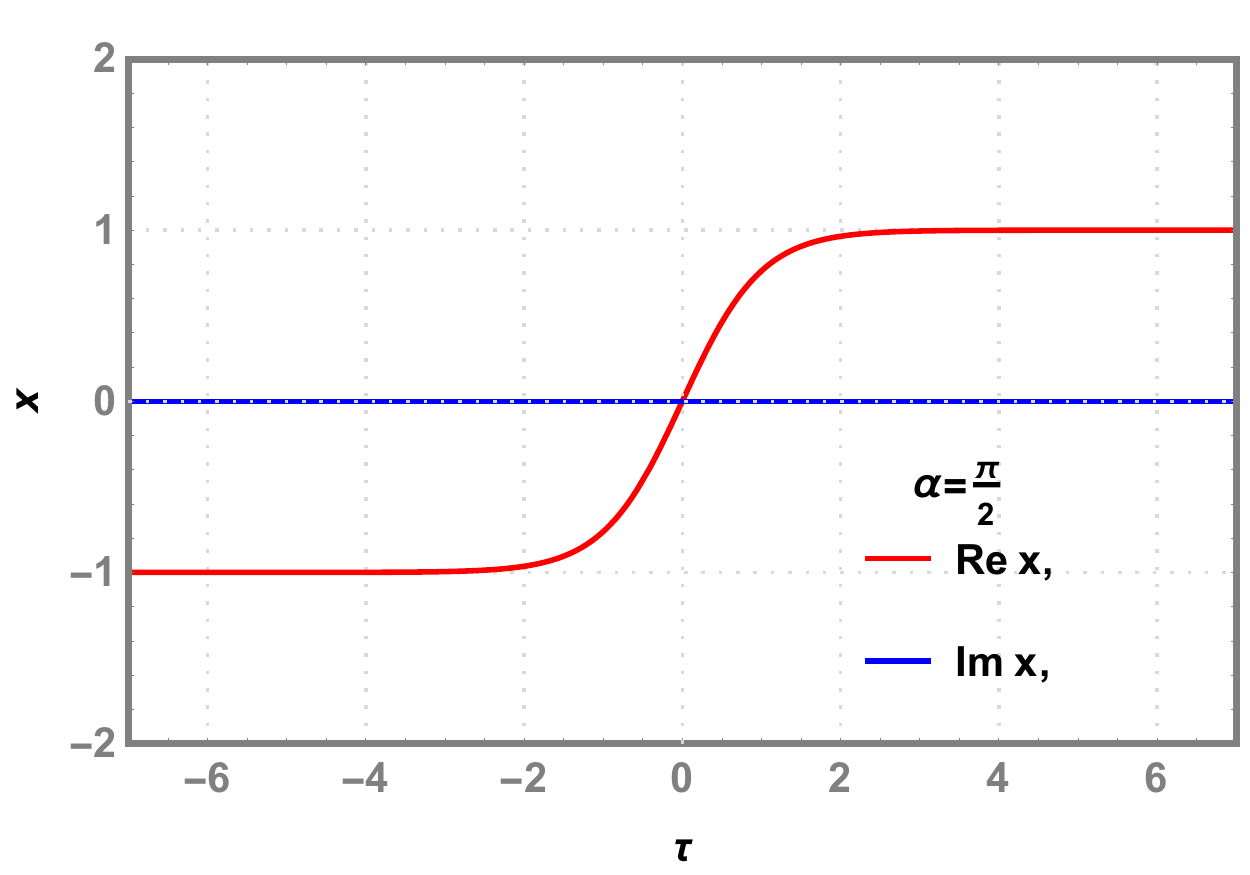} 
\includegraphics[scale=.5]{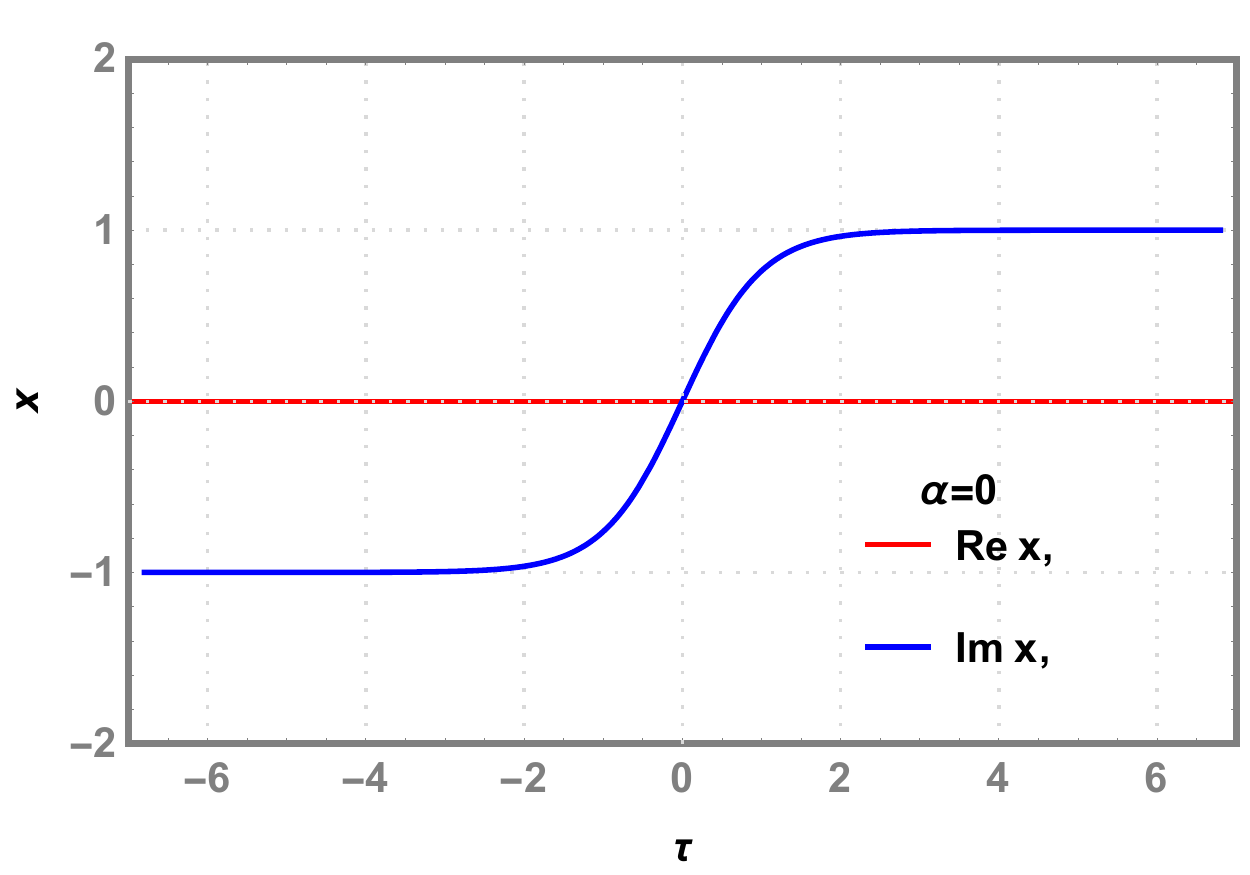}    
\includegraphics[scale=.51]{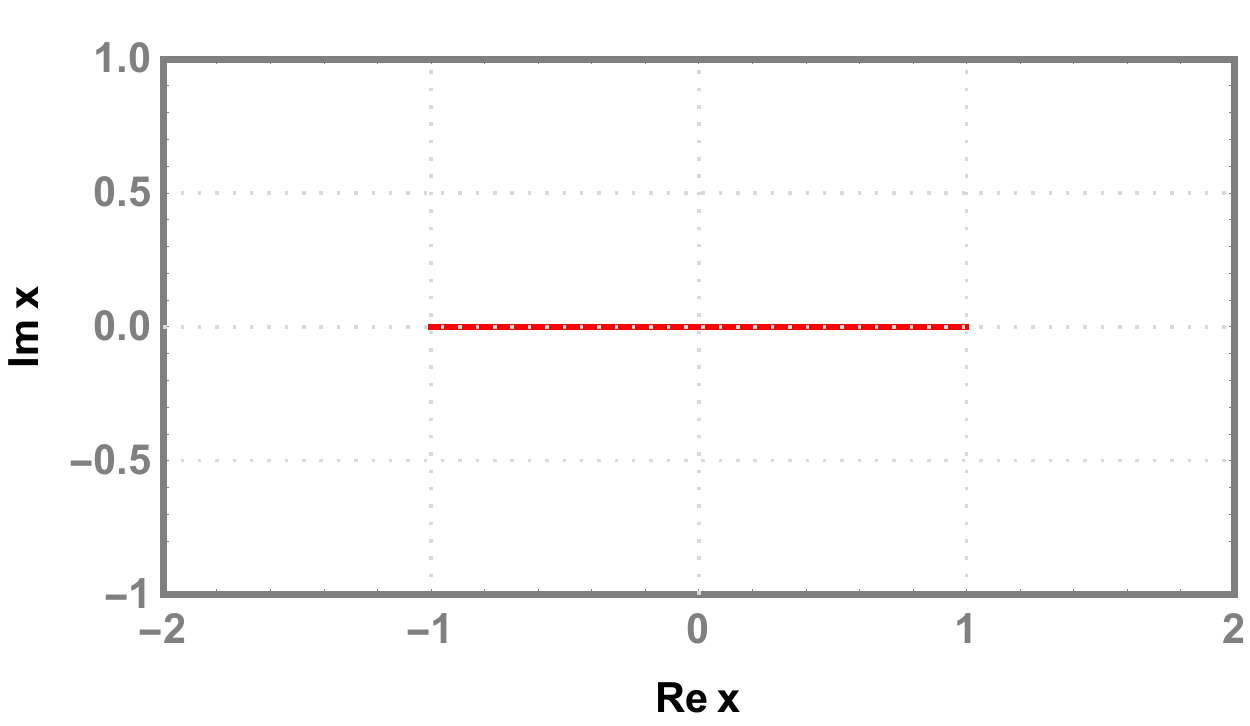}
\hspace{1.7cm} 
\includegraphics[scale=.31]{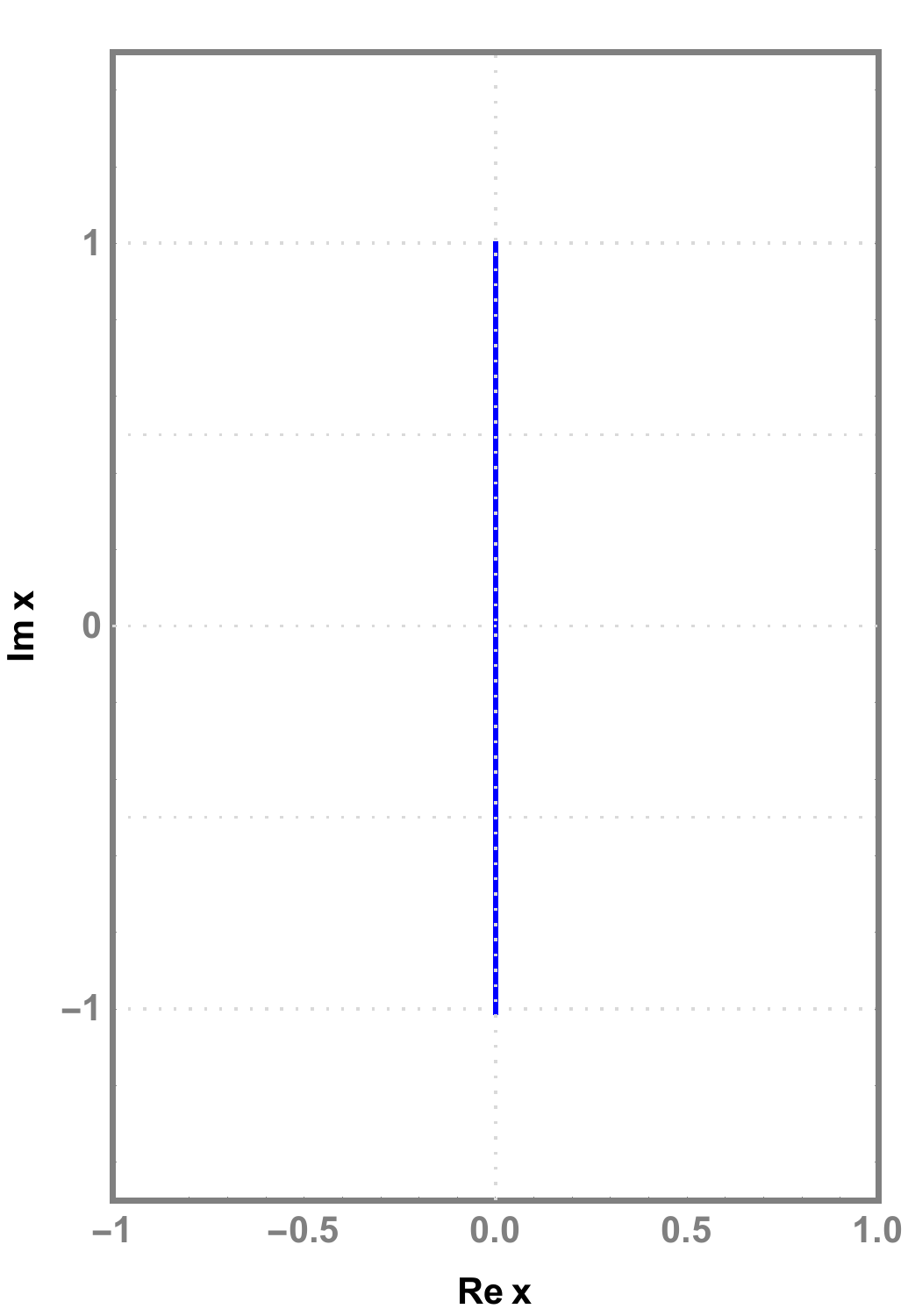}
\hspace{1.7cm}    
\caption{The complexified instanton configuration. The Euclidean instanton domination is shown in the left column, while the real-time instanton domination is shown in the right column.}
\label{F3.6} 
\end{figure} 
\begin{figure} [h!]
\centering
\includegraphics[scale=.5]{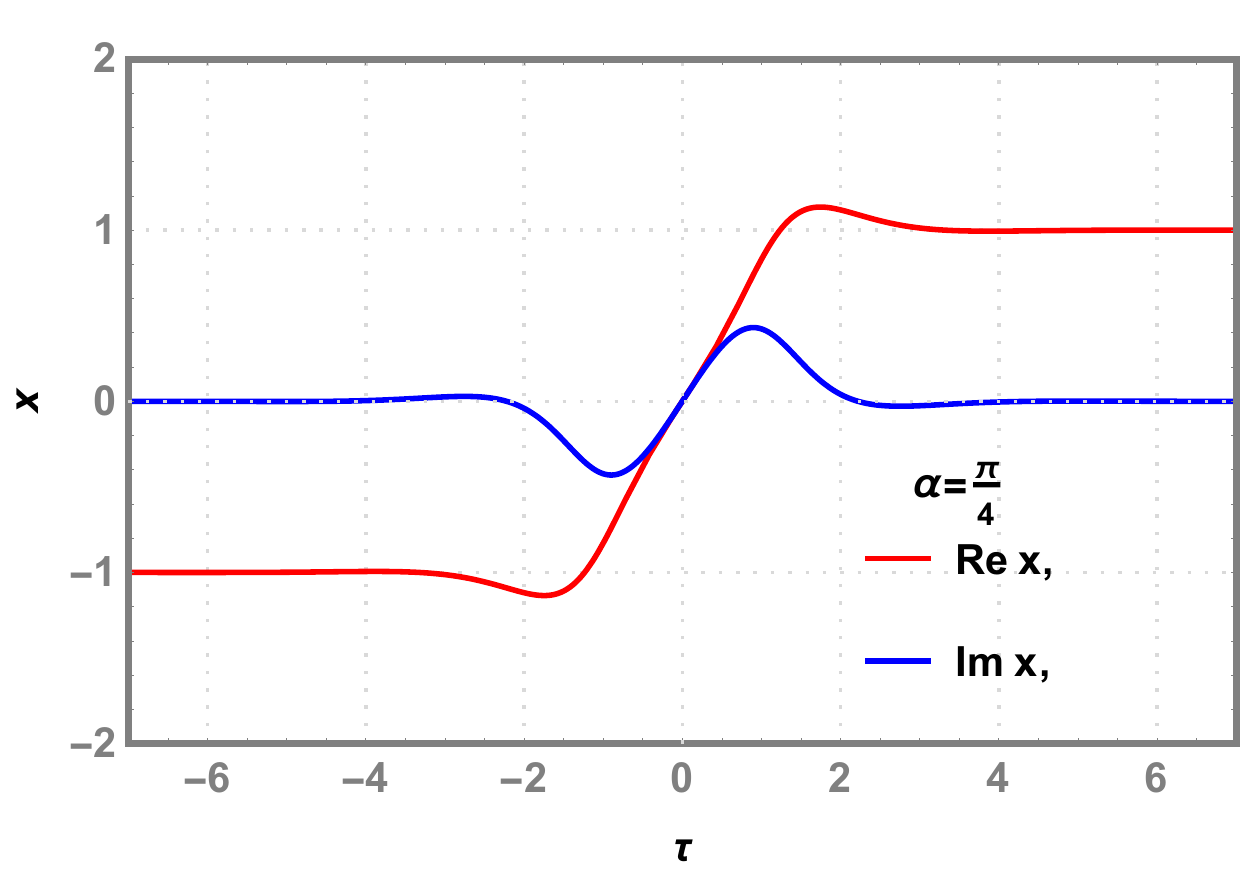} 
\includegraphics[scale=.5]{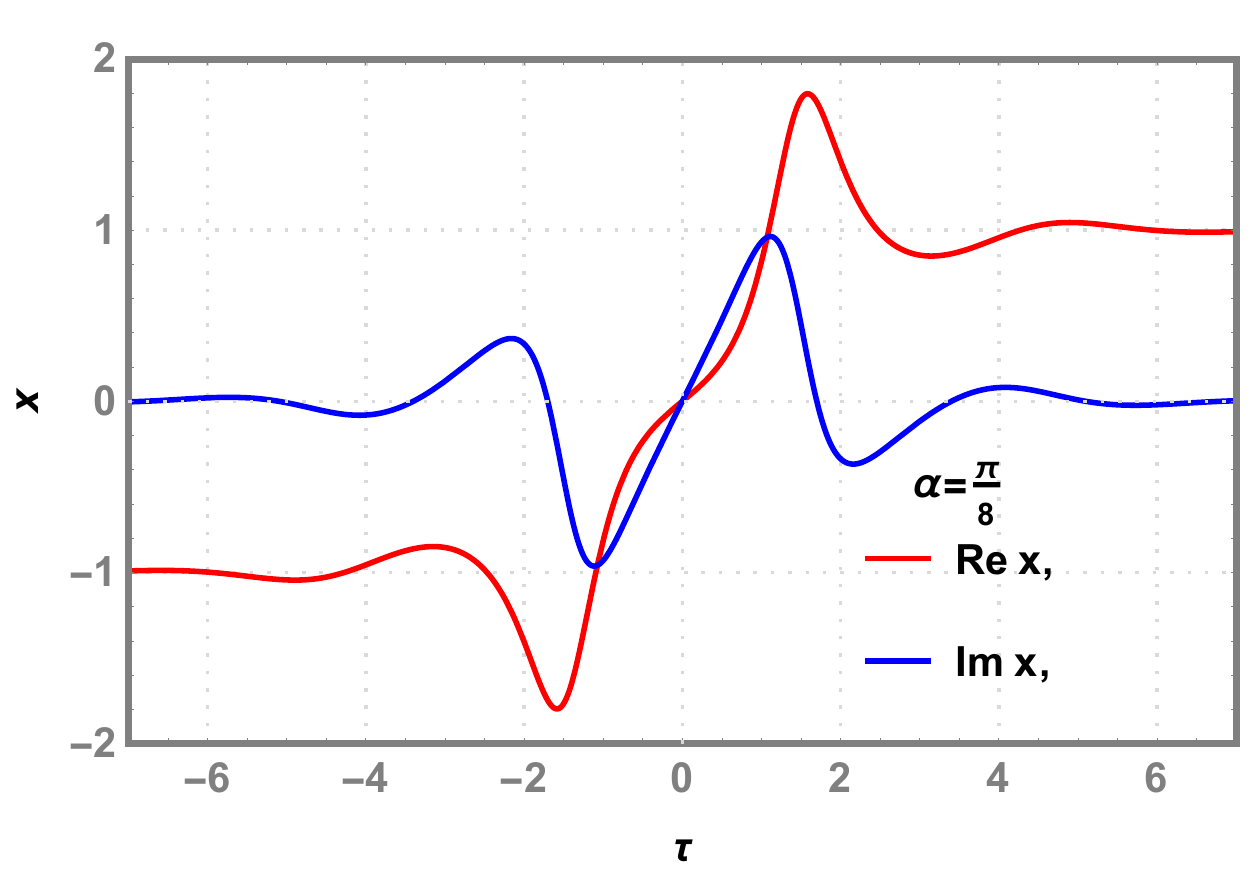}    
\includegraphics[scale=.51]{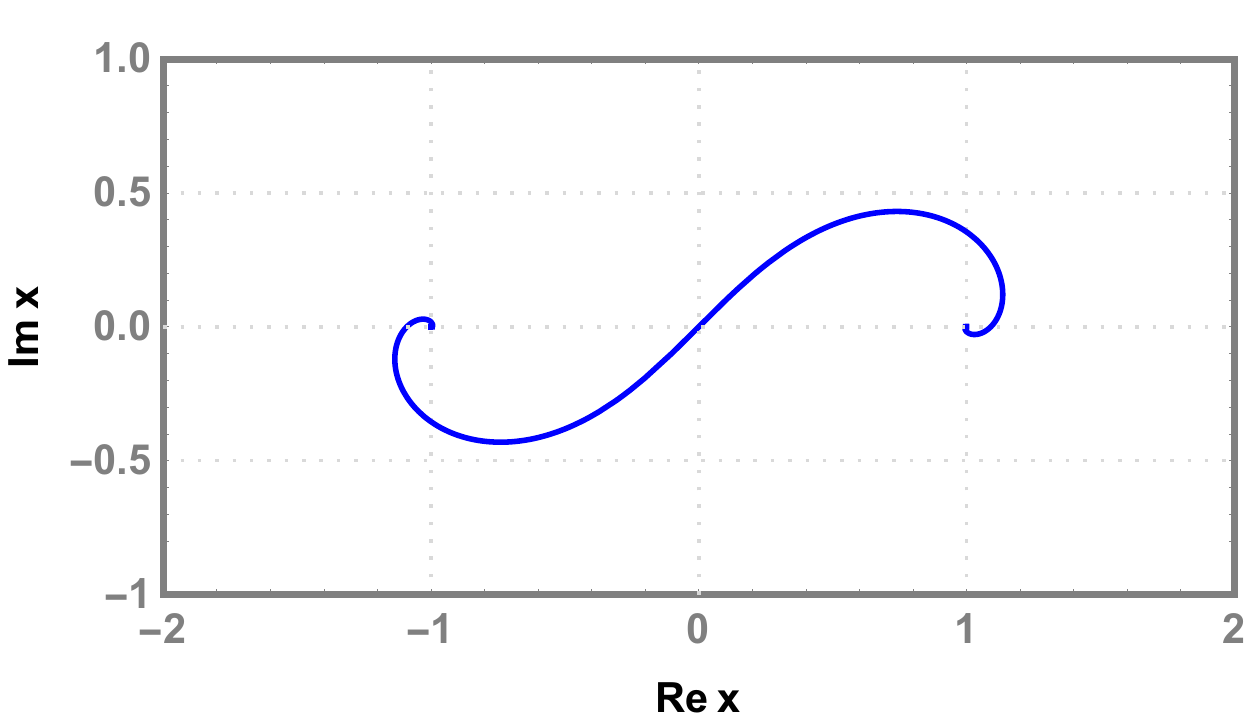} 
\includegraphics[scale=.51]{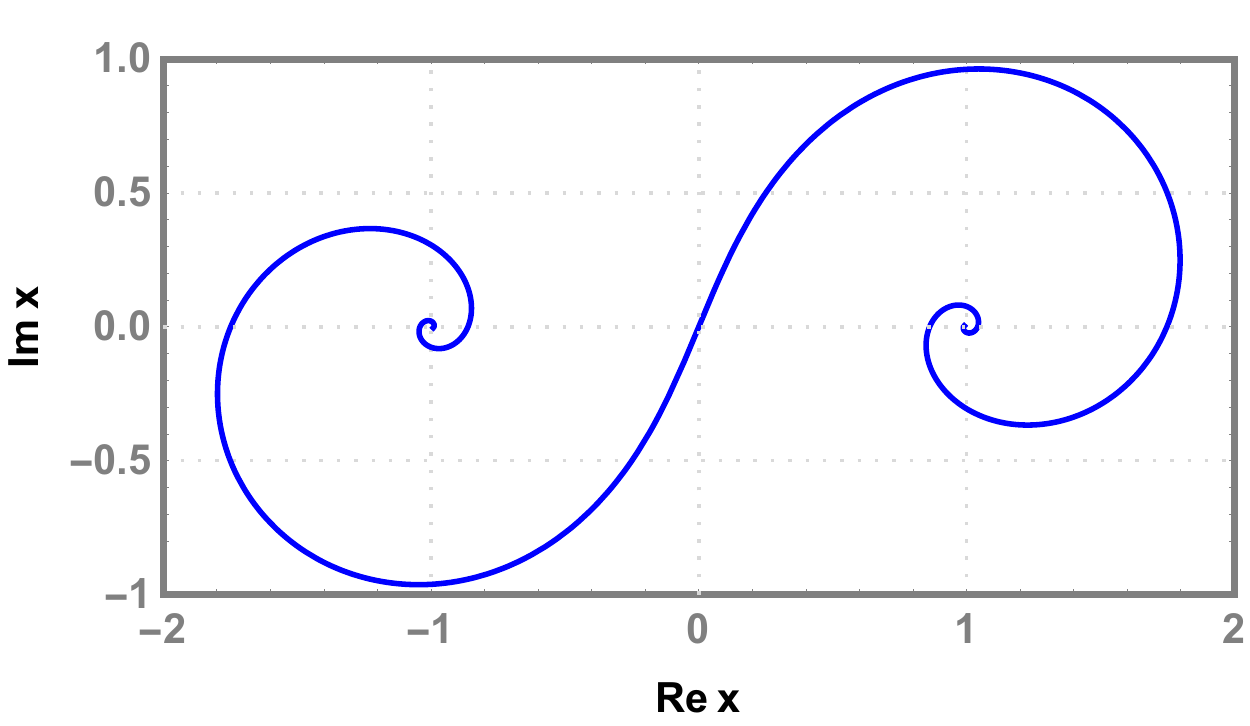}    
\caption{The complexfied instanton configuration for $0< \alpha< \frac{\pi}{2}$.}
\label{F3.7} 
\end{figure} 
\item[(iv)] Consider the situation for infinitesimal $\alpha$ and $\vert \tau \vert \rightarrow \infty$, the behaviour of the complexfied instanton in this situation is shown in Fig. (\ref{F3.8}). If the instanton staring at $x=-a$ at $\tau=-\infty$, it is supposed to spend more and more time wiping around in the $\operatorname{Re} x < 0$ region in a huge number of more and more tightly space increasingly large arcs for $\tau < \tau_0$. After a while, the instanton jumps over to the $\operatorname{Re} x > 0$ region and winds closer and closer to $x=a$ in the same way, until it reaches $x=a$ and remains there  at $\tau=\infty$.
 \begin{figure} [h!]
\centering
\includegraphics[scale=.5]{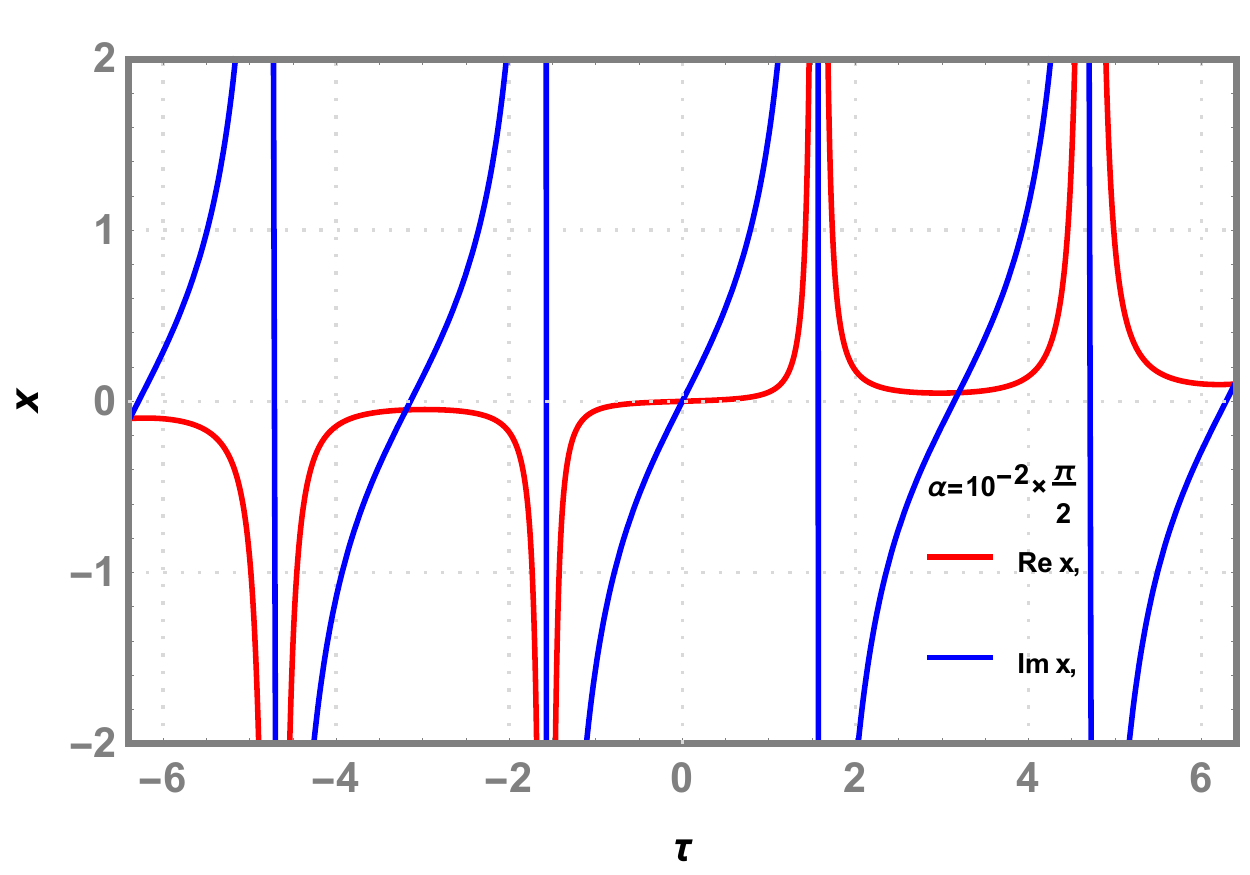} 
\includegraphics[scale=.5]{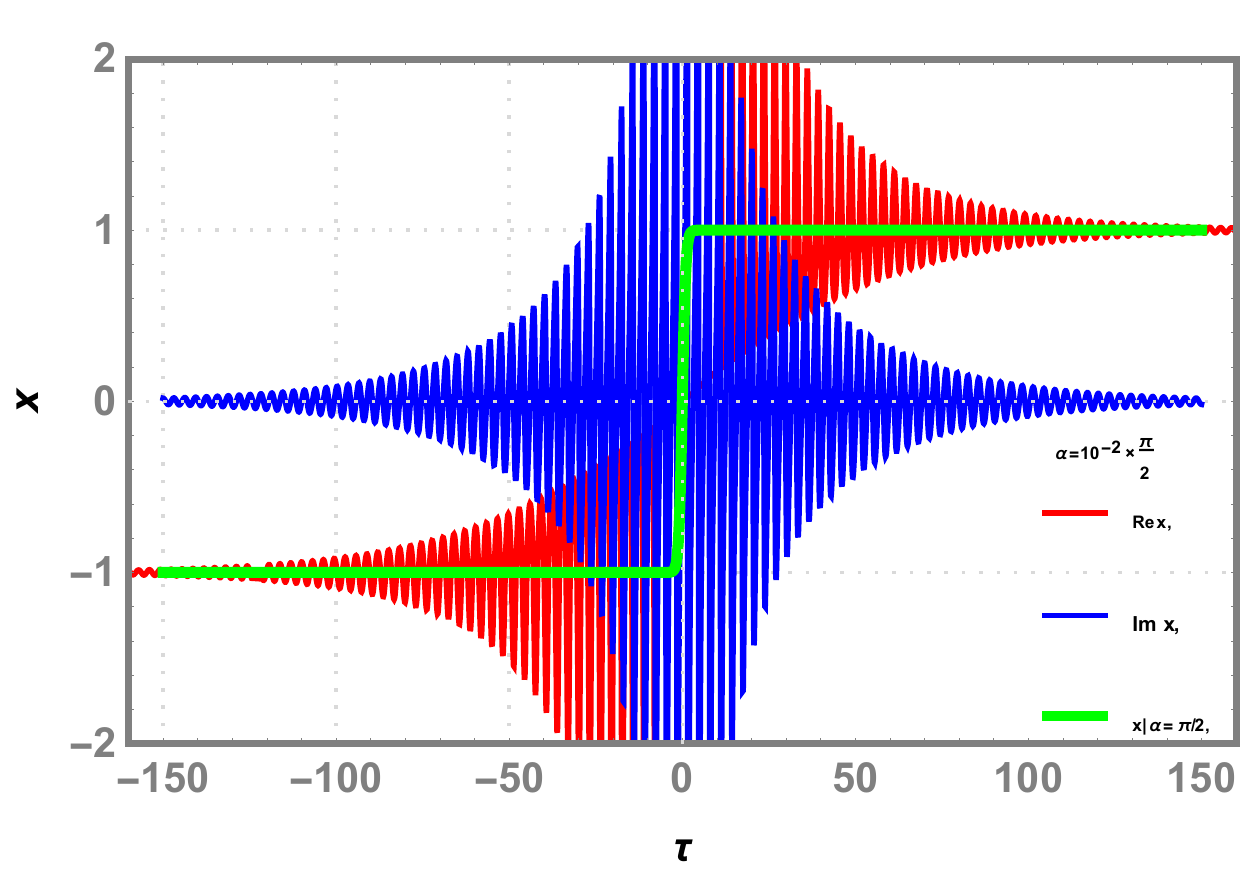}    
\includegraphics[scale=.5]{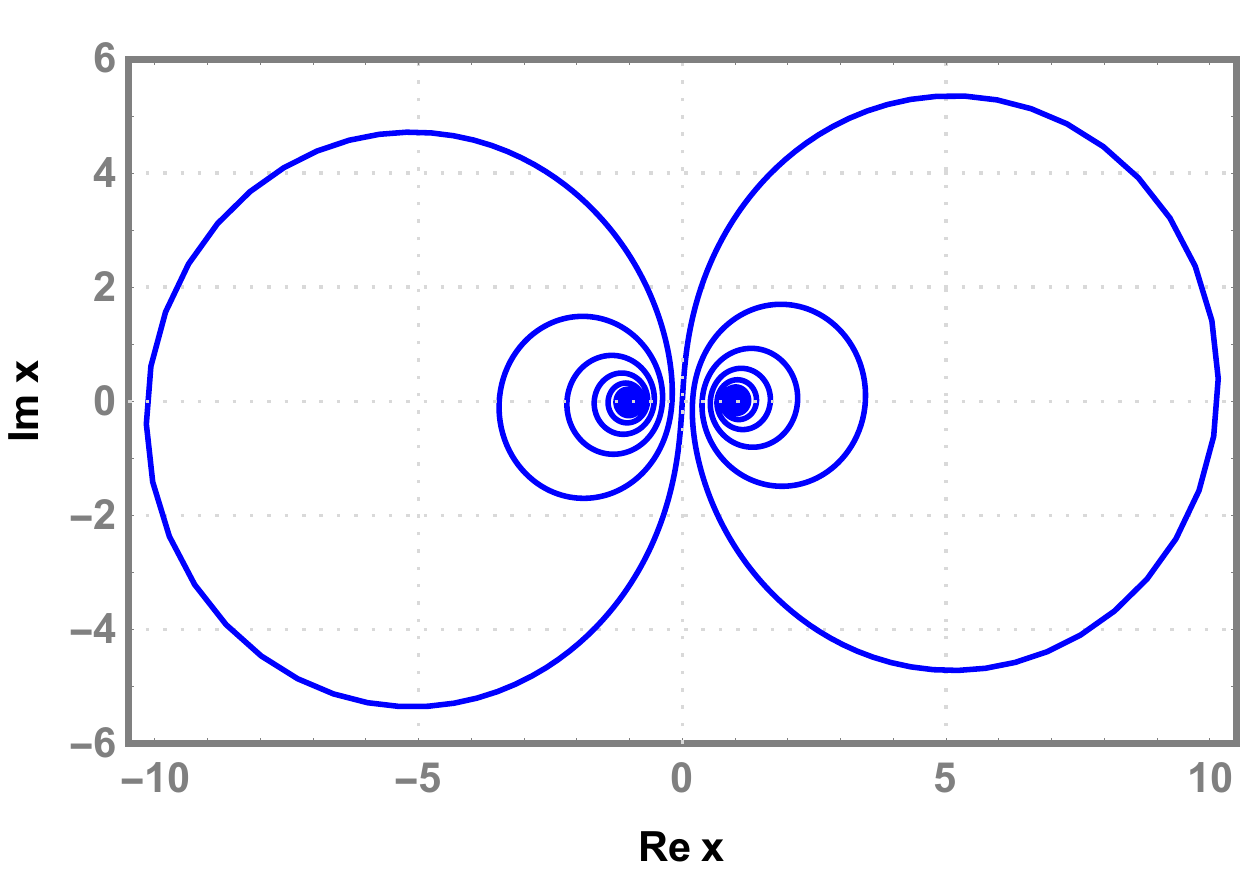} 
\includegraphics[scale=.5]{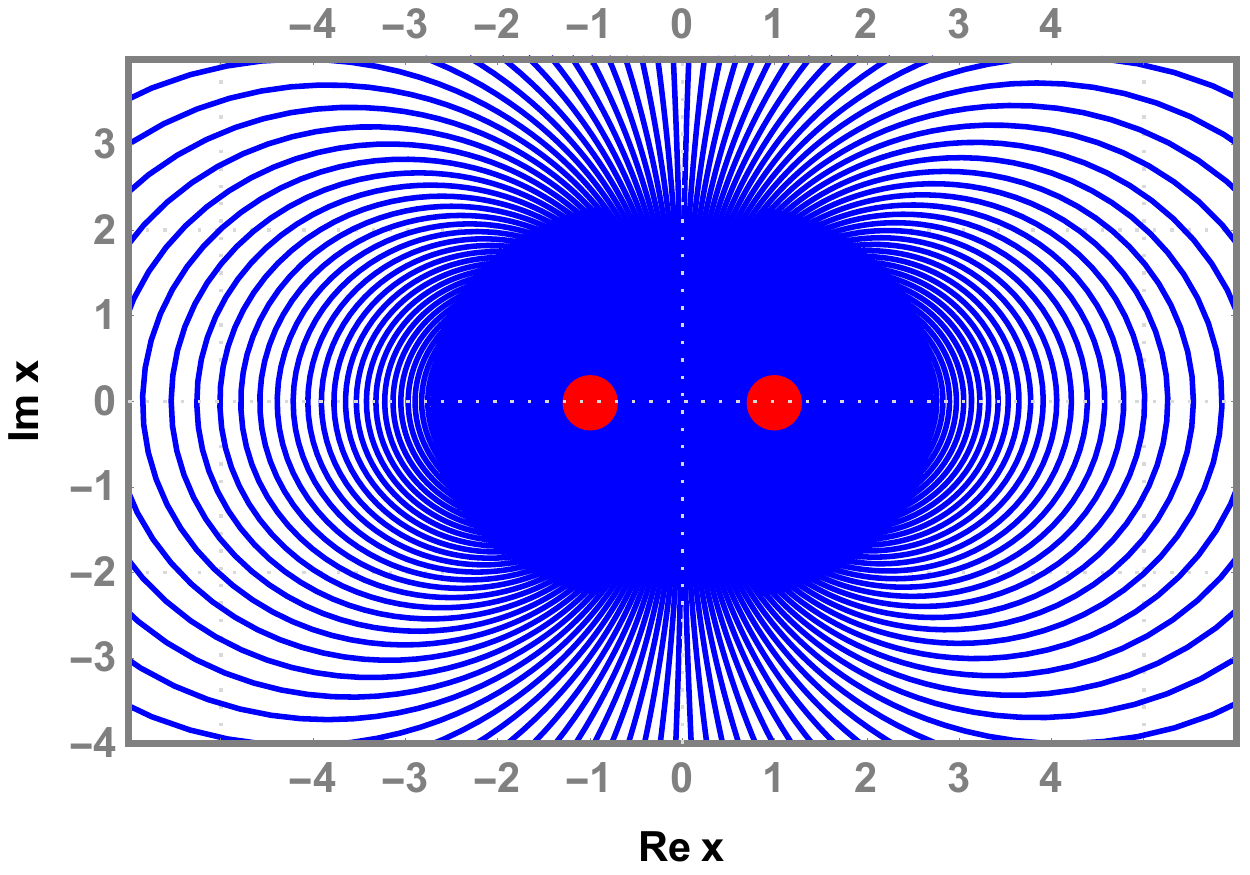}    
\caption{The complexfied instanton configuration for infinitesimal $\alpha$ and $\vert \tau \vert \rightarrow \infty $.}
\label{F3.8} 
\end{figure} 
\end{itemize}   

}
\par{

According to the previous discussion, it seems that the only reasonable description of real-time tunneling is the use of highly oscillatory complex solutions. We see that the tunneling amplitude can be given by a real-time instanton, which requires a complexification of the space of paths and the addition of an infinitesimal imaginary part to the time coordinate as a regulator for its definition as a smooth configuration. In the real-time limit, the instanton trajectory takes a wild ride through complex configuration space on its way between the minima of the potential, with the path resembling a space-filling curve.  Thus,  it is demonstrated that quantum tunneling is possible to be described based on a semi-classical treatment of the real-time Feynman path integral formalism.
} 
\chapter{\textbf{Discussion and Conclusion}} \label{ch4}

\par{In this dissertation, we reviewed the basic aspects of supersymmetric quantum mechanics. We started with introducing the algebra of Grassmann variables, we then looked into quantum mechanics of the supersymmetric harmonic oscillator, which includes fermionic as well as bosonic fields. Afterward, we investigated the algebraic structure of supersymmetric quantum mechanics. We started by investigating the superalgebra using Dirac brackets. Then, we introduced the concept of the supercharge operators $\hat{\mathit{Q}}$ and $\hat{\mathit{Q}}^\dagger$. In general, the supersymmetry is constructed by introducing supersymmetric transformations which are generated by the supercharge operators, where the role of the supercharges is to change the bosonic state into the fermionic state and vice versa, while the Lagrangian remains invariant. Moreover, we have presented the basic properties of supersymmetric quantum mechanics.}
 
\par{Furthermore, we illustrated for a supersymmetric quantum mechanical system that, the energy spectrum is degenerate except for the ground state, which must have a zero eigenvalue in order for the system to have good supersymmetry. Also, we have explained that, if there is a supersymmetric state, it is the zero-energy ground state. If such a zero-energy ground state exists, it is said that the supersymmetry is unbroken. So far there has been no unbroken supersymmetry observed in nature, and if nature is described by supersymmetry, of course, it must be broken.}

\par{In fact, supersymmetry may be broken spontaneously at any order of perturbation theory, or dynamically due to non-perturbative effects. To examine this statement, we studied the normalization of the ground state of the supersymmetric harmonic oscillator. Then we used perturbation theory to calculate the corrections to the ground state energy. We found that the perturbation does not affect the energy of the ground state at second order in perturbation theory, but this result can be expanded to any finite order. This means that the supersymmetry breaking is not seen in perturbation theory, and it must be due to the non-perturbative effects.}    

\par{We also briefly reviewed the main aspects of the path integral formulation of quantum mechanics. In order to see how this method work, the simple example of quantum mechanics of the harmonic oscillator was discussed.}

\par{The most interesting aspect of this dissertation was introducing the Euclidean-time instantons and using this method to study the quantum mechanics of the symmetric double-well potential. However, the more interesting part was the calculation of the determinant for the problem of the double-well potential. Actually, we have explained this point in detail. Moreover, we used the instantons method to calculate the transition amplitude between the two minima of the double-well potential. In addition, we  successfully calculated the corrections to the energy of the ground state and the first excited state.} 

\par{In quantum mechanics, the quantum tunneling restores the symmetry and solves the degeneracy between the classical minima of the potential, so it is a non-perturbative phenomena. We have investigated this fact by studying the tunneling using the description of the real-time part of the highly oscillatory complex solution of instanton.}
 
\par{ As a result, we can see that the perturbation technique gives us incorrect results for both the wave function as well as the energy spectrum and fails to give an explanation to the supersymmetry breaking. However, the instantons, which are a non-perturbative effect in quantum mechanics and can not be seen in perturbation theory, leads us  to calculate the corrections to the ground state energy and gives us a possible explanation for the supersymmetry breaking.}

\par{Finally, this dissertation went through simple and basic but important aspects of our two main topics, supersymmetric quantum mechanics and path integral realization of instantons. By providing calculations and proofs that may allow someone to further understand the basic properties of both supersymmetry and path integrals. That helps the simplification of calculations needed by anyone new in these two fields.}
\chapter*{\textbf{Appendix A}}
\addcontentsline{toc}{chapter}{\textbf{Appendix A}}
\begin{appendices}
\renewcommand{\thechapter}{\arabic{chapter}}
\renewcommand{\thesection}{A.\arabic{section}}
\renewcommand{\theequation}{A.\arabic{equation}}
\setcounter{equation}{0}

\section{The Generalised Hamilton's Equations of Motion} \label{AppendixA.1}
\par{
Recall the Hamiltonian (\ref{a2.31})  
\begin{equation} \label{A.1}
\mathit{H} = \frac{1}{2} \left( p^2 + \omega^2 q^2 \right) - i \omega \psi_1 \psi_2.
\end{equation}
We showed in subsection \ref{SubSec.2.3.3}, that the previous Hamiltonian could be written in the form (\ref{a2.36})
\begin{equation} \label{A.2}
\mathit{H} = \frac{1}{2} \left( p^2 + \omega^2 q^2 \right) +4 i \omega \pi^1 \pi^2.
\end{equation}
Also consider the momenta $\pi^{\alpha}$, and the primary constraint $\phi^{\alpha}$,  respectively define as
\begin{align} \label{A.3}
\pi^{\alpha} &= \dfrac{\partial \mathit{L}}{\partial \dot{\psi}_{\alpha}} = - \frac{i}{2} \delta^{\alpha \beta} \psi_{\beta}, \nonumber \\
\phi^{\alpha} &= \pi^{\alpha} + \frac{i}{2} \delta^{\alpha \beta} \psi_{\beta}.
\end{align}
Downward we explain how to use Poisson bracket to get the generalised Hamilton's equations of motion, which we have only write it down in Eq. (\ref{b2.44}).}

\begin{itemize}
\item[(i)]
\par{ 
The extended Hamiltonian (\ref{a2.41}) is written as
\begin{equation}
\mathit{H} = \dot{q}_i p^i + \dot{\psi}_{\alpha} \pi^{\alpha} - \phi^A \lambda_A - \mathit{L}.
\end{equation}
Therefore
\begin{equation}
\dfrac{\partial \mathrm{H}}{\partial p^i} = \dot{q}_i - \dfrac{\partial \phi^A}{\partial p^i} \lambda_A.
\end{equation}
Thus
\begin{equation} \label{A.6}
\dot{q}_i = \dfrac{\partial \mathrm{H}}{\partial p}  + \dfrac{\partial \phi^A}{\partial p} \lambda_A.
\end{equation}
Using Poisson brackets. we find that
\begin{align} \label{A.7}
\{ q_i, \mathit{H}+ \phi^A \lambda_A \}_P & = \left( \dfrac{\partial q_i}{ \partial q_i} \dfrac{ \partial (\mathit{H} + \phi^A \lambda_A)}{ \partial p^i} - \dfrac{ \partial (\mathit{H} + \phi^A \lambda_A)}{ \partial q_i} \dfrac{\partial q_i}{\partial p^i} \right) \nonumber \\ 
& \quad - \left( \dfrac{\partial q_i}{\partial \psi_\alpha} \dfrac{\partial (\mathit{H} + \phi^A \lambda_A)}{\partial \pi^\alpha} - \dfrac{\partial (\mathit{H} + \phi^A \lambda_A)}{\partial \psi_\alpha} \dfrac{\partial q_i}{\partial \pi^\alpha} \right) \nonumber \\ 
& = \dfrac{\partial \mathit{H}}{\partial p^i} + \dfrac{\partial \pi^A}{\partial p^i} \lambda_A. 
\end{align} 
From Eqs. (\ref{A.6}, \ref{A.7}), we obtain
\begin{equation}
\dot{q}_i = \dfrac{\partial \mathrm{H}}{\partial p}  + \dfrac{\partial \phi^A}{\partial p} \lambda_A = \{ q_i, \mathrm{H}+ \phi^A \lambda_A \}_P.
\end{equation}

}
\item[(ii)]
\par{
\begin{align}
\dfrac{\partial \mathit{H}}{\partial q_i} & = - \dfrac{\partial \mathit{L}}{\partial q_i}- \dfrac{ \partial \phi^A}{\partial q_i} \lambda_A  
= - \dfrac{d}{dt} \dfrac{\partial \mathit{L}}{\partial \dot{q_i}}- \dfrac{ \partial \phi^A}{\partial q_i} \lambda_A \nonumber \\
& = - \dot{p}^i- \dfrac{ \partial \phi^A}{\partial q_i} \lambda_A. 
\end{align}
Thus
\begin{equation}\label{A.10}
\dot{p}^i = - \dfrac{\partial \mathit{H}}{\partial q_i} - \dfrac{ \partial \phi^A}{\partial q_i} \lambda_A.
\end{equation}
Using Poisson brackets. we find that
\begin{align} \label{A.11}
\{ p^i, \mathit{H}+ \phi^A \lambda_A \}_P &= \left( \dfrac{\partial p^i}{ \partial q_i} \dfrac{ \partial (\mathit{H} + \phi^A \lambda_A)}{ \partial p^i} - \dfrac{ \partial (\mathit{H} + \phi^A \lambda_A)}{ \partial q_i} \dfrac{\partial p^i}{\partial p^i} \right) \nonumber \\ 
& \quad - \left( \dfrac{\partial p^i}{\partial \psi_\alpha} \dfrac{\partial (\mathrm{H} + \phi^A \lambda_A)}{\partial \pi^\alpha} - \dfrac{\partial (\mathit{H} + \phi^A \lambda_A)}{\partial \psi_\alpha} \dfrac{\partial p^i}{\partial \pi^\alpha} \right) \nonumber \\
& = - \dfrac{\partial \mathrm{H}}{\partial q_i} - \dfrac{\partial \phi^A}{\partial q_i}.
\end{align}
From Eqs. (\ref{A.10}, \ref{A.11}), we obtain
\begin{equation}
\dot{p}^i = - \dfrac{\partial \mathit{H}}{\partial q_i} - \dfrac{ \partial \phi^A}{\partial q_i} \lambda_A = \{ p^i, \mathit{H}+ \phi^A \lambda_A \}_P.
\end{equation}

}
\item[(iii)]
\par{
\begin{align}
\dfrac{\partial \mathit{H}}{\partial \pi^\alpha} & = - \dot{\psi_\alpha}- \dfrac{ \partial \phi^A}{\partial q\pi^\alpha} \lambda_A. 
\end{align}
Thus
\begin{equation} \label{A.14}
\dot{\psi_\alpha} = - \dfrac{\partial \mathit{H}}{\partial \pi^\alpha} - \dfrac{ \partial \phi^A}{\partial q\pi^\alpha} \lambda_A.
\end{equation}
Using Poisson brackets, we find that
\begin{align} \label{A.15}
\{ \psi_\alpha, \mathit{H}+ \phi^A \lambda_A \}_P & = \left( \dfrac{\partial \psi_\alpha}{ \partial q_i} \dfrac{ \partial (\mathit{H} + \phi^A \lambda_A)}{ \partial p^i} - \dfrac{ \partial (\mathit{H} + \phi^A \lambda_A)}{ \partial q_i} \dfrac{\partial p^i}{\partial \psi_\alpha} \right)  \nonumber \\ 
& \quad - \left( \dfrac{\partial \psi_\alpha}{\partial \psi_\alpha} \dfrac{\partial (\mathit{H} + \phi^A \lambda_A)}{\partial \pi^\alpha} - \dfrac{\partial (\mathit{H} + \phi^A \lambda_A)}{\partial \psi_\alpha} \dfrac{\partial \psi_\alpha }{\partial \pi^\alpha} \right)  \nonumber \\
& = - \dfrac{\partial \mathit{H}}{\partial \pi^\alpha} - \dfrac{\partial \phi^A}{\partial \pi^\alpha }.
\end{align}
From Eqs. (\ref{A.14}, \ref{A.15}), we obtain
\begin{equation}
\dot{\psi_\alpha} = - \dfrac{\partial \mathit{H}}{\partial \pi^\alpha} - \dfrac{ \partial \phi^A}{\partial q\pi^\alpha} \lambda_A = \{ \psi_\alpha, \mathit{H}+ \phi^A \lambda_A \}_P.
\end{equation}

}
\item[(iv)]
\par{
\begin{align}
\dfrac{\partial \mathit{H}}{\partial \psi_\alpha} & = - \dfrac{\partial \mathit{L}}{\partial \psi_\alpha}- \dfrac{ \partial \phi^A}{\partial q \psi_\alpha} \lambda_A \nonumber \\
&= - \dfrac{d}{dt} \left(\dfrac{\partial \mathit{L}}{\partial \dot{\psi}_\alpha}\right) - \dfrac{ \partial \phi^A}{\partial \psi_\alpha} \lambda_A \nonumber \\
& = - \dot{\pi}^\alpha - \dfrac{ \partial \phi^A}{\partial \psi_\alpha} \lambda_A.
\end{align}
Thus
\begin{equation} \label{A.18}
\dot{\pi}^{\alpha} = - \dfrac{\partial \mathit{H}}{\partial \psi_{\alpha}} - \dfrac{ \partial \phi^A}{\partial \psi_\alpha} \lambda_A.
\end{equation}
Using Poisson brackets, we obtain
\begin{align} \label{A.19}
\{ \pi_\alpha, \mathit{H}+ \phi^A \lambda_A \}_P &= \left( \dfrac{\partial \pi_\alpha}{ \partial q_i} \dfrac{ \partial (\mathit{H} + \phi^A \lambda_A)}{ \partial p^i} - \dfrac{ \partial (\mathit{H} + \phi^A \lambda_A)}{ \partial q_i} \dfrac{\partial p^i}{\partial \pi_\alpha} \right) \nonumber \\ 
& \quad - \left( \dfrac{\partial \pi_\alpha}{\partial \psi_\alpha} \dfrac{\partial (\mathit{H} + \phi^A \lambda_A)}{\partial \pi^\alpha} - \dfrac{\partial (\mathit{H} + \phi^A \lambda_A)}{\partial \psi_\alpha} \dfrac{\partial \pi_\alpha }{\partial \pi^\alpha} \right) \nonumber \\
&=  - \dot{\pi}^\alpha - \dfrac{ \partial \phi^A}{\partial \psi_\alpha} \lambda_A.
\end{align}
From Eqs. (\ref{A.18}, \ref{A.19}), we obtain
\begin{equation}
\dot{\pi}^{\alpha} = - \dfrac{\partial \mathit{H}}{\partial \psi_{\alpha}} - \dfrac{ \partial \phi^A}{\partial \psi_\alpha} \lambda_A = \{ \pi_\alpha, \mathit{H}+ \phi^A \lambda_A \}_P. 
\end{equation}

}
\item[(v)]
\par{
By definition from Eqs. (\ref{A.3})
\begin{align}
\phi^A &= \pi^A + \frac{i}{2} \psi_A \nonumber \\
& = - \frac{i}{2} \psi_A + \frac{i}{2} \psi_A \nonumber \\
&= 0.
\end{align}

}
\end{itemize}

\newpage

\section{Dirac Bracket and the Superalgebra} \label{AppendixA.2}
\par{
From Appendix (\ref{AppendixA.1}) recall the expressions of the Hamiltonian (\ref{A.1}) and the extended Hamiltonian (\ref{A.2})
\begin{equation} \label{A.22}
\mathit{H} = \frac{1}{2} \left( p^2 + \omega^2 q^2 \right) - i \omega \psi_1 \psi_2.
\end{equation}
As well take into account  the  for the constraint (\ref{A.3}) and suppose $\{ \alpha ,\beta\}$ can only take the values $\{1,2\}$, then we have
\begin{align} \label{A.23}
\phi^1 &= \pi^1 + \frac{i}{2} \psi_1, \nonumber \\
\phi^2 &= \pi^2 + \frac{i}{2} \psi_2. 
\end{align}  
Furthermore, consider the supercharge formula, and once again suppose $\{ \alpha ,\beta\}$ can only take the values $\{1,2\}$, then we have 
\begin{align}\label{A.24}
\mathit{Q}_\alpha  = p \psi_\alpha + \omega q \epsilon_{\alpha \beta} \delta^{\beta \gamma} \psi_\gamma \quad \Rightarrow \quad 
\begin{split} 
\mathit{Q}_1  &= p \psi_1 + \omega q \psi_2  \\
\mathit{Q}_2  &= p \psi_2 - \omega q \psi_1,
\end{split}
\end{align}
where, $\epsilon_{\alpha \beta}$ and $\delta^{\alpha \beta}$ are the Levi-Civita symbol and Kronecker delta function, respectively.}
\par{
To verify that the supercharges (\ref{A.24}) together with the Hamiltonian (\ref{A.22}) and the constraints (\ref{A.23}) satisfy the Dirac bracket super-algebra (\ref{aa2.73}), we need as a first step to calculate the Dirac brackets of the supercharges
\begin{align} \label{A.25}
\{Q_\alpha, Q_\beta \}_D & = \{Q_1, Q_1\}_D + \{Q_1, Q_2 \}_D + \{Q_2, Q_1\}_D + \{Q_2, Q_2 \}_D 
\end{align}
To simplify the calculation of Eq. (\ref{A.25}), let us start with calculating the Poisson bracket of the super-charges and the constraints.}
\par{
\begin{align} \label{A.26}
\{Q_1,Q_1\}_P &= \left( \frac{\partial Q_1 }{\partial q_i} \frac{\partial Q_1 }{\partial p^i} - \frac{\partial Q_1 }{\partial p^i} \frac{\partial Q_1}{\partial q_i} \right) \nonumber \\
& \quad - \left( \frac{\partial Q_1}{\partial \psi_1} \frac{\partial Q_1}{\partial \pi^1} + \frac{\partial Q_1}{\partial \pi^1} \frac{\partial Q_1}{\partial \psi_1} \right)
- \left( \frac{\partial Q_1}{\partial \psi_2} \frac{\partial Q_1}{\partial \pi^2} + \frac{\partial Q_1}{\partial \pi^2} \frac{\partial Q_1}{\partial \psi_2} \right) \nonumber \\
& = 0.
\end{align}
\begin{align}
\{Q_1,Q_2\}_P & = \left( \frac{\partial Q_1 }{\partial q_i} \frac{\partial Q_2 }{\partial p^i} - \frac{\partial Q_1 }{\partial p^i} \frac{\partial Q_2}{\partial q_i} \right) \nonumber \\
& \quad - \left( \frac{\partial Q_1}{\partial \psi_1} \frac{\partial Q_2}{\partial \pi^1} + \frac{\partial Q_1}{\partial \pi^1} \frac{\partial Q_2}{\partial \psi_1} \right)
- \left( \frac{\partial Q_1}{\partial \psi_2} \frac{\partial Q_2}{\partial \pi^2} + \frac{\partial Q_1}{\partial \pi^2} \frac{\partial Q_2}{\partial \psi_2} \right) \nonumber \\
& =  \omega \left( \psi_1^2 + \psi_2^2 \right).
\end{align}
\begin{align}
\{Q_2,Q_1\}_P &= \left( \frac{\partial Q_2 }{\partial q_i} \frac{\partial Q_1 }{\partial p^i} - \frac{\partial Q_2 }{\partial p^i} \frac{\partial Q_1}{\partial q_i} \right) \nonumber \\
& \quad - \left( \frac{\partial Q_2}{\partial \psi_1} \frac{\partial Q_1}{\partial \pi^1} + \frac{\partial Q_2}{\partial \pi^1} \frac{\partial Q_1}{\partial \psi_1} \right)
- \left( \frac{\partial Q_2}{\partial \psi_2} \frac{\partial Q_1}{\partial \pi^2} + \frac{\partial Q_2}{\partial \pi^2} \frac{\partial Q_1}{\partial \psi_2} \right) \nonumber \\
& = - \omega \left(\psi_1^2 + \psi_2^2 \right).
\end{align}
\begin{align}
\{Q_2,Q_2\}_P &= \left( \frac{\partial Q_2 }{\partial q_i} \frac{\partial Q_2 }{\partial p^i} - \frac{\partial Q_2 }{\partial p^i} \frac{\partial Q_2}{\partial q_i} \right) \nonumber \\
& \quad - \left( \frac{\partial Q_2}{\partial \psi_1} \frac{\partial Q_2}{\partial \pi^1} + \frac{\partial Q_2}{\partial \pi^1} \frac{\partial Q_2}{\partial \psi_1} \right)
- \left( \frac{\partial Q_2}{\partial \psi_2} \frac{\partial Q_2}{\partial \pi^2} + \frac{\partial Q_2}{\partial \pi^2} \frac{\partial Q_2}{\partial \psi_2} \right) \nonumber \\
&= 0.
\end{align}
\begin{align}
\{Q_1, \phi^1\}_P &= \left( \frac{\partial Q_1 }{\partial q_i} \frac{\partial \phi^1 }{\partial p^i} - \frac{\partial Q_1 }{\partial p^i} \frac{\partial \phi^1}{\partial q_i} \right) \nonumber \\
& \quad - \left( \frac{\partial Q_1}{\partial \psi_1} \frac{\partial \phi^1}{\partial \pi^1} + \frac{\partial Q_1}{\partial \pi^1} \frac{\partial \phi^1}{\partial \psi_1} \right)
- \left( \frac{\partial Q_1}{\partial \psi_2} \frac{\partial \phi^1}{\partial \pi^2} + \frac{\partial Q_1}{\partial \pi^2} \frac{\partial \phi^1}{\partial \psi_2} \right) \nonumber \\
&= - p.
\end{align}
\begin{align}
\{\phi^1, Q_1\}_P &= \left( \frac{\partial \phi^1 }{\partial q_i} \frac{\partial Q_1 }{\partial p^i} - \frac{\partial \phi^1 }{\partial p^i} \frac{\partial Q_1}{\partial q_i} \right) \nonumber \\
& \quad - \left( \frac{\partial \phi^1}{\partial \psi_1} \frac{\partial Q_1}{\partial \pi^1} + \frac{\partial \phi^1}{\partial \pi^1} \frac{\partial Q_1}{\partial \psi_1} \right)
- \left( \frac{\partial \phi^1}{\partial \psi_2} \frac{\partial Q_1}{\partial \pi^2} + \frac{\partial \phi^1}{\partial \pi^2} \frac{\partial Q_1}{\partial \psi_2} \right) \nonumber \\
&= -p.
\end{align}
\begin{align}
\{Q_1, \phi^2\}_P & = \left( \frac{\partial Q_1 }{\partial q_i} \frac{\partial \phi^2 }{\partial p^i} - \frac{\partial Q_1 }{\partial p^i} \frac{\partial \phi^2}{\partial q_i} \right) \nonumber \\
& \quad - \left( \frac{\partial Q_1}{\partial \psi_1} \frac{\partial \phi^2}{\partial \pi^1} + \frac{\partial Q_1}{\partial \pi^1} \frac{\partial \phi^2}{\partial \psi_1} \right)
- \left( \frac{\partial Q_1}{\partial \psi_2} \frac{\partial \phi^2}{\partial \pi^2} + \frac{\partial Q_1}{\partial \pi^2} \frac{\partial \phi^2}{\partial \psi_2} \right) \nonumber \\
&= - \omega q.
\end{align}
\begin{align}
\{\phi^2 ,Q_1 \}_P &= \left( \frac{\partial \phi^2 }{\partial q_i} \frac{\partial Q_1 }{\partial p^i} - \frac{\partial \phi^2 }{\partial p^i} \frac{\partial Q_1}{\partial q_i} \right) \nonumber \\
& \quad - \left( \frac{\partial \phi^1}{\partial \psi_2} \frac{\partial Q_1}{\partial \pi^1} + \frac{\partial \phi^2}{\partial \pi^1} \frac{\partial Q_1}{\partial \psi_1} \right)
- \left( \frac{\partial \phi^2}{\partial \psi_2} \frac{\partial Q_1}{\partial \pi^2} + \frac{\partial \phi^2}{\partial \pi^2} \frac{\partial Q_1}{\partial \psi_2} \right) \nonumber \\
&= - \omega q.
\end{align}
\begin{align}
\{Q_2, \phi^1 \}_P &= \left( \frac{\partial Q_2 }{\partial q_i} \frac{\partial \phi^1 }{\partial p^i} - \frac{\partial Q_2 }{\partial p^i} \frac{\partial \phi^1}{\partial q_i} \right) \nonumber \\
& \quad - \left( \frac{\partial Q_2}{\partial \psi_1} \frac{\partial \phi^1}{\partial \pi^1} + \frac{\partial Q_2}{\partial \pi^1} \frac{\partial \phi^1}{\partial \psi_1} \right)
- \left( \frac{\partial Q_2}{\partial \psi_2} \frac{\partial \phi^1}{\partial \pi^2} + \frac{\partial Q_2}{\partial \pi^2} \frac{\partial \phi^1}{\partial \psi_2} \right) \nonumber \\
&=  \omega q.
\end{align}
\begin{align}
\{\phi^1,Q_2 \}_P &= \left( \frac{\partial \phi^1 }{\partial q_i} \frac{\partial Q_2 }{\partial p^i} - \frac{\partial \phi^1 }{\partial p^i} \frac{\partial Q_2}{\partial q_i} \right) \nonumber \\
& \quad - \left( \frac{\partial \phi^1}{\partial \psi_1} \frac{\partial Q_2}{\partial \pi^1} + \frac{\partial \phi^1}{\partial \pi^1} \frac{\partial Q_2}{\partial \psi_1} \right)
- \left( \frac{\partial \phi^1}{\partial \psi_2} \frac{\partial Q_2}{\partial \pi^2} + \frac{\partial \phi^1}{\partial \pi^2} \frac{\partial Q_2}{\partial \psi_2} \right) \nonumber \\
&=  \omega q.
\end{align}
\begin{align}
\{Q_2, \phi^2 \}_P &= \left( \frac{\partial Q_2 }{\partial q_i} \frac{\partial \phi^2 }{\partial p^i} - \frac{\partial Q_2 }{\partial p^i} \frac{\partial \phi^2}{\partial q_i} \right) \nonumber \\
& \quad - \left( \frac{\partial Q_2}{\partial \psi_1} \frac{\partial \phi^2}{\partial \pi^1} + \frac{\partial Q_2}{\partial \pi^1} \frac{\partial \phi^2}{\partial \psi_1} \right)
- \left( \frac{\partial Q_2}{\partial \psi_2} \frac{\partial \phi^2}{\partial \pi^2} + \frac{\partial Q_2}{\partial \pi^2} \frac{\partial \phi^2}{\partial \psi_2} \right) \nonumber \\
&= - p.
\end{align}
\begin{align} \label{A.37}
\{\phi^2,Q_2 \}_P &= \left( \frac{\partial \phi^2 }{\partial q_i} \frac{\partial Q_2 }{\partial p^i} - \frac{\partial \phi^2 }{\partial p^i} \frac{\partial Q_2}{\partial q_i} \right) \nonumber \\
& \quad - \left( \frac{\partial \phi^2}{\partial \psi_1} \frac{\partial Q_2}{\partial \pi^1} + \frac{\partial \phi^2}{\partial \pi^1} \frac{\partial Q_2}{\partial \psi_1} \right)
- \left( \frac{\partial \phi^2}{\partial \psi_2} \frac{\partial Q_2}{\partial \pi^2} + \frac{\partial \phi^2}{\partial \pi^2} \frac{\partial Q_2}{\partial \psi_2} \right) \nonumber \\
&= - p.
\end{align}

}
\par{
Using Dirac bracket definition and Eq.s ( \ref{A.26}-\ref{A.37}), then we get
\begin{align} \label{A.38}
\{Q_1, Q_1 \}_D &= \{Q_1,Q_1\}_P - \{Q_1,\phi^A \}_P C_{AB} \{ \phi^B, Q_1\}_P \nonumber \\
&= \{Q_1,Q_1\}_P - \{Q_1,\phi^1 \}_P C_{11} \{ \phi^1, Q_1\}_P - \{Q_1,\phi^2 \}_P C_{22} \{ \phi^2, Q_1\}_P \nonumber \\
&= - i ( p^2 + \omega^2 q^2). 
\end{align}
\begin{align} \label{A.39}
\{Q_1, Q_2 \}_D &= \{Q_1,Q_2 \}_P - \{Q_1,\phi^A \}_P C_{AB} \{ \phi^B, Q_2 \}_P \nonumber \\
&=  \{Q_1,Q_2\}_P - \{Q_1,\phi^1 \}_P C_{11} \{ \phi^1, Q_2\}_P - \{Q_1,\phi^2 \}_P C_{22} \{ \phi^2, Q_2\}_P \nonumber \\
&= \omega (\psi_1^2 + \psi_2^2).
\end{align}
\begin{align} \label{A.40}
\{Q_2, Q_1 \}_D &= \{Q_2,Q_1\}_P - \{Q_2,\phi^A \}_P C_{AB} \{ \phi^B, Q_1\}_P \nonumber \\
&=  \{Q_2,Q_1\}_P - \{Q_2,\phi^1 \}_P C_{11} \{ \phi^1, Q_1\}_P - \{Q_2,\phi^2 \}_P C_{22} \{ \phi^2, Q_1\}_P  \nonumber \\
&= - \omega (\psi_1^2 + \psi_2^2).
\end{align}
\begin{align} \label{A.41}
\{Q_2, Q_2 \}_D &= \{Q_2,Q_2 \}_P - \{Q_2,\phi^A \}_P C_{AB} \{ \phi^B, Q_2 \}_P \nonumber \\
&= \{Q_2,Q_2\}_P - \{Q_2,\phi^1 \}_P C_{11} \{ \phi^1, Q_2\}_P - \{Q_2,\phi^2 \}_P C_{22} \{ \phi^2, Q_2\}_P \nonumber \\
&= - i ( p^2 + \omega^2 q^2). 
\end{align}

}
\par{
Substitute from Eqs. (\ref{A.38}, \ref{A.39}, \ref{A.40}, \ref{A.41})  into Eq. (\ref{A.25}), thus we find
\begin{align} \label{A.42}
\{Q_\alpha, Q_\beta \}_D &= \{Q_1, Q_1 \}_D + \{Q_1, Q_2 \}_D + \{Q_2, Q_1 \}_D + \{Q_2, Q_2 \}_D \nonumber \\
& = - i ( p^2 + \omega^2 q^2) + \omega (\psi_1^2 + \psi_2^2) - \omega (\psi_1^2 + \psi_2^2) - i ( P^2 + \omega^2 q^2) \nonumber \\
&= - 2i ( P^2 + \omega^2 q^2).
\end{align}

On the other hand
\begin{align} \label{A.43}
-2 i \delta_{\alpha \beta} \mathit{H} &=  -2i \delta_{11} \mathrm{H} - 2i \delta_{22} \mathit{H} \nonumber \\ 
&= - 2i ( p^2 + \omega^2 q^2) - 2i ( P^2 + \omega^2 q^2) \nonumber \\
&= - 2i ( p^2 + \omega^2 q^2). 
\end{align}

Eqs. (\ref{A.42}) and (\ref{A.43}) imply that
\begin{equation} \label{A.44}
\{Q_\alpha, Q_\beta \}_D = -2i \delta_{\alpha \beta} \mathit{H}.
\end{equation}

}
\par{
So far,  as a second step to verify that the supercharges (\ref{A.24}) together with the Hamiltonian (\ref{A.22}) and the constraints (\ref{A.23}) satisfy the Dirac bracket superalgebra (\ref{aa2.73}), we need to calculate the Dirac brackets of the supercharges and the Hamiltonian
\begin{align} \label{A.45}
\{Q_\alpha, \mathit{H} \}_D & = \{Q_1, \mathit{H} \}_D + \{Q_2, \mathit{H} \}_D.
\end{align}
To simplify the calculation of Eq. (\ref{A.45}), let us calculate the Poisson bracket of the supercharges and the Hamiltonian.}
\par{
\begin{align} \label{A.46}
\{Q_1, \mathit{H} \}_P &= \left( \frac{\partial Q_1}{\partial q_i} \frac{\partial \mathit{H}}{\partial p^i} - \frac{\partial Q_1}{\partial p^i} \frac{\partial \mathit{H}}{\partial q_i} \right) 
- \left( \frac{\partial Q_{1}}{\partial \psi_{\alpha}} \frac{\partial \mathit{H}}{\partial \pi^{\alpha}} 
+ \frac{\partial Q_1}{\partial \pi^{\alpha}} \frac{\partial \mathit{H}}{\partial \psi_{\alpha}} \right) \nonumber \\
&=  \omega p \psi_2 - \omega^2 q \psi_1.
\end{align}
\begin{align}\label{A.47}
\{Q_2, \mathit{H} \}_P &= \left( \frac{\partial Q_2}{\partial q_i} \frac{\partial \mathit{H}}{\partial p^i} - \frac{\partial Q_2}{\partial p^i} \frac{\partial \mathit{H}}{\partial q_i} \right) 
- \left( \frac{\partial Q_{2}}{\partial \psi_{\alpha}} \frac{\partial \mathit{H}}{\partial \pi^{\alpha}} 
+ \frac{\partial Q_2}{\partial \pi^{\alpha}} \frac{\partial \mathit{H}}{\partial \psi_{\alpha}} \right) \nonumber \\
&=  - \omega p \psi_1 - \omega^2 q \psi_2.
\end{align}
\begin{align}\label{A.48}
\{\phi_1, \mathit{H} \}_P &= \left( \frac{\partial \phi_1}{\partial q_i} \frac{\partial \mathit{H}}{\partial p^i} - \frac{\partial \phi_1}{\partial p^i} \frac{\partial \mathit{H}}{\partial q_i} \right) 
- \left( \frac{\partial \phi_1}{\partial \psi_{\alpha}} \frac{\partial \mathit{H}}{\partial \pi^{\alpha}} 
+ \frac{\partial \phi_1}{\partial \pi^{\alpha}} \frac{\partial \mathit{H}}{\partial \psi_{\alpha}} \right) \nonumber \\
&=  i \omega \psi_2.
\end{align}
\begin{align} \label{A.49}
\{\phi_2, \mathit{H} \}_P &= \left( \frac{\partial \phi_2}{\partial q_i} \frac{\partial \mathit{H}}{\partial p^i} - \frac{\partial \phi_2}{\partial p^i} \frac{\partial \mathit{H}}{\partial q_i} \right) 
- \left( \frac{\partial \phi_2}{\partial \psi_{\alpha}} \frac{\partial \mathit{H}}{\partial \pi^{\alpha}} 
+ \frac{\partial \phi_2}{\partial \pi^{\alpha}} \frac{\partial \mathit{H}}{\partial \psi_{\alpha}} \right) \nonumber \\
&=  -i \omega \psi_1.
\end{align}

}
\par{
Therefore, using Eq.s ( \ref{A.46},\ref{A.47}, \ref{A.48}, \ref{A.49}) we calculate the Dirac bracket of the supercharges and the Hamiltonian.
\begin{align} \label{A.50}
\{Q_1, \mathit{H}\}_D & = \{Q_1, \mathit{H}\}_P - \{ Q_1, \phi^A \}_P C_{AB} \{ \phi^B, \mathrm{H} \}_P \nonumber \\
& = \{Q_1, \mathit{H}\}_P - \{ Q_1, \phi^1 \}_P C_{11} \{ \phi^1, \mathit{H} \}_P - \{ Q_1, \phi^2 \}_P C_{22} \{ \phi^2, \mathit{H} \}_P \nonumber \\
& =   \omega p \psi_2 - \omega^2 q \psi_1 - \omega p \psi_2 + \omega^2 q \psi_1.
\end{align}
\begin{align} \label{A.51}
\{Q_2, \mathit{H}\}_D & = \{Q_2, \mathit{H}\}_P - \{ Q_2, \phi^A \}_P C_{AB} \{ \phi^B, \mathit{H} \}_P  \nonumber \\
& = \{Q_2, \mathit{H}\}_P - \{ Q_2, \phi^1 \}_P C_{11} \{ \phi^1, \mathit{H} \}_P - \{ Q_2, \phi^2 \}_P C_{22} \{ \phi^2, \mathit{H} \}_P \nonumber \\
& = - \omega p \psi_1 - \omega^2 q \psi_2 + \omega^2 q \psi_2 + \omega p \psi_1.
\end{align}
Substitute from Eqs. (\ref{A.50}, \ref{A.51})  into equation (\ref{A.45}), thus we find
\begin{align} \label{A.52}
\{Q_\alpha, \mathit{H} \}_D & = \{Q_1, \mathit{H} \}_D + \{Q_2, \mathit{H} \}_D \nonumber \\
& = \left( \omega p \psi_2 - \omega^2 q \psi_1 - \omega p \psi_2 + \omega^2 q \psi_1 \right) + \left( - \omega p \psi_1 - \omega^2 q \psi_2 + \omega^2 q \psi_2 + \omega p \psi_1 \right) \nonumber \\
&= 0.
\end{align}

}
\par{
It is clear from Eqs. (\ref{A.44},\ref{A.52}), that the supercharge (\ref{A.24}) together with the Hamiltonian (\ref{A.22}) and the constraints (\ref{A.23}) satisfy the two conditions (\ref{aa2.73}) of the Dirac bracket superalgebra.}

\end{appendices}
\chapter*{\textbf{Appendix B}}
\addcontentsline{toc}{chapter}{\textbf{Appendix B}}
\begin{appendices}
\renewcommand{\thechapter}{\arabic{chapter}}
\renewcommand{\thesection}{B.\arabic{section}}
\renewcommand{\theequation}{B.\arabic{equation}}
\setcounter{equation}{0}

\section{The First-Order correction} \label{AppendixB.1}

\par{ In this appendix we use conventional perturbation theory to compute the first-order correction to the ground state energy of the supersymmetric harmonic oscillator.  Good discussion of the perturbation theory can be found in \cite{goldstein1965classical,griffiths2005introduction,book:17492,sakurai2011modern,shankar2012principles}. In the non-degenerate time-independent perturbation theory, the first-order correction to the energy of the $n^{th}$' state is given by
\begin{equation} \label{B.1}
\mathit{E}_n^{(1)} = \langle \psi_n^0 \vert \mathit{\hat{H}}' \vert \psi_n^0 \rangle.
\end{equation}
We need now to use this formula to compute the first-order correction to the ground state energy of our system. As we have shown in Section \ref{Sec2.6}, the supersymmetric quantum mechanics Hamiltonian is given in the general form
\begin{equation}
\mathit{\hat{H}} = \frac{1}{2} (\hat{p}^2 + \mathit{W}'^2) \mathbbmtt{I}_2 + \frac{1}{2} \hbar W'' \sigma_3.
\end{equation}  
Let us here consider the superpotential $\mathit{W}(x)$ defined as
\begin{equation}
\mathit{W}= \frac{1}{2} \omega \hat{q}^2 + \mathit{g} \hat{q}^3 \quad \Rightarrow \quad \begin{matrix}  \;\;\; \mathit{W}'  =  \omega \hat{q}+ 3 \mathit{g} \hat{q}^2 \\ \mathit{W}''  = \omega + 6 \mathit{g} \hat{q}, \end{matrix} 
\end{equation}
where $\mathit{g}$ is a perturbation.If $\mathit{g}$ is zero the Hamiltonian $\mathit{\hat{H}}$ reduce to the supersymmetric harmonic oscillator Hamiltonian. Therefore, when $\mathit{g}$ is small, the Hamiltonian $\mathit{\hat{H}}$  can be written as 
\begin{equation}
\mathit{\hat{H}} = \mathit{\hat{H}}^0 + \mathit{\hat{H}}',
\end{equation}
where $\mathit{\hat{H}}^{0}$ is the Hamiltonian of the unperturbed supersymmetric harmonic oscillator and $\mathit{\hat{H}}'$ is the perturbation, 
\begin{equation}
\mathit{\hat{H}}'= (\frac{9}{2} \mathit{g}^2 \hat{q}^4 + 3 \omega \mathit{g} \hat{q}^3) \mathbbmtt{I}_2 + 3 \hbar \mathit{g} \hat{q} \sigma_3.
\end{equation}

}
\par{As shown in Section \ref{Sec2.5}, for the supersymmetric harmonic oscillator, expect the ground energy state, all the energy levels degenerate into two energy states. Therefore, if we have an energy level degenerate to the two energy states $\Psi_n=\left( \begin{matrix} \psi_n^a \\ 0 \end{matrix} \right)$ and $\Psi_n= \left( \begin{matrix} 0\\ \psi_n^b \end{matrix} \right)$, we can write
\begin{equation} 
\langle \psi_n \vert \mathit{\hat{H}}' \vert \psi_m \rangle = \Big\langle\left( \begin{matrix} \psi_n^a \\ \psi_n^b \end{matrix} \right) \Big\vert f_1(\hat{q}) \mathbbmtt{I}_2 + f_2(\hat{q}) \sigma_3 \Big\vert \left( \begin{matrix} \psi_m^a \\ \psi_m^b \end{matrix} \right)  \Big\rangle  = 0 .
\end{equation} 
If we interest in a ground state of the form $\left(\begin{matrix} 0 \\ \psi_0 \end{matrix}\right)$, the previous equation reduces to
\begin{equation}
\langle \psi_0 \vert \mathit{\hat{H}}' \vert \psi_m \rangle = \langle \psi_0 \vert f_1 (\hat{q}) - f_2 (\hat{q}) \vert \psi_m \rangle  = 0.
\end{equation}
Based on this argument, we can see for this combination that the impact of the Hamiltonian $\mathit{\hat{H}}'$ on the wave function $\Psi_0^{(0)}$ is only due to the component,
\begin{equation} \label{B.8}
\mathcal{\hat{H}}'= \frac{9}{2} \mathit{g}^2 \mathit{q}^4 + 3 \omega \mathit{g} \hat{q}^3 - 3 \hbar \mathit{g} \hat{q}.
\end{equation}
Actually, this allows us to compute the first and second corrections to the ground state energy for this system just using the Hamiltonian $\mathcal{\hat{H}}'$. As well, we can use the non-degenerate time-independent perturbation theory to calculate the corrections to the ground state energy of our harmonic oscillator system even it has degenerate energy levels.}

\par{Thus, taking into account the supersymmetric harmonic oscillator ground state wave function in the form (\ref{f2.142}) and the Hamiltonian (\ref{B.8}), then we can now use the formula (\ref{B.1}) to compute the first-order correction to the energy of the ground state as follows
\begin{align}
\mathit{E}_0^{(1)} & = \langle \psi_0^0 \vert \mathcal{\hat{H}}' \vert \psi_0^{(0)} \rangle \nonumber \\
& = \langle \psi_0^{(0)} \vert (\frac{9}{2} \mathit{g}^2 \hat{q}^4 + 3\omega \mathit{g} \hat{q}^3) \mathbbmtt{I}_2 + 3 \hbar \mathit{q} \hat{q} \sigma_3 \vert \psi_0^{(0)} \rangle  \nonumber  \\
&= \int_{-\infty}^{+\infty} \sqrt{\frac{\omega}{\pi \hbar}} e^{- \frac{\omega \hat{q}^2}{\hbar}} \left( \frac{9}{2} \mathit{g}^2 \hat{q}^4 + 3 \omega \mathit{g} \hat{q}^3 - 3 \hbar \mathit{g} \hat{q} \right) d q \nonumber \\
&= \frac{9 \mathit{g}^2}{2} \sqrt{\frac{\omega}{\pi \hbar}} \int_{-\infty}^{+\infty} \hat{q}^4 e^{- \frac{\omega \hat{q}^2}{\hbar}} dq + 3 \omega \mathit{g} \sqrt{\frac{\omega}{\pi \hbar}} \int_{-\infty}^{+\infty} \hat{q}^3 e^{- \frac{\omega \hat{q}^2}{\hbar}} dq  \nonumber \\
&{\quad}- 3 \hbar \mathit{g} \sqrt{\frac{\omega}{\pi \hbar}} \int_{-\infty}^{+\infty} \hat{q} e^{- \frac{\omega \hat{q}^2}{\hbar}} dq.
\end{align}

}
\par{
Both the second term and the third term of the previous integration vanish since the integral functions $\hat{q}^3 e^{- \frac{\omega \hat{q}^2}{\hbar}}$ and $\hat{q} e^{- \frac{\omega \hat{q}^2}{\hbar}}$ are odd functions and there integrations from $- \infty$ to $+ \infty$ are equal to zero. Therefore the previous integration reduces to
\begin{align}\label{B.10}
\mathit{E}_0^{(1)} &=  \frac{9 g^2}{2} \sqrt{\frac{\omega}{\pi \hbar}} \int_{-\infty}^{+\infty} \hat{q}^4 e^{- \frac{\omega \hat{q}^2}{\hbar}} dq = 9 \mathit{g}^2 \sqrt{\frac{\omega}{\pi \hbar}} \int_{0}^{+\infty} e^{- \frac{\omega \hat{q}^2}{\hbar}} \hat{q}^2 dq \nonumber \\ 
&= 9 \mathit{q}^2 \sqrt{\frac{\omega}{\pi \hbar}} \times \frac{3}{8} \sqrt{\frac{\pi \hbar^5}{\omega^5}} = \frac{27}{8}\frac{\hbar^2}{\omega^2} \mathit{g}^2 \simeq \mathcal{O}(g^2).
\end{align}
The last equation implies that the first-order correction to the ground state energy reduces to zero. Notice that to solve the integration which appeared in the last equation, we have used the integral formula
\begin{equation}
\int_{0}^{\infty}  x^{2n} \exp(- a x^2)dx = \frac{1 \cdot 3 \cdots(2n-1)}{2^{n+1}} \sqrt{\frac{\pi}{a^{2n+1}}}.
\end{equation}

}

\section{The Second-Order Correction } \label{AppendixB.2}

\par{As described in \cite{goldstein1965classical,griffiths2005introduction,book:17492,sakurai2011modern,shankar2012principles}, the second-order correction to the energy in the time-independent non-degenerate perturbation theory, is given by the form
\begin{equation} \label{B.12}
\mathit{E}_0^{(2)} = \sum_{m \neq n} \frac{\vert \langle \psi_n^{(0)} \vert \mathcal{\hat{H}}' \vert \psi_m^{(0)} \rangle \vert^2}{\mathit{E}_n^{(0)} - \mathit{E}_m^{(0)}}.
\end{equation}

}
\par{
Furthermore, the harmonic oscillator wave function is orthogonal polynomial, and the harmonic oscillator wave function of degree $n$ can be obtained by the following recursion formula    
\begin{equation} \label{B.13}
\psi_n^{(0)} (\hat{q})= \frac{1}{\sqrt{2^n} n!} \left( \frac{\omega}{\pi \hbar} \right)^{\frac{1}{4}} e^{- \frac{\omega \hat{q}^2}{2 \hbar}} \mathit{\hat{H}}_n \left(\sqrt{\frac{\omega}{\hbar}} \hat{q} \right), 
\end{equation} 
where $\mathit{\hat{H}}_n$ is the Hermite Polynomial of degree $n$. Also, it is helpful to consider the recursion relation of the Hermite Polynomials
\begin{equation} \label{B.14}
\hat{q} \mathit{\hat{H}}_n = \frac{1}{2} \mathit{\hat{H}}_{n+1} + n \mathit{\hat{H}}_{n-1}.
\end{equation}

Moreover, for two wave functions $\psi_n$ and $\psi_m$ we have the relation
\begin{equation} \label{B.15}
\langle \psi_n \vert \psi_m \rangle = \delta_{nm}.
\end{equation} 
Eqs. (\ref{B.13}, \ref{B.14}, \ref{B.15}) imply that
\begin{equation}
\langle \psi_0 \vert q \vert \psi_m \rangle = \frac{1}{2} C_{11} \delta_{0(m+1)} + C_{12} n \delta_{0(m-1)} = C_{12} n \delta_{1m}.
\end{equation}

}

\par{
This argument leads us to know that all the terms of the summation with $m>4$ in Eq. (\ref{B.12}) are vanish. So that to compute the second-order correction to the energy of the ground state using Eq. (\ref{B.12}), we only need to consider the non-zero terms with $m \leq 4$.}
\par{
Now we are able to compute the first term with $m=1$. First considering the ground and the first excited state wave functions
\begin{equation}
\psi_0^{(0)} =(\frac{\omega}{\pi \hbar})^{\frac{1}{4}} e^{- \frac{\omega q^2}{2 \hbar}}, \quad \texttt{and} \quad
\psi_1^{(0)} = (\frac{4\omega^3}{\pi \hbar^3})^{\frac{1}{4}} q e^{- \frac{\omega q^2}{2 \hbar}}.
\end{equation} 
Then,
\begin{align}\label{B.18}
\MoveEqLeft \frac{\vert \langle \psi_0^{(0)} \vert 3\omega \mathit{g} \hat{q}^3 -3 \hbar \mathit{g} \hat{q}  \vert \psi_1^{(0)} \rangle \vert^2}{\mathit{E}_0^{(0)} - \mathit{E}_1^{(0)}} = \frac{\Big\vert 3\omega \mathit{g} \langle \psi_0^{(0)} \vert \hat{q}^3 \vert \psi_1^{(0)} \rangle - 3 \hbar \mathit{g} \langle \psi_0^{(0)} \vert \hat{q} \vert \psi_1^{(0)} \rangle \Big\vert^2}{ - \hbar \omega} \nonumber \\ 
& = \frac{-1}{\hbar \omega} \left[ 6 \omega \hat{g} \sqrt{\frac{2 \omega^2}{\pi \hbar^2}} \int_0^{\infty} \hat{q}^4 e^{\frac{-\omega \hat{q}^2}{\hbar}} dq - 6 \hbar \mathit{g} \sqrt{\frac{2 \omega^2}{\pi \hbar^2}} \int_0^{\infty} \hat{q}^2 e^{\frac{-\omega \hat{q}^2}{\hbar}} dq \right]^2 \nonumber \\ 
& = \frac{-\mathit{g}^2}{\hbar \omega} \left[\frac{9}{4} \sqrt{\frac{2 \hbar^3}{\omega}} - \frac{3}{2} \sqrt{\frac{2 \hbar^3}{\omega}} \right]^2 = - \frac{9}{8}\frac{\hbar^2}{\omega^2} \mathit{g}^2.
\end{align}

}

\par{Similarly, we can compute the second term with $m=2$. The second excited state wave functions is given by the form
\begin{equation}
\psi_2^{(0)}= \left( \frac{\omega}{ 4 \pi \hbar}\right)^{\frac{1}{4}}\left( \frac{ 2 \omega}{\hbar} \hat{q}^2 - 1 \right) e^{- \frac{\omega q^2}{2 \hbar}}.
\end{equation} 
Then,
\begin{align}\label{B.20}
\frac{\vert \langle \psi_0^{(0)} \vert \frac{9}{2} \mathit{g}^2 \hat{q}^4  \vert \psi_2^{(0)} \rangle \vert^2}{\mathit{E}_0^{(0)} - \mathit{E}_2^{(0)}} 
& = \frac{\Big\vert \Big\langle (\frac{\omega}{\pi \hbar})^{\frac{1}{4}} e^{- \frac{\omega \hat{q}^2}{2 \hbar}} \Big\vert \frac{9}{2} \mathit{g}^2 \hat{q}^4 \Big\vert \left(\frac{\omega}{4 \pi \hbar}\right)^{\frac{1}{4}}\left( \frac{2\omega}{\hbar} \hat{q}^2 -1\right) e^{- \frac{\omega \hat{q}^2}{2 \hbar}} \Big\rangle \Big\vert^2}{0 - 2 \hbar \omega} \nonumber \\
&= \frac{-81 \mathit{g}^4}{16 \pi \hbar^2} \left[ \Big\vert \frac{4\omega}{\hbar} \int_{0}^{+\infty} \hat{q}^6 e^{-\frac{\omega \hat{q}^2}{\hbar}} dq -2 \int_{0}^{+\infty} \hat{q}^4 e^{-\frac{\omega \hat{q}^2}{\hbar}} dq \Big\vert^2 \right] \nonumber \\
&= \frac{-81 \mathit{g}^4}{16 \pi \hbar^2} \left( \frac{15}{4} \sqrt{\frac{\pi \hbar^5}{\omega^5}} - \frac{3}{4} \sqrt{\frac{\pi \hbar^5}{\omega^5}}  \right)^2 = - \frac{729}{16} \frac{\hbar^3}{\omega^5} \mathit{g}^4.
\end{align}

}
\par{Furthermore, the same way we can compute the second term with $m=3$. The  third excited state wave functions take the form
\begin{equation}
\psi_3^{(0)}= (\frac{\omega^3}{9 \pi \hbar^3})^{\frac{1}{4}}(2 \frac{\omega}{\hbar} \hat{q}^3-3 \hat{q}) e^{- \frac{\omega \hat{q}^2}{2 \hbar}}.
\end{equation} 
Then,
\begin{align}\label{B.22}
\frac{\vert \langle \psi_0^{(0)} \vert 3 \omega \mathit{g} \hat{q}^3  \vert \psi_3^{(0)} \rangle \vert^2}{\mathit{E}_0^{(0)} - \mathit{E}_3^{(0)}}
& = \frac{\Big\vert \Big\langle (\frac{\omega}{\pi \hbar})^{\frac{1}{4}} e^{- \frac{\omega q^2}{2 \hbar}} \Big\vert 3 \omega gq^3  \Big\vert (\frac{\omega^3}{9 \pi \hbar^3})^{\frac{1}{4}}(2 \frac{\omega}{\hbar} \hat{q}^3-3 \hat{q}) e^{- \frac{\omega \hat{q}^2}{2 \hbar}} \Big\rangle \Big\vert^2}{0 - 3 \hbar \omega} \nonumber \\
& = \frac{- \mathit{g}^2 \omega^3}{\pi \hbar^3} \left[ \Big\vert \Big\langle e^{- \frac{\omega \hat{q}^2}{2 \hbar}} \Big\vert \hat{q}^3 \Big\vert (2 \frac{\omega}{\hbar} \hat{q}^3-3 \hat{q}) e^{- \frac{\omega \hat{q}^2}{2 \hbar}} \Big\rangle \Big\vert^2 \right] \nonumber \\ 
& = \frac{- \mathit{g}^2 \omega^3}{\pi \hbar^3} \left[ \Big\vert \frac{2\omega}{\hbar} \int_{-\infty}^{+\infty} \hat{q}^6 e^{- \frac{\omega \hat{q}^2}{\hbar}} dq - 3 \int_{-\infty}^{+\infty} \hat{q}^4 e^{- \frac{\omega \hat{q}^2}{\hbar}} dq \Big\vert^2 \right] \nonumber \\
&= \frac{- \mathit{g}^2 \omega^3}{\pi \hbar^3} \left[ \Big\vert \frac{4\omega}{\hbar} \int_{0}^{+\infty} \hat{q}^6 e^{- \frac{\omega \hat{q}^2}{\hbar}} dq - 6 \int_{0}^{+\infty} \hat{q}^4 e^{- \frac{\omega \hat{q}^2}{\hbar}} dq \Big\vert^2 \right] \nonumber \\
& = \frac{- \mathit{g}^2 \omega^3}{\pi \hbar^3} \left[ \Big\vert \frac{15}{4} \sqrt{\frac{\pi \hbar^5}{\omega^5}} - \frac{9}{4} \sqrt{\frac{\pi \hbar^5}{\omega^3}} \Big\vert^2 \right] = \frac{- \mathit{g}^2 \omega^3}{\pi \hbar^3} \left(\frac{3}{2} \sqrt{\frac{\pi \hbar^5}{\omega^3}} \right)^2  \nonumber \\
& = - \frac{9}{4}\frac{\hbar^2}{\omega^2} \mathit{g}^2 .
\end{align}

}
\par{
Also similarly we can compute the term with $m=4$. The fourth excited state wave function is
\begin{equation}
\psi_4^{(0)}= \left(\frac{\omega}{576 \pi \hbar}\right)^{\frac{1}{4}}\left( \frac{4\omega^2}{\hbar^2} \hat{q}^4 -\frac{12\omega}{\hbar} \hat{q}^2 +3\right) e^{- \frac{\omega \hat{q}^2}{2 \hbar}}.
\end{equation} 
Then,
\begin{align}\label{B.24}
\MoveEqLeft \frac{\vert \langle \psi_0^{(0)} \vert \frac{9}{2} \mathit{g}^2 \hat{q}^4  \vert \psi_4^{(0)} \rangle \vert^2}{\mathit{E}_0^{(0)} - \mathit{E}_4^{(0)}}
 = \frac{\Big\vert \Big\langle (\frac{\omega}{\pi \hbar})^{\frac{1}{4}} e^{- \frac{\omega \hat{q}^2}{2 \hbar}} \Big\vert \frac{9}{2} \mathit{g}^2 \hat{q}^4 \Big\vert \left(\frac{\omega}{576 \pi \hbar}\right)^{\frac{1}{4}}\left( \frac{4\omega^2}{\hbar^2} \hat{q}^4 -\frac{12\omega}{\hbar} \hat{q}^2 +3\right) e^{- \frac{\omega \hat{q}^2}{2 \hbar}} \Big\rangle \Big\vert^2}{0 - 4 \hbar \omega} \nonumber \\
& = \frac{-27 \mathit{g}^4}{128 \pi \hbar^2} \left[ \Big\vert \Big\langle e^{-\frac{\omega \hat{q}^2}{2\hbar}} \Big\vert \hat{q}^4 \Big\vert \left( \frac{4\omega^2}{\hbar^2} \hat{q}^4 -\frac{12\omega}{\hbar} \hat{q}^2 +3\right) e^{- \frac{\omega \hat{q}^2}{2 \hbar}} \Big\rangle \Big\vert^2 \right] \nonumber \\ 
& = \frac{-27 \mathit{g}^4}{128 \pi \hbar^2} \left[ \Big\vert \frac{4\omega^2}{\hbar^2} \int_{-\infty}^{+\infty} \hat{q}^8 e^{-\frac{\omega \hat{q}^2}{\hbar}} dq -\frac{12\omega}{\hbar} \int_{-\infty}^{+\infty} \hat{q}^6 e^{-\frac{\omega \hat{q}^2}{\hbar}} dq +3 \int_{-\infty}^{+\infty} \hat{q}^4 e^{-\frac{\omega \hat{q}^2}{\hbar}} dq \Big\vert^2 \right] \nonumber \\
& = \frac{-27 \mathit{g}^4}{32 \pi \hbar^2} \left[ \Big\vert \frac{4\omega^2}{\hbar^2} \int_{0}^{+\infty} \hat{q}^8 e^{-\frac{\omega \hat{q}^2}{\hbar}} dq -\frac{12\omega}{\hbar} \int_{0}^{+\infty} \hat{q}^6 e^{-\frac{\omega \hat{q}^2}{\hbar}} dq +3 \int_{0}^{+\infty} \hat{q}^4 e^{-\frac{\omega \hat{q}^2}{\hbar}} dq \Big\vert^2 \right] \nonumber \\
& = \frac{-27 \mathit{g}^4}{32 \pi \hbar^2} \left[ \Big\vert\frac{105}{8} \sqrt{\frac{\pi \hbar^5}{\omega^5}} - \frac{45}{4} \sqrt{\frac{\pi \hbar^5}{\omega^5}} + \frac{9}{8} \sqrt{\frac{\pi \hbar^5}{\omega^5}} \Big\vert^2 \right] = \frac{-27 \mathit{g}^4}{32 \pi \hbar^2} \left[ \Big\vert 3 \sqrt{\frac{\pi \hbar^5}{\omega^5}} \Big\vert^2 \right] \nonumber \\
& = - \frac{243}{32} \frac{\hbar^3}{\omega^5} \mathit{g}^4.
\end{align}

}

\par{
Using Eqs. (\ref{B.18}, \ref{B.20}, \ref{B.22}, \ref{B.24}), we are able to compute the second-order correction to the ground state energy for the supersymmetric harmonic oscillator as following

\begin{align} \label{B.25}
\mathit{E}_2^{(0)} & = \frac{\vert \langle \psi_0^{(0)} \vert - 3 \hbar \mathit{g} \hat{q}  \vert \psi_1^{(0)} \rangle \vert^2}{\mathit{E}_0^{(0)} - \mathit{E}_1^{(0)}} + \frac{\vert \langle \psi_0^{(0)} \vert \frac{9}{2} \mathit{g}^2 \hat{q}^4  \vert \psi_2^{(0)} \rangle \vert^2}{\mathit{E}_0^{(0)} - \mathit{E}_2^{(0)}} + \frac{\vert \langle \psi_0^{(0)} \vert 3 \omega \mathit{g}  \hat{q}^3  \vert \psi_3^{(0)} \rangle \vert^2}{\mathit{E}_0^{(0)} - \mathit{E}_3^{(0)}} \nonumber \\
& \quad +  \frac{\vert \langle \psi_0^{(0)} \vert \frac{9}{2} \mathit{g}^2 \hat{q}^4  \vert \psi_4^{(0)} \rangle \vert^2}{\mathit{E}_0^{(0)} - \mathit{E}_4^{(0)}} \nonumber \\
& = - \frac{9}{8}\frac{\hbar^2}{\omega^2} \mathit{g}^2 - \frac{729}{16} \frac{\hbar^3}{\omega^5} \mathit{g}^4 - \frac{9}{4}\frac{\hbar^2}{\omega^2} \mathit{g}^2  - \frac{243}{32} \frac{\hbar^3}{\omega^5} \mathit{g}^4 \nonumber \\
&= - \frac{27}{8}\frac{\hbar^2}{\omega^2} \mathit{g}^2  - \frac{1701}{32} \frac{\hbar^3}{\omega^5} \mathit{g}^4.
\end{align}    

}
\par{
Finally, from Eqs. (\ref{B.10}) and (\ref{B.25}), we realize that up to the second-order the energy corrections to the ground state energy of the supersymmetric quantum mechanical harmonic oscillator vanish. This is concluded as
\begin{align}
\mathcal{O}(g)&=0, \nonumber \\
\mathcal{O}(g^2) &= \frac{27}{8}\frac{\hbar^2}{\omega^2} - \frac{27}{8}\frac{\hbar^2}{\omega^2}=0.
\end{align}

}

\end{appendices}

\chapter*{\textbf{Appendix C}}
\addcontentsline{toc}{chapter}{\textbf{Appendix C}}
\begin{appendices}
\renewcommand{\thechapter}{\arabic{chapter}}
\renewcommand{\thesection}{C.\arabic{section}}
\renewcommand{\theequation}{C.\arabic{equation}}
\setcounter{equation}{0}

\section{The Double-Well Potential Eigenvalue Problem} \label{AppendixC.1}
\par{
Consider the same problem with a symmetric double-well potential, which we have discussed in section (\ref{Sec3.3}). Taking into account Eqs. (\ref{3.38}, \ref{3.39}), then we rewrite the potential in the form
\begin{equation}
V(x) = \frac{\omega^2}{8 a^2} \left( x^2- a^2 \right)^2.
\end{equation}
It is simple to get the first and second derivative of $V(x)$ respect to $x$, they are
\begin{align}
V'(x) &=  \frac{\omega^2}{2 a^2} x\left( x^2-a^2 \right), \nonumber \\
V''(x) &= \frac{\omega^2}{2 a^2} \left( 3 x^2-a^2 \right).
\end{align}
For instance, recall the classical solution Eq. (\ref{3.50})
\begin{equation} \label{C.3}
X= \pm a \tanh \left( \frac{\omega}{2} (\tau-\tau_c) \right).
\end{equation}
Therefore, using Eq. (\ref{C.3}) we are able to find the explicit form for the second derivative of the potential, as follows 
\begin{align}
V''(X) &= \frac{\omega^2}{2 a^2} \left(3 a^2 \tanh^2\left( \frac{\omega}{2} (\tau-\tau_c) \right) - a^2 \right) \nonumber \\
&= \frac{\omega^2}{2} \left( 3 \left[ 1 - \frac{1}{\cosh^2  \frac{\omega}{2} (\tau -\tau_2)} \right] - 1 \right) \nonumber \\
&= \omega^2 \left( 1 - \frac{3}{2 \cosh^2  \frac{\omega}{2} (\tau -\tau_2)}  \right).
\end{align}
We have used the hyperbolic identity
\begin{equation}
\tanh^2 (x) + \sech^2(x)= 1 \quad \& \quad \sech(x) = \frac{1}{\cosh(x)}.
\end{equation}

}
\section{Derivation of Equation (\ref{3.68})} \label{AppendixC.2}
\par{
Recall Eqs. (\ref{3.66}, \ref{3.67})
\begin{equation} \label{C.6}
\dfrac{d}{d \xi} (1-\xi^2) \dfrac{d}{d \xi} x_n + \left( a + \frac{b}{1-\xi^2} \right) x_n = 0,
\end{equation}
\begin{equation} \label{C.7}
x = \left(1-\xi^2\right)^c \chi.
\end{equation}
For instance, substituting Eq. (\ref{C.7}) into Eq. (\ref{C.6}), then we get
\begin{equation*}
\dfrac{d}{d \xi} (1-\xi^2)  \dfrac{d}{d \xi} \left(1-\xi^2\right)^c \chi + \left( a + \frac{b}{1-\xi^2} \right) \left(1-\xi^2\right)^c \chi =0.
\end{equation*}
Solving the second derivative in the first factor 
\begin{equation*}
\dfrac{d}{d \xi} (1-\xi^2) \left( (-2 \xi c) (1-\xi^2)^{c-1} \chi + ( 1- \xi^2)^c \frac{d \chi}{ d \xi} \right) + \left( a + \frac{b}{1-\xi^2} \right) (1-\xi^2)^c \chi =0.
\end{equation*}
This gives us
\begin{equation*}
\dfrac{d}{d \xi}  \left( (-2 \xi c) (1-\xi^2)^c \chi + ( 1- \xi^2)^{c+1} \frac{d \chi}{ d \xi} \right) + \left( a + \frac{b}{1-\xi^2} \right) (1-\xi^2)^c \chi =0.
\end{equation*}
Subsequently, solving the derivative $\dfrac{d}{d \xi}$
\begin{align*}
& -2c (1-\xi^2)^c \chi + 4 \xi^2 c^2 (1-\xi^2)^{c-1} \chi -2\xi c (1-\xi^2)^c \dfrac{d \chi}{d \xi}  - 2\xi (c+1) (1-\xi^2)^c \dfrac{d \chi}{d \xi} \\ &+ (1-\xi^2)^{c+1} \dfrac{d^2 \chi}{d \xi^2} + \left( a+ \frac{b}{1-\xi^2} \right) (1-\xi^2)^c \chi =0.
\end{align*}
This equation can be simplified to give us
\begin{align*}
-2c \chi + \frac{ 4 \xi^2 c^2}{1-\xi^2} \chi - -2 c \xi  \frac{d \chi}{d \xi} - (2 \xi c + 2 \xi) \dfrac{d \chi}{d \xi} + (1- \xi^2) \dfrac{d^2 \chi}{d \xi^2} + \left(a+ \frac{b}{1-\xi^2}  \right) \chi &=0.
\end{align*}
Furthermore
\begin{align}
(1-\xi^2)  \dfrac{d^2 \chi}{d \xi^2} - (4c+2) \xi \dfrac{d \chi}{d \xi} + \left( a-2c+\frac{4c^2 \xi^2}{1-\xi^2}+\frac{b}{1-\xi^2} \right) \chi &=0\nonumber \\
(1-\xi^2) \dfrac{d^2 \chi}{d \xi^2} - (4c+2) \xi \dfrac{d \chi}{d \xi} + \left( a-2c-4c^2+ \frac{b+4c^2}{1-\xi^2} \right) \chi&=0.
\end{align}
This is Eq. (\ref{3.68}). }
\end{appendices} 
\renewcommand{\bibname}{\textbf{References}}
\nocite{*}
\bibliographystyle{unsrt}
\bibliography{References}
\addcontentsline{toc}{chapter}{\textbf{References}}
\end{document}